\documentstyle [aps,prd,preprint]{revtex}
\tighten
\begin{document}
\preprint{CLNS 97/1481}
\title{Inflaton Decay and Heavy Particle Production with Negative Coupling}

\author{Brian R. Greene\footnote{On leave from: F.\ R.\ Newman
         Laboratory of Nuclear Studies, Cornell University, Ithaca, NY
         14853.  E-mail: greene@math.columbia.edu}}
         
\address{Departments of Physics and Mathematics, Columbia University,
         New York, NY 10027}

\author{Tomislav Prokopec\footnote{E-mail:
         tomislav@mail.lns.cornell.edu} and Thomas
         G. Roos\footnote{E-mail: roost@mail.lns.cornell.edu}}
         
\address{F.\ R.\ Newman Laboratory of Nuclear Studies, Cornell
         University, Ithaca, NY 14853} 

\date{May 19, 1997} 

\maketitle

\begin{abstract}
We study the decay of the inflaton in a general ${\cal Z}2\times {\cal
Z} 2$ symmetric two scalar theory. Since the dynamics of the system is
dominated by states with large occupation numbers which admit a
semiclassical description, the decay can be studied by solving the
classical equations of motion on the lattice. Of particular interest
is the case when the cross-coupling between the inflaton and the
second scalar field is negative, which is naturally allowed in many
realistic models. While the inflaton decays {\em via} parametric
resonance in the positive coupling case we find that for negative
coupling there is a new mechanism of particle production which we call
{\em negative coupling instability}. Due to this new mechanism the
variances of the fields grow significantly larger before the
production is shut off by the backreaction of the created particles,
which could have important consequences for symmetry restoration by
nonthermal phase transitions. We also find that heavy particles are
produced much more efficiently with negative coupling, which is of
prime importance for GUT baryogenesis. Using a simple toy model for
baryogenesis and the results of our lattice simulations we show that
for natural values of the cross-coupling enough $10^{14}\hbox{GeV}$
bosons are created to produce a baryon to entropy ratio consistent
with observation. This is to be contrasted with the situation for
positive coupling, where the value of the cross-coupling required to
produce such massive particles is unnaturally large. In addition to
our numerical results we obtain analytical estimates for the maximum
variances of the fields in an expanding universe for all cases of
interest: massive and massless inflaton, positive and negative
cross-coupling, with and without significant self interactions for the
second field.
\end{abstract}



\section{Introduction}

It has recently been realized
\cite{TraschenBrandenberger,KofmanLindeStarobinskii} that an
oscillating scalar field can transfer its energy efficiently to other
bosonic degrees of freedom \cite{dolgovkirilova} via a process
analogous to the classical phenomenon of parametric resonance. The
most important application of this mechanism is to reheating after
inflation. When the inflationary stage terminates the universe is
essentially devoid of matter and the inflaton field oscillates with
large amplitude about the minimum of its potential. Its coupling to
other bosonic fields can then lead to explosive particle production
via the non-perturbative resonance mechanism. This is sometimes called {\it
preheating} in the literature because the produced particles are far
from thermal equilibrium. The simplest model for studying this
phenomenon is a theory with two real scalar fields and effective
potential
\begin{equation}
V(\phi,\chi) = \frac{1}{2} m^2_\phi\phi^{2} + 
\frac{1}{2}m^2_\chi\chi^{2}
+\frac{\lambda_{\phi}}{4}\phi^{4} +
\frac{\lambda_{\chi}}{4}\chi^{4} + \frac{g}{2}
\phi^{2}\chi^{2}.
\label{eq:V}
\end{equation}
Such models exhibit chaotic inflation when one of the fields, called
the {\it inflaton}, acquires a large expectation value (compared to the
Planck mass) \cite{chaotic}.  The energy density of the universe
then becomes dominated by the potential energy of the inflaton and the
universe starts inflating. This continues until the amplitude of the
inflaton expectation value becomes of order the Planck mass. At this
point the inflationary stage terminates and the inflaton starts oscillating.

An intuitive picture of the parametric resonance mechanism emerges by
assuming that the inflation field $\phi$ oscillates sinusoidally about
its minimum \cite{oscillation}.  With $\phi(\vec x,t)\rightarrow\Phi_0\cos
(\omega_\phi^0 t)$ and neglecting the $\lambda_\chi$ term, the mode
equations for the $\chi$ field can be written as the Mathieu equation
\begin{eqnarray}
\frac{d^2\chi_k}{dz^2} &+& \bigl [A_0(k)-2q_0\cos(2 z)\bigr 
]\chi_k=0\,,
\label{eq:mathieu}\\
A_0(k) &=& \frac{\omega_\chi(k)^2}{(\omega_\phi^0)^2}
+2q_0\,,\qquad
q_0=\frac{g\Phi_0^2}{4(\omega_\phi^0)^2} 
\label{eq:A0,q0}
\end{eqnarray}
where $z=\omega_\phi^0 t$ and $\omega_\chi(k)^2=k^2+m^2_\chi$ is the
frequency squared of $\chi_k$.  It is well known that the Mathieu
equation possesses unstable solutions for which the modes grow as
$\chi_k\propto \exp [\mu_k\omega_\phi^0 t]$. This corresponds to
exponentially growing occupation numbers $n_\chi(k)\propto \exp
[2\mu_k\omega_\phi^0 t]$ and is interpreted as particle production.
The most important features of the solutions to Eq.~(\ref{eq:mathieu})
are readily understood from the stability chart depicted in
figures~1(a) and~(b). The $\mu=0$ curves divide the chart into stable
(dark) and unstable (light) regions. Some curves of constant positive
$\mu$ are also shown.  For clarity figure~1(a) contains only a limited
range of $A_0$ and $q_0$, although the physically interesting values
are $|q_0|\gg 1$ \cite{KofmanLindeStarobinskii,ProkopecRoos}. The
$q_0\gg 1$ regime is called ``broad resonance'' in the
literature. Figure~1(b) shows a piece of the instability chart for
large $|q_0|$ around the line $A_0=2q_0$.  Notice that, if one makes
the usual assumption $g>0$, then for $m^2_\chi\ge 0$ the allowed $A_0$
lie above the line $A_0=2q_0$ ({\it cf.\/} Eq.~(\ref{eq:A0,q0})).  As
a consequence the regions of large $\mu$ are excluded, and the maximum
value $\mu$ reaches is $\lesssim 1$, attained for $A_0\approx 2q_0$
and $q_0\gtrsim 1$.  Above $A_0=2q_0$, $\mu$ is a rapidly decreasing
function of $A_0$.  Since the separation between neighboring
instability bands for large $q_0$ is of order $q_0^{1/2}$, this
implies that the dominant production of $\chi$ particles occurs for
$k\sim k_{\rm res}\simeq q_0^{1/4}\omega_\phi^0$.  It is important to
note that the inflaton also decays into its own fluctuations {\it
via\/} parametric resonance \cite{KofmanLindeStarobinskii}. However,
the value of $\mu$ corresponding to this decay channel is typically
a few times smaller than the value of $\mu$ for decay into $\chi$
fluctuations as long as $q_0$ is large. Hence, unless the decay into
$\chi$ particles is somehow shut off, the decay into $\phi$ particles
is subdominant.

While the above picture is simple and intuitive, in reality the
situation is much more complex.  First of all, the equations of
motion (EOM), and hence the dynamics, are strongly modified in an
expanding universe. In particular, the amplitude of the inflaton oscillations
decreases even in the absence of particle production.  Secondly, as the
inflaton decays, the backreaction of created particles alters the
parameters in the Mathieu equation. The amplitude decreases as energy is
drained away and the masses get contributions of the form $\delta
m_\phi^2=g\langle(\delta\chi)^2\rangle
+3\lambda_\phi\langle(\delta\phi)^2\rangle$, $\delta
m_\chi^2=g\langle(\delta\phi)^2\rangle
+3\lambda_\chi\langle(\delta\chi)^2\rangle $, with
$\langle(\delta\phi)^2\rangle=\langle\phi^2\rangle-
\langle\phi\rangle^2$ and
$\langle(\delta\chi)^2\rangle=\langle\chi^2\rangle-\langle\chi
\rangle^2$ denoting the variances of the fields. One can still gain insight
into the decay by replacing the initial values $A_0$
and $q_0$ in Eq.~(\ref{eq:mathieu}) by time dependent parameters 
\begin{equation}
A(k) = \frac{k^2+m^2_{\chi\; \rm eff}}{\omega_{\phi\; \rm eff}^2}
+2q\,,\qquad
q=\frac{g\Phi^2}{4\omega_{\phi\; \rm eff}^2} 
\label{eq:A,q}
\end{equation}
where
\begin{eqnarray}
m^2_{\chi\;\rm eff} &=& m^2_{\chi} + 
g\langle(\delta\phi)^2\rangle
+3\lambda_\chi\langle(\delta\chi)^2\rangle\,,
\nonumber\\
\omega^2_{\phi\; \rm eff} &=& \omega^2_{\phi}+
g\langle(\delta\chi)^2\rangle
+3\lambda_\phi\langle(\delta\phi)^2\rangle\,.
\label{eq: mass and frequency}
\end{eqnarray}
Here $\Phi$ is the slowly varying amplitude of the oscillating
inflaton expectation value $\phi_0(t)\equiv \langle\phi(\vec
x,t)\rangle$, and $\omega_{\phi}$ is the frequency of $\phi_0(t)$ in
the absence of backreaction. We distinguish $\omega_{\phi}$ from its
initial value $\omega_{\phi}^0$ because due to the $\phi^4$ term in
Eq.~(\ref{eq:V}) the frequency is time dependent even without decay in
an expanding universe. The reason it is useful to think about the
decay in terms of the parameters defined in Eq.~(\ref{eq: mass and
frequency}) is that the basic features of the resonance mechanism are
extremely robust. The amplification of the modes depends only on the
fact that one is dealing with oscillators with a time dependent
frequency, and in what follows we will only make use of the most
prominent features of the stability chart when obtaining analytical
estimates. For large $q$ these are independent of the details of the
potential and the exact time dependence of the parameters. The
description in terms of an equation of the form of
Eq.~(\ref{eq:mathieu}) finally breaks down completely when the
resonant mode amplitudes grow large.  For then the nonlinear terms in
the EOM become important and scatterings become fast. The result is a
slowly evolving ``scattering regime'' characterized by smooth power
spectra
\cite{KhlebnikovTkachevI,ProkopecRoos,KhlebnikovTkachevIII}. The
resonant decay of the inflaton has been treated in various
approximations by many authors
\cite{TraschenBrandenberger,KofmanLindeStarobinskii,Shtanov Traschen
Brandenberger,KhlebnikovTkachevII,allother,yoshiandco,boyanovsky}.
The full nonlinear problem has been studied in detail using lattice
simulations. The idea is that one can evolve the system using the
classical equations of motion because the dynamics is dominated by
states with large occupation numbers which admit a semiclassical
description. This approach was pioneered in \cite{KhlebnikovTkachevI},
where it was applied to the decay of a massless field into its own
fluctuations. The decay into a second massless field with and without
self-coupling was treated in \cite{ProkopecRoos}, and the decay of a
massive inflaton was discussed in \cite{KhlebnikovTkachevIII}.

Our main interest in this paper is to investigate what happens when
the coupling $g$ in Eq.~(\ref{eq:V}) is negative, which has not been
previously studied.  The point is that parameters like $g$ that couple
two or more scalar fields appear in almost all extensions of the
standard model, and that there is generally no reason whatsoever that
these parameters should be positive. Of course the effective potential
should be bounded from below in order to ensure the existence of a
sensible vacuum. For the potential given in Eq.~(\ref{eq:V}) this
requires
\begin{equation}
r\equiv \frac{\lambda_\phi\lambda_\chi}{g^2}>1\,.
\label{eq: r}
\end{equation}
We will assume that this stability bound is satisfied.

Naively, as one can see from figure~1(a), the negative $g$ case is
dramatically different from the positive $g$ case. The resonant
momenta are now limited to the region above the line $A_0=-2|q_0|$,
instead of above the line $A_0=2|q_0|$.  When $2\vert q_0\vert\ge
\vert A_0\vert \gg 1 $ \cite{Strutt}
\begin{eqnarray}
\cosh{2\pi\mu} &=& \cosh
\Im\left[ \int_0^{2\pi} dz 
(A_0-2q_0\cos(2z))^{\frac{1}{2}} \right]
\nonumber\\
&\times& \cos \Re\left[ \int_0^{2\pi} dz 
(A_0-2q_0\cos(2z))^{\frac{1}{2}} \right]\; .
\label{eq: mu for large A q}
\end{eqnarray}
%
Along $A_0=-2|q_0|$ this evaluates to $\mu=(4/\pi)|q_0|^{1/2}$ (see also
\cite{negg mu}). For large $|q_0|$ this is easily orders of magnitude
larger than the corresponding positive $g$ value, which is always
$\lesssim 1$.  Since the decay time is proportional to $\mu^{-1}$ one
might hence expect that particle production and consequently inflaton decay
is much faster for negative $g$.

This conclusion, however, is premature since we have implicitly
assumed that initially the expectation value $\chi_0\equiv
\langle\chi(\vec x,t)\rangle=0$.  To see whether this is the case one
has to examine the EOM for the fields during inflation. Note that the
large expectation value $\phi_0(t)$ of the inflaton induces a
$\phi_0$-dependent minimum in the potential for the second field:
\begin{equation}
\tilde\chi_0^2 =  \cases{{-m_\chi^2-g\phi_0^2 \over\lambda_\chi} 
            & \hbox{for} $\quad m_\chi^2+g\phi_0^2<0$ \cr 
0 & \hbox{otherwise}} 
\label{eq: chi_0}
\end{equation}
It is not hard to see analytically that if $\chi_0\approx\tilde\chi_0$
initially, then it remains so during inflation, as long as
$|g|/\lambda_\phi \gg 1/2$ and $m_\chi$ is not too large\cite{why-not
chi-not}.  Our numerical simulations show that even if $\chi_0$ is not
close to $\tilde\chi_0$ initially, it will become so by the end of
inflation, provided the above conditions are satisfied.  Since the
physically important broad resonance case with $\lambda_\phi \ll |g|$
is our main concern, this means that the correct initial condition for
the oscillatory regime is given by Eq.~(\ref{eq: chi_0}).

To illustrate the effect of $\chi_0\neq 0$, consider the linearized
mode equation for $\delta\chi_k$, the Fourier transform of
$\chi-\chi_0$. Neglecting the expansion of the Universe we obtain
\begin{equation}
\frac{d^2\delta\chi_k}{d^2t}+
\biggl(k^2+m_{\chi}^2+3\lambda_\chi\chi_0^2+
g\phi_0^2\biggr)\delta\chi_k+
2g\phi_0\chi_0\delta\phi_k = 0
\label{eq: correct chi mode equation}
\end{equation}
where $\phi_0$ is the oscillating expectation value of the inflaton
field. Note that {\it if\/} $\chi_0\approx\tilde\chi_0$ even during
the oscillatory regime, then the term in parentheses of (\ref{eq:
correct chi mode equation}) is positive ({\it cf.\/} (\ref{eq:
chi_0})).  This would mean that the evolution of $\delta\chi_k$
resembles that of a positive $g$ case.  Neglecting the last term for
the moment, we see that in the simple case when $m_\chi=0$ and
$\chi_0=\tilde\chi_0$, Eq.~(\ref{eq: correct chi mode equation})
reduces to Eq.~(\ref{eq:mathieu}) with
$A_0=(k^2/(\omega_\phi^0)^2)+2q_0$ and
$q_0=2|g|\Phi_0^2/(2\omega_\phi^0)^2$. In the case $m_\chi\neq 0$ there is
no such simple correspondence, but the effective $q$ is still
positive. Note that the effect of the last term in Eq.~(\ref{eq: correct
chi mode equation}) is that it produces resonant growth of
$\delta\chi_k$ when the amplitude of $\delta\phi_k$ is large and
$\chi_0\neq 0$.  Since most of the time $\delta\chi_k\gg\delta\phi_k$
for the resonant modes, the effect of this term will be
small. The analogous term in the mode equation for $\delta\phi_k$
speeds up its growth when $\delta\chi_k$ is large.

We have just seen that if $\chi_0$ follows its instantaneous
equilibrium value $\tilde\chi_0$ the fact that $g$ is negative should
not make much difference. The situation changes drastically if
$\chi_0$ is not ``enslaved'' by $\phi_0$ in this way: there is then a
portion of the period for which the term in parentheses of (\ref{eq:
correct chi mode equation}) is negative. During this time the
$\delta\chi_k$ are unstable and are expected to grow exponentially
with large $\mu$, just as in the naive negative $g$ picture discussed
initially ({\it cf.\/} also \cite{negg mu}).  The physics hence
depends crucially on the detailed dynamics of the field expectation
values. This dynamics is very complicated due to the interactions with
the nonzero modes, especially once their amplitudes become large due
to the resonance. To fully investigate the negative $g$ case one must
hence resort to lattice simulations, and this is the subject we turn
to next. 

In section~\ref{sec:massless} we will discuss massless fields,
followed by a detailed study of massive fields in
section~\ref{sec:massive}. The relevance of our results for grand
unified theory (GUT) baryogenesis are discussed in
section~\ref{sec:baryogenesis}. Finally, our conclusions are presented
in section~\ref{sec:conclusion}. 

Before diving into the analysis we should point out that most of our
numerical calculations and analytical estimates will be concerned with
obtaining the maximum variances attained by the fields during
preheating. There are two reasons for focusing on these quantities:
first, the maximum variances determine the symmetry restoration
``power'' relevant to the occurrence of so-called nonthermal phase
transitions \cite{nonthermal,KasuyaKawasaki}. Second, for very massive
particles such as the bosons needed for GUT baryogenesis the variance
is directly proportional to the energy and particle number
densities. This will be discussed in more detail in
sections~\ref{sec:massive} and \ref{sec:conclusion}.

\section{Massless Fields}
\label{sec:massless}

\subsection{Numerical Results}
\label{sec:masslessI}
We begin our investigation by studying massless fields, which is
convenient for two reasons.  First, the positive $g$ massless case has
been studied extensively on the lattice in \cite{ProkopecRoos},
allowing for a detailed comparison of the positive and negative $g$
cases.  Second, massless fields ($m_\chi=m_\phi=0$) are the simplest
to treat numerically because in this case the EOM in a
Friedman-Robertson-Walker (FRW) universe are conformally equivalent to
those in Minkowski space. By this we mean the following: in terms of
the variables $\tau=\int dt/a(t)$, $\bar{\phi}=\phi a(\tau)/a(0)$, and
$\bar{\chi}=\chi a(\tau)/a(0)$, where $a(\tau)$ is the scale factor,
the expanding universe EOM reduce to those of two interacting scalar
fields in static space-time
\cite{KhlebnikovTkachevI,ProkopecRoos}. This is true provided the
energy density behaves as radiation dominated, which is indeed the
case for a massless inflaton in the oscillatory regime \cite{Shtanov
Traschen Brandenberger}. Since it is less time consuming to do
simulations in Minkowski space our numerics for the massless fields
were done in terms of the barred variables defined above, and our
lattice time was $\tau$. The conversion of lattice quantities to
physical quantities will be discussed in detail below.

The computational techniques involved are described in detail in
\cite{ProkopecRoos}. In short we evolve the classical EOM on a three
dimensional lattice starting from certain random initial condition for
the inhomogeneous modes \cite{ic}. The initial conditions for the zero
modes (field expectation values) are given by the values at the
breakdown of the slow roll inflationary stage: $\Phi_0 = \Phi(t=0)\sim
M_{\rm P}$, where $M_{\rm P}\equiv (8\pi G)^{-1/2} \simeq 2.4\times
10^{18}\hbox{GeV}$ is the reduced Planck mass, and $\chi_0 =
\tilde{\chi_0}$ as in Eq.~(\ref{eq: chi_0}).  The results presented in
this section were obtained on $128^3$ lattices.

We begin by comparing the dynamics of two theories with opposite signs
of $g$ but otherwise identical parameters. The chosen values are
$\lambda_\phi=10^{-12}$, $\lambda_\chi=10^{-7}$, and $|g|=10^{-10}$.
For a massless inflaton the zero mode $\phi_0$ oscillates with
frequency $\omega_{\phi}=c\lambda_\phi^{1/2}\Phi$, where $c\approx
0.85$. Hence these couplings correspond to $q_0\simeq 35$ and $r=10$
({\it cf.\/} Eqs.~(\ref{eq:A,q}), (\ref{eq: r})). The evolution of the
expectation values $\phi_0$ and $\chi_0$ for the negative and positive
$g$ cases are shown in figures~2 and~3, respectively. The
corresponding variances are shown in figures~4 and~5. Since the energy
density in the inflaton zeromode is proportional to $\Phi^4$ we see
from figure~2 that in the negative $g$ case 40\% of the initial energy
is in fluctuations by the time $\tau\approx
3\times10^8\Phi_0^{-1}$. Notice also that $\chi_0$ follows $\phi_0$
faithfully ({\it i.e.\/} $\chi_0\approx\tilde\chi_0$) until
$\tau\approx 8\times10^7\Phi_0^{-1}$. After this the two expectation
values are no longer in phase, even though the dynamics of $\chi_0$ is
clearly still strongly influenced by $\phi_0$. Figure~3 shows that it
takes until $\tau\approx 4.5\times 10^8\Phi_0^{-1}$ for 40\% of the
energy to decay in the positive $g$ case, and in contrast to the
situation for $g<0$ here almost all of the decayed energy goes into
$\phi$ fluctuations. After $\tau\approx 4.5\times 10^8\Phi_0^{-1}$ the
decay rate becomes very small, a fact that is also reflected in the
variances (figure\/~5), which are both changing very slowly beyond
this point. Similarly we see in figure~4 that for negative $g$ the
variances level off at $\tau\sim 9\times 10^7\Phi_0^{-1}$, although
the $\phi$ variance is still growing slowly until $\tau\sim 2.2\times
10^8\Phi_0^{-1}$. We will explain all of these features, and give
estimates for the magnitudes of the relevant quantities, in
section~\ref{sec:masslessII}.

Before doing so let us briefly discuss the conversion of lattice
quantities to physical quantities. As discussed at the beginning of
this section, the lattice fields are rescaled by a factor $a(\tau)$,
where $\tau$ is the conformal time (for convenience we define
$t=\tau=0$ to correspond to the beginning of our simulation (i.e.\ the
end of slow roll inflation) and choose $a(0)=1$). This means that one
must simply divide lattice quantities such as energy densities and
variances by the appropriate powers of $a(\tau)$ to get the physical
values. For example, $\langle\delta\chi^2\rangle =
\langle\delta\bar{\chi}^2\rangle/a(\tau)^2$.  In the  oscillatory 
regime of a massless inflaton the scale factor behaves as in a
radiation dominated universe, so we can write $a(\tau)=(1+H_0\tau)$
where $H_0=(\rho(0)/3)^{1/2}\, M_{\rm P}^{-1}$ is the Hubble constant
at the end of inflation. It is easy to show that at the end of slow
roll $\Phi_0\approx 1.8M_{\rm P}$ and the kinetic energy density is
about half the potential energy density \cite{rho}. Using the fact
that $\rho_{\rm PE}=\lambda_\phi\Phi_0^4/4$ for $g>0$ and $\rho_{\rm
PE}=(1-1/r)\lambda_\phi\Phi_0^4/4$ for $g<0$ we thus obtain $H_0
\approx \sqrt\lambda_\phi M_{\rm P}$ for $g>0$ and $H_0 \approx
(\lambda_\phi (1-1/r))^{1/2}M_{\rm P}$ for $g<0$.

With these estimates, and recalling that our lattice time is $\tau$,
it is simple to convert from lattice to physical quantities. This was
done in obtaining figure~6(b) from figure~6(a): figure~6(a) shows the
maximum lattice variances for negative $g$ as a function of $q_0$ for
two values of $r$. The corresponding physical variances are shown in
figure~6(b). Since the variances are fluctuating significantly (see,
for example, figure~4), both peak and valley values are given.  These
curves, too, will be explained in section~\ref{sec:masslessII}. Here we simply
note that when the lattice variances behave as shown in figures~4 and
5 it is easy to obtain the maximum physical variances: since we must
divide by $a(\tau)^2$ to obtain the latter from the former, the
maximum occurs at the time when the slowly varying state sets
in. After that the slow increase cannot keep up with the decrease due
to the expansion.

\subsection{Discussion and Analytical Estimates}
\label{sec:masslessII}

The most striking feature of figures~2--5 is the fact that after an
initial period of rapid exponential decay the system reaches a slowly
evolving state during which the rate of energy transfer from the
inflaton zeromode to the fluctuations is quite small. This effect has
been studied in detail in \cite{ProkopecRoos}, where it is shown that
for large $q_0$ or $r$ backreaction and scattering limits the maximum
amplitudes of the variances and essentially shuts down the
resonance. Throughout this paper we will refer to the slowly evolving
state as the {\it scattering regime} \cite{large q_0}. The variances
of the physical fields at the time the scattering regime begins are of
central interest since they are the maximum variances of the fields
reached during preheating. The reason that the decay of the inflaton
shuts off for large $q_0$ or $r$ while most of the energy is still in
the oscillating zeromode is the backreaction of the created particles
on the EOM. In particular the effective masses of the fields get
contributions of the form $\delta
m_\phi^2=g\langle(\delta\chi)^2\rangle
+3\lambda_\phi\langle(\delta\phi)^2\rangle$, $\delta
m_\chi^2=g\langle(\delta\phi)^2\rangle
+3\lambda_\chi\langle(\delta\chi)^2\rangle $. As was explained in the
introduction, the instability index $\mu_k$ is a rapidly decreasing
function of $A$ above the line $A=2|q|$ \cite{mu-decay} (see also
figure~1).  In fact the decay due to the second band above this line
is already very slow. Now the backreaction terms add a contribution
$\delta A = \delta m_\chi^2 / \omega_\phi^2$ to $A$, and for large $q$
the distance between the first two bands above $A=2\vert q\vert $ is
about $|q|^{1/2}$ ({\it cf.\/} figure~1(b)). Hence when the shift in
$A$ becomes so large that
\begin{equation}
A(k=0)+\delta A \ge  2|q|+{|q|}^{\frac{1}{2}}
\label{eq: delta A}
\end{equation}
the first band above $A=2\vert q \vert$ is rendered ineffective and
the decay slows down dramatically \cite{decaystop}. From this
condition one can obtain estimates of the variances at
the beginning of the slowly varying scattering regime. As explained in
\cite{ProkopecRoos}, the (lattice) variances are then kept roughly
constant at these maximum values by a feedback mechanism. It is
important to point out that scattering plays a crucial part in the
shut off mechanism. For example, we will see below that when
$\lambda_\chi$ is small the resonance is shut off by the
$g\langle(\delta\phi)^2\rangle$ term. This term grows large due to
scattering of resonant $\chi$ particles off the $\phi$ zeromode, an
effect that is completely lost in the Hartree type approximations
often used to study preheating. The fact that the $\phi$ variance is
responsible for shutting off the resonance for small $\lambda_\chi$ was
first realized in \cite{ProkopecRoos}.

Before deriving estimates for the variances based on Eq.~(\ref{eq:
delta A}) we wish to point out that there is always a substantial
``intrinsic'' uncertainty associated with such estimates. By this we
mean that physical quantities such as maximum variances, energies,
{\it et cetera\/}, are extremely sensitive functions of the parameters
in the model.
For example, we have seen above that for large $q$ a shift of order
$\vert q\vert ^{1/2}$ is enough to move the initial position of the
first resonant momentum above $A=2|q|$ from $k^2\approx 0$ to
$k^2\approx \vert q\vert ^{1/2}\omega_\phi^2$. Since $\mu_k$ is a
rapidly decreasing function of $k$ \cite{mu-decay} such a shift can
have a substantial effect on the early evolution of the system.  The
upshot is that for a small change $\delta g/g=\delta q/q\sim
|q|^{-1/2}\ll 1$, the time at which the scattering regime sets in
changes by a factor of a few. As a consequence the maximum physical
variances reached vary by about an order of magnitude. In this context
we point out that the maximum lattice variances reached are much less
sensitive to small changes in the parameters. The reason is simply
that the lattice variances do not decrease due to the expansion, so 
the time it takes to reach the scattering regime is immaterial. This
observation explains why the curves in figure~6(a) are relatively
smooth compared to those in figure~6(b). The sensitivity to small
changes in parameter values has also been observed in
\cite{KhlebnikovTkachevII}.

We will now estimate the maximum variances in the massless case based
on Eq.~(\ref{eq: delta A}).  Positive and negative coupling -- {\bf
Cases I\/} and {\bf II} -- will be treated separately. As explained
above, the maximum physical variances are reached at the beginning of
the slowly evolving scattering regime. Since the variances fluctuate
we distinguish between peak values and valley values. Estimates
obtained from Eq.~(\ref{eq: delta A}) correspond to peak values since
Eq.~(\ref{eq: delta A}) is the condition for shutting off the
resonance. This condition must be met during those parts of the period
of the inflaton zero mode when the resonance is active, but in between
the variances can drop to smaller (valley) values. The formulae below
will be given in terms of the quantity $\Phi_{\rm s}$, which is the
amplitude of oscillations at the time the maximum variances are
reached\cite{t max var}. We will discuss how to estimate $\Phi_{\rm
s}$ after deriving the expressions for the variances. Note that
Eq.~(\ref{eq: delta A}) involves the time dependent parameter $q$
rather than the initial value $q_0$.  Recalling from section
\ref{sec:masslessI} that in the massless case
$\omega_{\phi}=c\lambda_\phi^{1/2}\Phi$, with $c\approx 0.85$, a
glance at Eqs.~(\ref{eq:A,q}) and (\ref{eq: mass and frequency})
reveals that both the numerator and the denominator of $q$ scale as
$a(t)^{-2}$. Hence, while $q$ may change significantly due to the
decay of the inflaton, it does {\it not} change due to the expansion
in the massless case.

{\bf Case I.\/} For $g>0$, Eq.~(\ref{eq: delta A}) reduces to
\begin{equation}
\delta A\equiv \frac{m^2_{\chi\;\rm eff}}{\omega^2_{\phi\; \rm 
eff}}
\simeq q^{\frac 12}\; ,
\label{eq: delta A II}
\end{equation}
with $m^2_{\chi\;\rm eff}$ and $\omega^2_{\phi\; \rm eff}$ given by
Eq.~(\ref{eq: mass and frequency}).  When the self-coupling
$\lambda_\chi$ is small, in the sense that the
$3\lambda_\chi\langle(\delta\chi)^2\rangle$ term in $m^2_{\chi\;\rm
eff}$ may be neglected compared to $g\langle(\delta\phi^2)\rangle$,
Eq.~(\ref{eq: delta A II}) yields an estimate for the $\phi$ field
variance:
\begin{equation}
\langle(\delta\phi)^2\rangle_{\rm peak} 
\simeq\frac{1}{2}g^{-\frac{1}{2}}\Phi_{\rm s}\omega_{\phi\;\rm eff} 
=\frac{1}{4}\, q^{-\frac{1}{2}}\Phi_{\rm s}^2\,.
\label{eq:variance II}
\end{equation}
This can be turned into an estimate for $\langle(\delta\chi)^2\rangle$
by assuming approximate equipartition of energy between the
fluctuations of the two fields.  In \cite{ProkopecRoos} we have
argued that this is reasonable because the transfer of
energy between the two fields is efficient in the scattering
regime. We thus obtain
\begin{equation}
\langle(\delta\chi)^2\rangle_{\rm peak}\simeq
\frac{\omega_\phi^2}{4g}=\frac{1}{16}\, q^{-1}\Phi_{\rm s}^2 \, .
\label{eq:variance III}
\end{equation}
Our original assumption that $\lambda_\chi$ was small can be
translated into the constraint 
\begin{equation}
\langle(\delta\chi)^2\rangle <
\frac{1}{48 c^2} \, q^{-\frac{3}{2}} r^{-1} \Phi_{\rm s}^2\;.
\end{equation}
Hence Eqs.~(\ref{eq:variance II}) and (\ref{eq:variance III}) are
valid provided that $r<(3c^2q^{1/2})^{-1}$. This condition 
defines what is
meant by ``small $\lambda_\chi$''. Note that it translates roughly
into $\lambda_\chi < g^{3/2}/\lambda_\phi^{1/2}$. None of the runs
presented in this paper satisfy this condition, and the above
estimates are included here only for completeness. The ``small
$\lambda_\chi$'' case was treated in detail in \cite{ProkopecRoos}.

For ``large $\lambda_\chi$'', {\it i.e.\/} $rq^{1/2}>(3c^2)^{-1}$,
Eq.~(\ref{eq: delta A II}) immediately yields an estimate for the
$\chi$ field variance:
\begin{eqnarray}
\langle(\delta\chi)^2\rangle_{\rm peak} & \simeq & 
\frac{1}{6}\frac{g^{\frac{1}{2}}\Phi_{\rm s}
\omega_{\phi}}{\lambda_\chi}
\label{eq:variance I} \\
& = & \frac{1}{48 c^2} q^{-\frac{3}{2}}r^{-1}\Phi_{\rm s}^2\, .
\label{eq:variance Ib}
\end{eqnarray}
%
This is obtained using $\omega^2_{\phi\;\rm eff} \approx
\omega^2_{\phi}$, which is true because here $r>(12 c^2
q^{1/2})^{-1}$.  The run shown in figures~3 and~5 belongs in the large
$\lambda_\chi$, positive $g$ category. To see how well
Eq.~(\ref{eq:variance Ib}) works in this case note that we can replace
$(\delta\chi)^2$ with the lattice quantity $(\delta{\bar \chi})^2$ on
the left hand side if we do the same with $\Phi_{\rm s}$ on the right
hand side. From figure~3, ${\bar \Phi}_{\rm s} \approx \Phi_0$, so
Eq.~(\ref{eq:variance Ib}) predicts $\langle(\delta{\bar
\chi})^2\rangle_{\rm peak} = 1.4\times10^{-5}\Phi_0^2$. From figure~5
we see that this is in good agreement with the first peak
$\langle(\delta\bar\chi)^2\rangle_{\rm peak} \simeq 1\times
10^{-5}\Phi_0^2$ at $\tau\approx 0.65\times 10^8 \Phi_0^{-1}$. To test
the $q_0$ dependence in Eq.~(\ref{eq:variance Ib}) we ran our
code for several values of $q_0$ ranging from $3.5$ to $1000$. A
power law fit to the data gives $\langle(\delta\chi)^2\rangle_{\rm
peak} \propto q_0^{-1.41}$, in reasonable agreement with the
prediction.

Figure~5 also illustrates nicely that the inflaton decays into its own
fluctuations via parametric resonance, as discussed in the
introduction. Once the $\chi$ resonance is shut off this usually
subdominant process becomes important and the $\phi$ variance
continues to grow. After $\tau = 2\times 10^8 \Phi_0^{-1}$ the growth
is clearly exponential, and during this stage the inflaton decays
essentially as if it were not coupled to the $\chi$ field at all. The
decay finally stops when the $\phi$ field variance reaches its
scattering value $\langle(\delta\phi)^2\rangle_{\rm peak} \sim
10^{-1}\Phi^2(\tau^\phi_{\rm s})\sim 10^{-6}M_{\rm P}^2$ \cite{phi
variance}. This is discussed in more detail in
\cite{ProkopecRoos}, as is the amount of energy in the $\chi$
fluctuations at the beginning of scattering regime
\cite{energy in flucts}.

\bigskip

{\bf Case II.\/} When $g<0$, Eq.~(\ref{eq: delta A}) reduces to
\begin{equation}
\delta A\equiv \frac{m^2_{\chi\;\rm eff}}{\omega^2_{\phi\; \rm 
eff}}
\simeq 4\vert q\vert
\,.\label{eq: delta A III}
\end{equation}
Since we must now allow for the possibility that $\chi_0$ has an
appreciable amplitude, Eq.~(\ref{eq: mass and frequency}) for
$m^2_{\chi\;\rm eff}$ and $\omega^2_{\phi\; \rm eff}$ should really be
modified to
\begin{eqnarray}
m^2_{\chi\;\rm eff} &= & 
g\langle(\delta\phi)^2\rangle
+3\lambda_\chi\left(\langle(\delta\chi)^2\rangle+
\chi_0^2\right)\,,
\nonumber\\
\omega^2_{\phi\; \rm eff} &=& \omega^2_{\phi}+
g\left(\langle(\delta\chi)^2\rangle+\chi_0^2\right)
+3\lambda_\phi\langle(\delta\phi)^2\rangle
\,.
\label{eq: mass and frequency II}
\end{eqnarray}
In this analysis, however, we set $\chi_0$ to {\it zero\/}. The reason
is simply that we are interested in obtaining an estimate of the {\it
peak} variance. Once $\langle(\delta\chi)^2\rangle$ becomes
appreciable $\chi_0$ no longer follows $\tilde\chi_0$ of Eq.~(\ref{eq:
chi_0}) \cite{chi 0}, and there are certainly times when $\chi_0$ is
small ({\it cf.\/} figure~2). These times are precisely the times
during which $\langle(\delta\chi)^2\rangle$ reaches its maximum
value, as can be seen from Eqs.~(\ref{eq: delta A III}) and  (\ref{eq:
mass and frequency II}), as well as figure~4. 

With $\chi_0=0$ Eq.~(\ref{eq: delta A III}) simply says
\begin{equation}
m^2_{\chi\;\rm eff} \simeq \vert g\vert \Phi_{\rm s}^2
\,,\label{eq: delta A IV}
\end{equation}
independent on $\omega_\phi$. Assuming that
$\langle(\delta\chi)^2\rangle$ dominates $m^2_{\chi\;\rm eff}$
\cite{x^2 dominates} we obtain
\begin{equation}
\langle(\delta\chi)^2\rangle_{\rm peak}\simeq
\frac{|g|}{3\lambda_\chi}\Phi_{\rm s}^2=
\frac{1}{12c^2}\, \vert q\vert^{-1}r^{-1}\Phi_{\rm s}^2\,.\qquad
\label{eq:variance IV}
\end{equation}
This is a factor $4\vert q\vert ^{1/2}$ times larger than the
corresponding positive $g$ value Eq.~(\ref{eq:variance Ib}).  The
difference between positive and negative $g$ is nicely illustrated in
figure~1. In order to shut off the fast exponential growth in the
$g>0$ case, it is sufficient that the first instability band above
$A=2|q|$ is rendered ineffective, {\it i.e.\/} $\delta A\sim
q^{1/2}$. In the negative $g$ case, on the other hand, the instability
is shut off only when $A$ is shifted all the way from $A=-2|q|$ to above
the line $A=2\vert q\vert $, {\it i.e.\/} $\delta A\sim 4\vert
q\vert$.  After $\langle(\delta\chi)^2\rangle_{\rm peak}$ is reached,
$\langle(\delta\phi)^2\rangle$ wants to grow to its one field
scattering value $\simeq 0.1\Phi^2(\tau^\phi_{\rm s})$\cite{phi
variance}, just as in the positive $g$ case. However, due to the large
$\chi$ occupation numbers scatterings become fast before the $\phi$
variance can grow this large, and it is cut off at a somewhat smaller
value. If one assumes that, because of the efficient scattering,
equipartition is reasonably well satisfied, one obtains the estimate
\cite{phi chi equipartition}
\begin{equation}
\langle(\delta\phi)^2\rangle_{\rm peak}\simeq
\frac{|g|^{\frac{3}{2}}}{3\lambda_\chi}
\frac{\Phi_{\rm s}^3}{\omega_{\phi\;\rm eff}}=
\frac{1}{6c^2}\,\vert q\vert^{-\frac{1}{2}}r^{-1}\Phi_{\rm s}^2
\,.
\label{eq:variance V}
\end{equation}
Here we have taken the appropriate value of $\Phi$ to be $\Phi_{\rm
s}$ even though the $\phi$ variance really peaks somewhat later than
the $\chi$ variance. To get an order of magnitude estimate the
difference should not matter as long as scatterings terminate the
$\phi$ resonance soon after $\langle(\delta\chi)^2\rangle$ reaches its
maximum. While our simulations indicate that this formula holds
fairly well for moderate values of $q$ ($\lesssim 100$) we are at
present unsure how well it extrapolates to very large $q$ or $r$. The
reason is that we cannot reliably capture both the $\phi$ and $\chi$
resonances simultaneously if $q$ is too large \cite{phi chi
resonance}. We should point out that we expect Eq.~(\ref{eq:variance
V}) to fail for very large $r$ ({\it i.e.\/} large $\lambda_\chi$). The
reason is simply that in this case the $\chi$ resonance shuts off for
relatively small occupation numbers, so that $\phi \leftrightarrow \chi$
scatterings should not affect the $\phi$ resonance very much, and
hence $\langle(\delta\phi)^2\rangle$ should reach its one field value. 

Note that the variances in Eqs.~(\ref{eq:variance IV})
and~(\ref{eq:variance V}) do not significantly change $\omega_\phi$ at
the time the scattering regime is reached since
$g\langle(\delta\chi)^2\rangle_{\rm peak},
3\lambda_\phi\langle(\delta\phi)^2\rangle_{\rm peak} <
\omega^2_{\phi}$ is satisfied for $\vert q \vert, r >1$.  We can now
compare the predictions of Eqs.~(\ref{eq:variance IV})
and~(\ref{eq:variance V}) with the numerical results shown in
figure~4. Recall that we can turn the above estimates for physical
variances into estimates for lattice variables by replacing the fields
on both sides of the equation with the barred lattice fields. Then
noting that $\bar\Phi_{\rm s}\approx\Phi_0$ we read from figure~4:
$\langle(\delta\bar\chi)^2\rangle_{\rm peak} \simeq 1\times
10^{-3}\Phi^2_0$ and $\langle(\delta\bar\phi)^2\rangle_{\rm peak}
\simeq 5\times 10^{-3}\Phi^2_0$.  Eqs.~(\ref{eq:variance IV})
and~(\ref{eq:variance V}) predict
$\langle(\delta\bar\chi)^2\rangle_{\rm peak} =0.33\times
10^{-3}\Phi^2_0$ and $\langle(\delta\bar\phi)^2\rangle_{\rm peak}
=4\times 10^{-3}\Phi^2_0$, so that the agreement is quite good.

We will now address the question how to estimate the quantity
$\Phi_{\rm s}$ in terms of which our expressions for the variances,
Eqs.~(\ref{eq:variance II}), (\ref{eq:variance III}),
(\ref{eq:variance I}), (\ref{eq:variance Ib}), (\ref{eq:variance IV}), and (\ref{eq:variance
V}), are given. Recall that $\Phi_{\rm s}$ is the amplitude of $\phi_0$ at
the time when the scattering regime is reached and the variances peak
in the expanding universe. The first thing to note is that in the {\it
absence} of expansion the amplitude of the $\phi$ zeromode does not
decrease significantly by the time the scattering regime is
reached. This can be observed in figures~2 and 3. The reason is simply
that the energy density is proportional to $\Phi^4$, so that even a
substantial loss of energy corresponds to a rather small change in
amplitude. To obtain an estimate for $\Phi_{\rm s}$ we can thus take the
decrease to be due to the expansion alone. Then
\begin{equation}
\Phi_{\rm s} = \frac{\Phi_0}{a(\tau_{\rm s})}\,,
\label{eq:phi*}
\end{equation}
where $\tau_{\rm s}$ is the beginning of the
scattering regime. As already pointed out at the beginning of this
section, $\tau_{\rm s}$ depends sensitively on the initial position of the
resonance and is consequently difficult to  estimate accurately. Since
the occupation numbers grow as $n=n_0\exp(2\mu\omega_\phi^0\tau)$ we find 
\begin{equation}
\tau_{\rm s}=\frac{1}{2\mu\omega_\phi^0}\ln\left(\frac{n_{\rm scatt}}{n_0}\right)\;,
\label{eq:tau}
\end{equation}
where $n_{\rm scatt}$ is the occupation number of the resonant modes
at the beginning of the scattering regime.  Our initial conditions
correspond to ``$n_0$''$\sim 1/2$ \cite{ProkopecRoos}, and $n_{\rm
scatt}$ is roughly $g^{-1}$ for the small $\lambda_\chi$, $g>0$ case,
and $10/\lambda_\chi$ for the large $\lambda_\chi$, $g>0$ and $g<0$
cases \cite{n_scatt}. The difference arises because in the first case
$\langle(\delta\phi)^2\rangle$ is responsible for shutting off the
resonance while in the latter two cases the growth of
$\langle(\delta\chi)^2\rangle$ terminates the exponential regime (see
also \cite{ProkopecRoos}). This leaves us with the question what value
to take for $\mu$. From figure~1 we see that $\mu\sim0.2$ is
reasonable for the positive $g$ case, where we are limited to the
region above $A=2|q|$. This is also correct in the negative $g$ case:
it is not hard to show that $\chi_0\approx\tilde\chi_0$ until
$\langle(\delta\chi)^2\rangle \sim (12c^2r|q|^{3/2})^{-1}\Phi^2$
(\cite{stop follow}), which is already larger than the positive $g$
value given in Eq.~(\ref{eq:variance Ib}). As discussed in the
introduction, the negative $g$ case is equivalent to positive $g$ as
long as $\chi_0\approx\tilde\chi_0$, so during this initial period the
positive $g$ value for $\mu$ is appropriate. Once $\chi_0$ stops
following $\tilde\chi_0$ the variance rises rapidly to its negative
$g$ peak, in a time interval negligible compared to the already
elapsed time. Hence we can estimate $\tau_{\rm s}$ using the positive
$g$ value for $\mu$ \cite{tau massless}. For example, for the runs in
figures~2--5 we obtain $\tau_{\rm s}\sim 1\times 10^{8}\Phi_0^{-1}$,
in reasonable agreement with the numerical results.
Eqs.~(\ref{eq:phi*}) and (\ref{eq:tau}), together with the formula
$a(\tau)=(1+H_0\tau)\approx (1+[\lambda_\phi(1-1/r)]^{1/2}M_{\rm
P}\tau)$ given in section~\ref{sec:masslessI}, complete our estimate
of $\Phi_{\rm s}$.

The final task of this section will be to explain the evolution of the
variances in more detail.  A closer look at figures~4 and~5 reveals a
rather complex behavior of the variances.  The first prominent feature
is a slow modulation of the maximum $\phi$ variance.
A simple explanation of this phenomenon can be gotten by assuming that
the infrared (IR) modes oscillate in phase and can be modeled by one
oscillator of a definite frequency $\omega^2=k^2+\omega_{\phi\;{\rm
eff}}^2$.  For $k\ll\omega_{\phi\;{\rm eff}}$ we then have two weakly
coupled oscillators, the IR mode and the zero mode, with the same
natural frequency. These modes will transfer energy back and forth
much like two pendulums coupled by a spring. As can be seen in
figures~3 and~5, the maxima of the variance envelope correspond to
minima of the zeromode envelope, and vice versa.  Note that no such
modulation occurs for the $\chi$ field since no dominant $\chi$
zeromode develops in the scattering regime.

The second prominent feature of the variances is that in the
scattering regime they fluctuate between ``peak'' and ``valley''
values with frequency $2\omega_{\phi\;{\rm eff}}$.  Understanding this
behavior is potentially important for applications were precise values
of the variances are required, such as nonthermal phase transitions
and baryogenesis.  A first estimate of the amplitude of these
fluctuations has been given in Ref.~\cite{KhlebnikovTkachevIII}. Here
we will present an alternative derivation, leading to somewhat
different results.  We will attempt to explain the origin of the
variance {\it fluctuations\/} using a simple toy model which contains
some generic features of the resonant growth, and also teaches us
something about the slowly varying scattering regime. We will
illustrate our model on the $\chi$ field and first treat the case
$g>0$.  After neglecting the nonlinear (scattering) terms, the
equation of motion for a field mode $X=\chi_k$ can be written as
\begin{equation}
\ddot X + \omega_X^2 X=0\,,
\label{eq: toy model for X}
\end{equation}
where
\begin{equation}
\omega_X^2 =
k^2+g\phi_0^2(t)+g\langle (\delta\phi)^2\rangle
+3\lambda_\chi\langle 
(\delta\chi)^2\rangle\,.
\label{eq:freq toy}
\end{equation}
Since we are interested in IR modes that contribute significantly to
the variance we can neglect the momentum dependence of the
frequency. Using Eq.~(\ref{eq: delta A II}) we then find that in the
scattering regime $\omega_X^2$ varies between $\omega_{\rm
max}^2\simeq g \Phi^2 \simeq 4 q\omega_\phi^2$ and $\omega_{\rm
min}^2\simeq 3\lambda_\chi\langle (\delta\chi)^2\rangle +g\langle
(\delta\phi)^2\rangle \sim q^{1/2}\omega_\phi^2$.  Dropping $k^2$ in
Eq.~(\ref{eq:freq toy}) hence amounts to assuming that the variance is
dominated by modes with $k^2\le k_{\rm res}^2\sim 
q^{1/2}\omega_\phi$.  Consider now the solution to Eq.~(\ref{eq: toy
model for X}) if $\omega_X=\omega_0$ for $t<0$ and $\omega_X=\omega_1$
for $t>0$:
\begin{eqnarray}
X &=& X_0\cos\omega_0t+
\frac{\dot X_0}{\omega_0}
\sin\omega_0t\,,\quad
\hbox{\rm for}\quad t<0
\nonumber\\
X &=& X_0\cos\omega_1t+
 \frac{\dot X_0}{\omega_1}
\sin\omega_1t\,,\quad 
\;\hbox{\rm for}\quad t>0\,.
\label{eq: solution to X}
\end{eqnarray}
The important feature of Eq.~(\ref{eq: solution to X}) is that for
$\omega_0>\omega_1$ the amplitude of the "kinetic term" $\dot
X_0/\omega_1$ at $t>0$ is amplified by $\omega_0/\omega_1$ in
comparison to the kinetic term at $t<0$.  This amplification captures
the essence of the amplitude growth mechanism, and it may be used to
explain the variance fluctuations in the scattering regime, as
follows. Consider a sequence of $N$ matchings as in Eq.~(\ref{eq:
solution to X}), with $\omega_n/\omega_{n+1}={\rm
e}^\epsilon$. Assuming that at each matching the solution of the
previous time interval has a random phase \cite{random phase} one
obtains an amplification $X_N/X_0 = [(1+{\rm
e}^{2\epsilon})/2]^{N/2}$. Using the fact that $\omega_0/\omega_N
\equiv \omega_{\rm max}/\omega_{\rm min}={\rm e}^{N\epsilon}$ we find
for $\epsilon$ small that $X_N/X_0 \approx (\omega_{\rm
max}/\omega_{\rm min})^{1/2}$.  Since these are the modes that
dominate the variance, this translates into $\langle(\delta
\chi)^2\rangle_{\rm peak}/\langle(\delta \chi)^2\rangle_{\rm
valley}\sim X_N^2/{X_0^2} \sim \omega_{\rm max}/{\omega_{\rm
min}}$. Plugging in the values for the frequencies obtained below
Eq.~(\ref{eq:freq toy}) yields
\begin{equation}
\frac{\langle(\delta \chi)^2\rangle_{\rm peak}} {\langle(\delta
\chi)^2\rangle_{\rm valley}}\sim
\frac{\omega_{\chi\;\rm max}}{\omega_{\chi\rm\; min}}\approx 
q^{\frac{1}{4}}\qquad (g>0)\,.
\label{eq:variance growth}
\end{equation} 
This estimate agrees well with figure~5 after
$\tau\approx4.5\times10^8\Phi_0^2$. Before that the oscillations are
somewhat larger. Eq.~(\ref{eq:variance growth}) also agrees well with
the large $q$ runs presented in \cite{ProkopecRoos}.

For negative $g$ the fluctuations in the variance are of different
origin. For one thing, the peaks in $\langle(\delta\chi)^2\rangle$ now
occur when $\phi_0$ is {\it maximum}, in contrast to the positive $g$
case just discussed. As already pointed out following
Eq.~(\ref{eq: mass and frequency II}), the largest peaks occur when
$\chi_0 \approx 0$ and $\phi_0$  is large. The reason for this is
simply that when $\chi_0$ gets out of phase with $\tilde \chi_0$, the
negative $g\phi_0^2$ contribution to the $\chi$ mass tends to
destabilize the system. To compensate, the positive
$3\lambda_\chi\langle(\delta\chi)^2\rangle_{\rm peak}$ term must acquire a
magnitude $3\lambda_\chi\tilde \chi_0^2$, since, as shown in the
introduction, this is the value required to stabilize the mode
equations. We have argued above that $\chi_0$ falls out of phase
with $\tilde \chi_0$ when $\langle(\delta\chi)^2\rangle \sim
(12c^2r|q|^{3/2})^{-1}\Phi^2$ \cite{stop follow}. For example, in
figure~2 this occurs at $\tau\approx 9\times10^{7}\Phi_0^{-1}$. Figure~4
shows that the largest variance is reached a quarter of an oscillation
later, when $\phi_0$ reaches its maximum. During this first
half-period after $\chi_0$ stops following $\tilde\chi_0$ there is
genuine instability, causing the variance to grow. At later times the IR
modes have sufficient amplitude to compensate the negative mass term
without growing exponentially: they simply get pulled by $\phi_0$ and
keep the effective mass positive at all times. When $\phi_0$ passes through
zero the variance returns to its previous ``valley'' value.

We can estimate the ratio of peak to valley variances by again
appealing to our toy model of matching oscillatory solutions with
different frequencies, as in Eq.~(\ref{eq: solution to X}). There are,
however, significant differences between the case at hand and the
positive $g$ case discussed earlier. For negative $g$ the maximum
frequency of an infrared $\chi$ mode that contributes significantly to
the variance occurs when $\phi_0$ passes through zero and is 
approximately given by
$\omega^2_{\rm max} \simeq
3\lambda_\chi\langle(\delta\chi)^2\rangle_{\rm valley}$. This
frequency holds only for a small fraction of the oscillation: as soon
as $|g|\phi_0^2 \geq 3\lambda_\chi\langle(\delta\chi)^2\rangle_{\rm
valley}$ the $\chi$ mode gets dragged along with the $\phi$ zeromode
and its frequency changes to $\omega_{\rm min} \simeq \omega_{\phi\;
{\rm eff}}$. Hence there is no continuous change in frequency as in
the positive $g$ case, and consequently only one matching of solutions
{\it $\grave{a}$ la\/} Eq.~(\ref{eq: solution to X}). The single
matching gives $X_{\rm max}/X_{\rm min} \approx \omega_{\rm
max}/\omega_{\rm min}$, where we assumed that after the matching the
kinetic term dominates. This is the case unless $\dot X_0/\omega_0\ll
X_0$ in Eq.~(\ref{eq: solution to X}). Next we use
$\langle(\delta\chi)^2\rangle_{\rm peak}\sim\tilde \chi_0^2\approx
|g|\Phi^2/\lambda_\chi$ to obtain
\begin{equation}
\frac{\langle(\delta \chi)^2\rangle_{\rm peak}} {\langle(\delta
\chi)^2\rangle_{\rm valley}}\sim
\left(\frac{\omega_{\chi\;\rm max}}{\omega_{\chi\rm\; min}}\right)^2
\approx 
|q|^{\frac{1}{2}} \qquad (g<0)\,.
\label{eq: peak-valley II}
\end{equation} 
This prediction can be compared to the numerical results presented in
figure~6. As explained previously, the maximum lattice variances are
much smoother functions of the parameters than the physical variances,
and we will thus concentrate on figure~6(a). Note that even though we
have presented $\vert q_0\vert=0.35$ runs in figure~6, these runs just
barely reach the scattering regime, especially for $r=2$ ({\it cf.\/}
\cite{large q_0}).  After discarding the $\vert q\vert=0.35$ runs we
obtain as a best fit $\langle(\delta \bar\chi)^2\rangle_{\rm peak}
\simeq 0.16 |q|^{-0.92}r^{-0.82}\bar\Phi^2$ and
$\langle(\delta\bar\chi )^2\rangle_{\rm valley} \simeq 0.1
|q|^{-1.46}r^{-0.77}\bar\Phi^2$, and hence $\langle(\delta
\bar\chi)^2\rangle_{\rm peak}/ \langle(\delta \bar\chi)^2\rangle_{\rm
valley}\simeq 1.6 |q|^{0.54}r^{-0.05}$.  This agrees reasonably well
with Eqs.~(\ref{eq:variance IV}) and~(\ref{eq: peak-valley II}).  Note
that the slopes in figure~6(a) are increasing with $|q|$, which may
indicate that at $|q_0|=3.5$ the variances still have not reached the
full scattering regime values.  This could be the explanation of the
somewhat low slopes quoted above. Unfortunately we cannot test our
estimates reliably for $|q_0|>350$ since this would require enormous
computing resources.  The problem is essentially that as $g$ and
$\lambda_\chi$ get large the particles scatter readily into high
momentum states and one needs an enormous ultraviolet cutoff in order
to  accommodate  the decayed energy. This must be combined with
good IR resolution in order not to miss scatterings such as
$\chi(k_{\rm res})\phi(k=0)\rightarrow\chi(k)\phi(k)$ which involve
very IR momenta. The upshot is that we have found negative $g$
simulations with $|q_0|\gtrsim 500$ to be unreliable even on $128^3$
lattices (in the massless case).

\section{Massive Fields}
\label{sec:massive}

\subsection{Numerical Results}
\label{sec:massiveI}

For massive fields the expansion of the universe cannot be taken into
account by a simple rescaling as was done for massless fields in
section~\ref{sec:masslessI}. The equations of motion in a FRW universe
are 
\begin{eqnarray}
\frac{\partial^2\phi(\vec x,t)}{\partial
t^2}+3H\frac{\partial\phi(\vec x,t)}{\partial t} +
\left[m_\phi^2-\frac{\nabla^2}{a(t)^2}+g\chi^2(\vec
x,t)+\lambda_\phi\phi^2(\vec x,t)\right]\phi(\vec x,t) &=& 0 
\,,\nonumber \\
\frac{\partial^2\chi(\vec x,t)}{\partial
t^2}+3H\frac{\partial\chi(\vec x,t)}{\partial t} +
\left[m_\chi^2-\frac{\nabla^2}{a(t)^2}+g\phi^2(\vec
x,t)+\lambda_\chi\chi^2(\vec x,t)\right]\chi(\vec x,t) &=& 0\,,
\end{eqnarray}
where 
\begin{equation}
H(t) = \frac{\dot a(t)}{a(t)} = \sqrt\frac{\rho}{3M_{\rm P}^2}\,,
\end{equation}
and $\rho$ is the total energy density.  This set of partial
differential equations was solved without approximation on the
lattice. The initial conditions for the fields were chosen as follows:
the $\phi$ zeromode $\phi_0$ was set equal to $2M_{\rm P}$. As
explained in section~\ref{sec:masslessI}, this corresponds to a time
slightly before the end of inflation. $\chi_0$ was set equal to 
$\tilde\chi_0$ as given in Eq.~(\ref{eq: chi_0}), and the velocities
$\dot\phi_0$ and $\dot\chi_0$ were set equal to their slow roll
values. The initial conditions for the inhomogeneous modes were chosen
as described in  \cite{ProkopecRoos} and \cite{ic} .

Here we will present a brief survey of our main numerical results for
massive fields. In section~\ref{sec:massiveII} below we will discuss
the figures in detail and give analytical estimates for various
quantities of interest. We should mention that throughout this paper
we have chosen parameters that are ``realistic'' within the context of
chaotic inflation. In particular we use $\Phi_0 \sim M_{\rm P}$,
$\lambda_\phi \sim 10^{-12}$, and $m_\phi\sim 10^{13}$GeV. The 
latter two values are determined by the observed anisotropies in the
microwave background, as measured by the cosmic background explorer
(COBE) \cite{cobe}. 

All of the numerical results presented are for $g<0$.  In figure~7(a)
we show the maximum $\chi$ variances produced in the expanding
universe for a massless inflaton as a function of $m_\chi$ for various
values of $|q_0|$. Figure~7(b) is a blow-up of the $|q_0|=350$
curve. The important feature to observe is the extremely {\it spiky\/}
nature of the curve, with huge jumps in production for small changes
in the parameters. We have already discussed in
section~\ref{sec:masslessII} that the maximum variances are not smooth
functions of the parameters, but the variation here is more extreme
and of a different origin. It will be explained in
section~\ref{sec:massiveII}. Figures~8(a) and~(b) show the variances
and occupation numbers \cite{occup} for a set of parameters for which
there is very little growth, {\it i.e.\/} a point corresponding to one
of the valleys in figure~7 \cite{quantum var}. In figure~9 we show the same quantities
for parameters were significant production does occur.

Figure~10 shows the maximum peak and valley $\chi$ variances produced
in the expanding universe for a massive inflaton as a function of
$m_\chi$ for $|q_0|=350$. Again we observe extreme sensitivity to
small changes in the parameters. In figures~11(a) and~(b) we show
the variances and $\chi$ field zeromode as a function of time for
three runs with just slightly different values of $m_\chi$. As can be
seen in figure~11(a) the maximum variances reached differ by several
orders of magnitude. In section~\ref{sec:massiveII} we will explain
how the growth of the variance depends on the dynamics of the $\chi$
zeromode, which is illustrated in figure~11(b).

An important point which will be discussed in detail in
section~\ref{sec:massiveII} is that a theory with negative coupling
can produce massive particles for much smaller values of $|q_0|$. In
fact we will see that for most of the parameter range in figures~7
and~10 there would be no particle production at all for $g>0$. As will
be explained in section~\ref{sec:baryogenesis}, the fact that massive
particles are produced more readily with negative $g$ is crucial in
constructing GUT baryogenesis models with ``natural'' coupling
constants. This will be illustrated by figures~12(a) and~(b), where we
show the maximum peak and valley variances reached for
$m_\chi=10^{14}$GeV as a function of $|g|$. For the range of couplings
shown in the figures it is impossible to produce such heavy particles
with $g>0$.
 
\subsection{Discussion and analytical estimates}
\label{sec:massiveII}

In this section we explain the main features of inflaton decay into
massive particles, with particular emphasis on the resonant growth and
its subsequent shut-off.  We will also give analytical estimates for
various quantities of interest.  The physics of the shut-off mechanism
is rather simple.  For positive $g$ parametric resonance produces
particles with {\it physical\/} momenta of order $k^{\rm phys}_{\rm
res} \sim q^{1/4}\omega_\phi$, so that $k^{\rm phys}_{\rm
res}\gg\omega_\phi$ when $q\gg 1$.  When the particles mass exceeds the
resonant momenta, the production shuts off.  For comparison we note
that perturbative decays are kinematically forbidden if the total mass
of the decay products is greater than the energy of the decaying
particles.
As a consequence, parametric
resonance can produce much heavier particles 
($m_\chi\sim q^{1/4}\omega_\phi$) than 
perturbative decays
($m_\chi\sim \omega_\phi$). 
(When $q\ll1$ the resonant and perturbative scales coincide.)
This observation could have important consequences in that parametric
resonance provides a mechanism to produce the heavy gauge and Higgs
bosons necessary for GUT baryogenesis.
This will be discussed in some detail in
section~\ref{sec:baryogenesis}. Before presenting our results
regarding the production of massive particles we wish to point out
that from the outset one might expect the sign of the coupling $g$ to
play an important role in this context. The reason is that in the
negative $g$ case the maximum resonant momenta are of order $k_{\rm
res}^{\rm phys}\sim 2|q|^{1/2}\omega_\phi$, which is by a factor
$2|q|^{1/4}$ larger than in the positive $g$ case.  Therefore, one
expects that with negative $g$ the resonance is much more effective in
producing massive particles. This expectation is confirmed below.

For completeness we will study all cases of interest: massless and
massive inflaton, positive and negative cross-coupling $g$.  We will
then compare our estimates for positive $g$ with the numerical results
of \cite{KhlebnikovTkachevII}, and for negative $g$ with figures $7 -
12$.  The estimates we obtain in this section will be used in
section~\ref{sec:baryogenesis} to estimate baryon production.  In
connection with this we remark that very massive particles are created
only marginally relativistic and quickly become nonrelativistic due to
redshifting. In this situation the variance is simply
related to the number density of particles $n$ and energy density
$\rho$ of the field: $n_\chi\approx
m_\chi\langle(\delta\chi)^2\rangle$ and $\rho_\chi \approx
m_\chi^2\langle(\delta\chi)^2\rangle$.

Based on Eq.~(\ref{eq: delta A}) one can make a quantitative estimate
of when the resonance shuts off.  Indeed, when $A$ becomes larger than
about $2|q|+|q|^{1/2}$ the inflaton decay slows down dramatically,
essentially because the instability exponent $\mu$ decreases rapidly
above the $A=2|q|$ line ({\it cf.\/} figure~1).  If the increase in
$A$ is mostly due to the (tree level) mass $m_\chi$, we say that the
resonance is shut off by the $\chi$ mass. If, on the other hand, the
growth of $A$ can be attributed mainly to the backreaction effects
(the contribution of the growing variances to the effective mass), the
situation resembles the massless case in the sense that the variances
reach their scattering regime values of section~\ref{sec:masslessII}
and a slowly varying state sets in.  Before we begin studying the
details of each of the cases mentioned above, we point out that the
Universe expansion is crucial in shutting off the inflaton decay.
This is so in essence because the inflaton amplitude decreases as the
Universe expands, while the tree level mass stays constant. More
concretely, the resonance shuts off when the mass becomes greater than
the typical physical resonant momentum, which scales as $k_{\rm
res}^{\rm phys} \propto a^{-1}$ for a massless inflaton and $k_{\rm
res}^{\rm phys} \propto a^{-3/4}$ for a massive inflaton.  This means
that, as the Universe expands, the (physical) scale on which particles
are produced redshifts and hence the relative importance of the $\chi$
mass increases, leading eventually to the termination of particle
production. Once the resonance is shut off, the only energy exchange
mechanism that remains active is the perturbative scatterings.
We now turn to discuss the details of each of the cases
mentioned above.   

\subsubsection{Massless inflaton, $g>0$}
\label{sec:massiveII.1}
For positive $g$ and a {\it massless\/} inflaton  
the criterion for the $\chi$ resonance shut-off is
({\it cf.\/} Eq.~(\ref{eq: delta A II}))
\begin{eqnarray}
m^2_{\chi\;\rm eff} &\gtrsim &
q^{\frac{1}{2}}\omega_{\phi\rm\; eff}^2\simeq
q_0^{\frac{1}{2}}(\omega_\phi^0)^2
\left(\frac{a_0}{a}\right)^2\,,
%
\end{eqnarray}
where $m^2_{\chi\;\rm eff} = m_\chi^2+g\langle(\delta\phi)^2\rangle
+3\lambda_\chi\langle(\delta\chi)^2\rangle $.  Here we have explicitly
written the dependence on the scale parameter $a$, used the fact that
$\Phi\propto a^{-1}$ for a massless inflaton, and assumed
$\omega_{\phi\rm\; eff}\simeq \omega_\phi$.  For massless fields the
maximum variances reached were given in Eqs.~(\ref{eq:variance II})
and~(\ref{eq:variance III}) for ``small'' $\lambda_\chi$
($rq^{1/2}<(3c^2)^{-1}$) and in Eqs.~(\ref{eq:variance I})
and~(\ref{eq:variance Ib}) for ``large'' $\lambda_\chi$. It is clear
that these estimates still hold for a massive $\chi$ field if, by the
{\it conformal\/} time $\tau_{\rm s}$ at which the scattering regime
is reached, $m_\chi^2<q^{1/2}(\omega_\phi^0)^2 (a_0/a(\tau_{\rm
s}))^2$ is satisfied.  Masses that satisfy this condition are small in
the sense that the resonance is shut off by the variance rather than
the mass term in the lagrangian, and the maximum variances reached are
hence the same as in the massless case.  If, on the other hand, the
resonance is shut off by the $\chi$ mass before the scattering regime
sets in, the maximum variances reached are smaller by a factor
$\exp[-2\mu\omega_\phi^0 (\tau_{\rm s}-\tau_{\rm m})]$, where
$\tau_{\rm m}$, the conformal time at which the mass term kills the
resonance, is defined by $m_\chi^2= q_0^{1/2}(\omega_\phi^0)^2
(a_0/a(\tau_{\rm m}))^2$.  The borderline case is the maximum $\chi$
mass for which the scattering regime is still reached:
\begin{equation}
m_\chi\simeq 
q_0^{\frac{1}{4}}\omega_\phi^0\frac{a_0}{a(\tau_{\rm s})}
\,,\qquad {\rm (massless\quad inflaton)}\,. 
\label{eq: m_chi massless}
\end{equation}
This can be rewritten as an estimate for the minimum initial value of
$q$ required to reach the scattering regime for a given value of
$m_\chi$. Using 
\begin{equation}
a_0\tau_{\rm s}\simeq \frac{1}{2\omega_\phi^0\mu}
\ln\frac{n_{\rm scatt}^\chi}{n_0^\chi}\,,
\label{eq: tau}
\end{equation}
and 
\begin{equation}
\frac{a}{a_0}=1+H_0a_0\tau\,,\qquad
H_0\approx\frac{1}{2\sqrt{2}\, c}\omega_\phi^0
\frac{\Phi_0}{M_{\rm P}}\,,
\label{eq: a and H_0}
\end{equation}
where $H_0$ is the 
Hubble parameter at the 
beginning
of the oscillatory stage, one obtains
%
%
\begin{equation}
q_0^{\rm min}\simeq
\left(\frac{m_\chi}{\omega_\phi^0}
\frac{a}{a_0}
\right)^4
\simeq \left(
\frac{1}{4c\sqrt{2}}
\frac{m_\chi}{\omega_\phi^0}
\frac{\Phi_0}{M_{\rm P}}
\frac{1}{\mu}\ln \frac{n_{\rm scatt}^\chi}{n_0^\chi} 
\right)^4\,,
\label{eq: minimum q massless}
\end{equation}
where $n_0^\chi\approx 1/2$, and $n_{\rm scatt}^\chi\sim g^{-1}$ for small
$\lambda_\chi$ and $\sim 10/\lambda_\chi$ for large
$\lambda_\chi$\cite{n_scatt}.  For $q_0\geq q_0^{\rm min}$ the
variances peak at their scattering regime values.  For a typical
choice of parameters $\Phi_0/M_{\rm P}=1.8$,
$\mu=0.1$, and taking $\lambda_\chi=0$, Eq.~(\ref{eq: minimum q
massless}) yields $q_0^{\rm min}\sim 2\times 10^6$ for
$m_\chi=\omega_\phi^0$, and $q_0^{\rm min}\sim 8\times 10^8$ for
$m_\chi=10\omega_\phi^0$ \cite{massless frequency}.
This is roughly in agreement with
Ref.~\cite{KhlebnikovTkachevII}, where the authors estimate that no
significant resonant production occurs for $q_0\le 10^6$ if
$m_\chi\le\omega_\phi^0$. Of course the choice of $\mu$ introduces
significant uncertainty into the estimates: using $\mu=0.2$, for
example, lowers the values of $q_0^{\rm min}$ by a factor of $5 -
10$. To test the validity of Eq.~(\ref{eq: minimum q massless}) we ran
our code with $m_\chi= 1.07\omega^0_\phi$, $\lambda_\chi=0$, and
$q_0=3.65\times 10^6$ and found that the scattering
regime is marginally reached when the resonance shuts off at $t\sim
1.5\times 10^8 M_{\rm P}^{-1}$. The peak and valley values of the
$\chi$ variance are $\langle(\delta\chi)^2\rangle_{\rm peak} \approx
4\times 10^{-10}M_{\rm P}^2$ and $\langle(\delta\chi)^2\rangle_{\rm
valley} \approx 6\times 10^{-12}M_{\rm P}^2$. This is in good
agreement with the prediction for the scattering regime peak and
valley values of Eqs.~(\ref{eq:variance III}) and~(\ref{eq:variance
growth}): $\langle(\delta\chi)^2\rangle_{\rm peak} \approx 4\times
10^{-10}M_{\rm P}^2$ and $\langle(\delta\chi)^2\rangle_{\rm valley}
\approx 1.3\times 10^{-11}M_{\rm P}^2$, where we used for $\Phi_{\rm
s}=\Phi_0(a_0/a(\tau_s))\simeq 0.08 M_{\rm P}$.  Note that even though
these variances are rather small the scattering regime is reached
since large $q$ corresponds to large $g$.  For $q_0=10^5$ we find
that the resonant growth shuts off well before the scattering regime
is reached. The $\chi$ variance grows to about $1\%$ of its scattering
regime values (\ref{eq:variance III}) and~(\ref{eq:variance growth}).

\subsubsection{Massive inflaton, $g>0$}
\label{sec:massiveII.2}
Before discussing the effect of the $\chi$ field mass on the resonance
shut off we will discuss briefly the simpler situation with $m_\chi=0$
but massive inflaton.  Since for a {\it massive\/} inflaton
$\Phi\propto a^{-3/2}$, we find that $q=q_0(a_0/a(t))^3$. There are
then two possible mechanisms for terminating the resonance: either the
resonance is shut off as usual by the variances, or $q$ becomes $\ll
1$ due to the expansion before this happens.  In the former case the
maximum variances can be estimated as usual from Eq.~(\ref{eq: delta
A}). Hence Eqs.~(\ref{eq:variance II}) and~(\ref{eq:variance III}) for
``small'' $\lambda_\chi$ ($rq^{1/2}<(3c^2)^{-1}$) and
Eq.~(\ref{eq:variance I}) for ``large'' $\lambda_\chi$ apply {\em also
in the massive inflaton case} if the variances shut off the resonance,
{\it i.e.\/} if the scattering regime is reached \cite{var eq massive}.
The difference is that $\Phi_{\rm s}$ is {\it not} obtained from
Eq.~(\ref{eq:phi*}). Instead one must use
\begin{equation}
\Phi_{\rm s}=\Phi_0\left[\frac{a_0}{a(t_{\rm s})}\right]^{3/2}\quad {\rm
(massive\, inflaton)}
\label{eq:phi*,massive}
\end{equation}
where
\begin{equation}
\frac{a(t)}{a_0}=\left(
1+\frac{3}{2}H_0 t
\right)^{\frac{2}{3}}\,,\quad
H_0\approx\frac{1}{2}
\frac{\Phi_0}{M_{\rm P}}m_\phi\,,\quad
t_{\rm s}\simeq \frac{1}{2 m_\phi\mu}
\ln{\frac{n_{\rm scatt}^\chi}{n_0^\chi}}\,,
\label{eq: a, H_0, and t}
\end{equation}
and $t_{\rm s}$ is the {\it physical\/} time at the beginning of the
scattering regime.  The estimates for the
$n^{\chi}_{\rm scatt}$ and $n^{\chi}_0$ are given below Eq.~(\ref{eq:
minimum q massless}).
 
On the other hand, if $q\ll 1$ before the variances reach the values
obtained from the above estimates then the resonance shuts off simply
because there is no strong instability for small $q$ ({\it cf.\/}
figure~1(a)). The regime with $q\ll 1$ is often called ``narrow
resonance'' in the literature because the width of the instability
bands goes to zero as $q$ becomes small. Using the fact that $q\propto
a^{-3}$ and the estimate for $t_{\rm s}$ above it is easy to find the
minimum initial value of $q$ such that $q(t_{\rm s})\gtrsim 1$. One
obtains
\begin{equation} 
q_0^{\rm min} \approx \left(\frac{3}{8} 
\frac{\Phi_0}{M_{\rm P}}\frac{1}{\mu} \ln\frac{n_{\rm
scatt}}{n_0}\right)^2 \quad , \qquad \qquad(m^2_\chi=0). 
\label{eq:qmin,massless}
\end{equation} 
For typical values $\Phi_0=1.8 M_P$ and $\mu\sim0.1$ this evaluates to
$q_0^{\rm min}\approx 10^4$ in the small $\lambda_\chi$ case and
$q_0^{\rm min} \approx 46\ln^2(\frac{8\pi}{\lambda_\chi})$ in the
large $\lambda_\chi$ case.  If $q_0\ll q_0^{\rm min}$ then only a
small fraction of the inflaton energy decays before $q(t)\ll 1$ and
the fast decay shuts off because $\mu\ll 1$. However, if $q_0\approx
q_0^{\rm min}$ then the inflaton can decay almost completely before
entering either the scattering regime or the narrow resonance
regime. This is the situation for which the variances of the fields
reach their maximum possible values.

When the $\chi$ field is massive, one requires larger values of $q_0$
for the variances to reach their scattering regime values.
The minimum value of $q_0$ for which the estimates of
section~\ref{sec:masslessII}, together with Eqs.~(\ref{eq:phi*,massive}) and
(\ref{eq: a, H_0, and t}), apply can be found as in the previous section by
requiring $m^2_\chi < q^{1/2}(t_{\rm s})m^2_\phi = q_0^{1/2}(a_0/a(t_{\rm
s}))^{3/2}m^2_\phi$. The result is 
\begin{equation}
q_0^{\rm min}\simeq \left(
\frac{m_\chi}{m_\phi}
\right)^4
\left(
\frac{a}{a_0}
\right)^3\simeq \left(
\frac{m_\chi}{m_\phi}
\right)^4
\left(
\frac{3}{8}
\frac{\Phi_0}{M_{\rm P}}
\frac{1}{\mu}\ln \frac{n_{\rm scatt}^\chi}{n_0^\chi} 
\right)^2\; .
\label{eq: minimum q massive}
\end{equation}
If $q_0$ is larger than this, the resonance is terminated by the
variance.
Taking the same parameter values as in the massless inflaton case of
section~\ref{sec:massiveII.1}
we obtain  $q_0^{\rm min}\sim 10^4$ 
for $m_\chi=m_\phi$
and $q_0^{\rm min}\sim 3\times 10^7$ 
for $m_\chi=10m_\phi$.
Hence we conclude that for a massive inflaton
the values of $q_0^{\rm min}$ are typically 
{\it one to two\/} orders of magnitude smaller  
than in the massless inflaton case. 
A comparison of Eqs.~(\ref{eq: minimum q massless}) and 
(\ref{eq: minimum q massive}) leads to 
$q_{\rm massive}/q_{\rm massless}^{1/2}\simeq (9c^2/2)
(m_\chi/\omega_\phi^0)^2$,
where we assumed equal occupation numbers, $\mu$'s,
$\chi$ masses, $\Phi_0$'s, and $\omega_\phi^0=m_\phi$.
We should point out that for a massive inflaton the slow roll 
ends somewhat earlier and that hence the appropriate values of
$\Phi_0$ are perhaps $20\%$ larger than in the massless case.
To summarize, the main reason why the massive inflaton is 
more efficient in producing massive particles is that the 
typical resonant momenta redshift slower. This gives the resonance
more time to create particles before it is shut off.

It is interesting to compare our analytical formula for the maximum
variance reached with the estimate given
in~\cite{KhlebnikovTkachevIII}. As explained above, the expressions
given in section~\ref{sec:masslessII} apply as long as $q_0$ is large
enough so that the scattering regime is reached. For small
$\lambda_\chi$ the appropriate formula is then Eq.~(\ref{eq:variance
III}), which is particularly easy to apply for a massive inflaton
since the combination $\Phi(t)^2/q(t)=4m_\phi^2/g$ does not redshift
due to the expansion. Hence $\langle(\delta\chi)^2\rangle_{\rm peak}
\approx \Phi_0^2/16q_0$.  For $q_0=10^8$, which as shown above is
sufficient to reach the scattering regime even when $m_\chi/m_\phi=10$,
we thus obtain $\langle(\delta\chi)^2\rangle_{\rm peak} \approx
10^{-9}M^2_{\rm P}$.  This is in good agreement
with~\cite{KhlebnikovTkachevIII}, where the authors find
$\langle(\delta\chi)^2\rangle_{\rm peak} \sim 10^{-9}M^2_{\rm
P}$ \cite{planckmass}. Note, however, that the parametric dependence 
of the estimate for $\langle(\delta\chi)^2\rangle_{\rm peak}$ given
in~\cite{KhlebnikovTkachevIII} seems to be different from our
Eq.~(\ref{eq:variance III}). 

\subsubsection{Negative Coupling}
\label{sec:massiveII.3}
The case of inflaton decay into  massive $\chi$ particles
with a negative cross-coupling $g$ is significantly 
more complex than the positive $g$ case. We hence begin by discussing
the main characteristics of the process.

The first notable feature is that the resonance is shut off by the
$\chi$ mass only when 
\begin{equation}
m_\chi^2\gtrsim |g|\Phi^2\approx 4|q|\omega_{\phi}^2
\, ,  
\label{eq: negative g shut off}
\end{equation}
which should be compared to the positive $g$ criterion
$m_\chi^2\gtrsim q^{1/2}\omega_{\phi}^2$.  Eq.~(\ref{eq: negative g
shut off}) is simply Eq.~(\ref{eq: delta A}) for
$A(k=0)=-2|q|+m^2_{\chi}/\omega_\phi^2$ and $\delta A/A \ll 1$, $q\gg
1$.  Hence in order to get {\it significant\/} resonant production,
one requires $m_\chi^2\ll |g|\Phi_0^2$, which is a much weaker
condition than for positive $g$. Since $\Phi$ scales with the
expansion of the Universe (as $\propto a^{-1}$ for a massless inflaton
and $\propto a^{-3/2}$ for a massive inflaton), there is then a time
interval during which resonant production is possible. The growth
mechanism is quite different from the positive $g$ case, and we will
describe its main features below. However, a complete analytical
understanding is extremely difficult to achieve. The reason is the
extremely ``chaotic'' nature of the process, which we have already
pointed out when describing the figures in section~\ref{sec:massiveI}.

It turns out that the positive $g$ resonant scale $k_{\rm res}^{\rm
phys}= |q|^{\frac{1}{4}}\omega_\phi$ plays an important role in
understanding the dynamics.  While $m_\chi^2\ll
|q|^{1/2}\omega_\phi^2$, the dynamics is equivalent to a positive $g$
case with $q_{\rm eff}=2|q|$ and $\mu_{\rm eff}\sim 0.1$, essentially
because $\chi_0$ evolves according to Eq.~(\ref{eq: chi_0}) (with
$m_\chi\approx 0$). Once $m_\chi> k_{\rm res}^{\rm phys}$, which
always occurs sooner or later since $k_{\rm res}^{\rm phys}$ redshifts
with the expansion of the Universe, the `positive $g$ resonance' shuts
off completely.  This does not mean that production stops
altogether. There is another effect which we refer to as the {\it
negative coupling instability\/}.

We now give a heuristic explanation of this new effect. In some sense
we already discussed it in section~\ref{sec:masslessII}, where we
argued that once the variances get so large that $\delta m_\chi \sim
|q|^{1/4}\omega_\phi$, $\chi_0$ falls out of phase with $\tilde\chi_0$
( {\it cf.\/} also \cite{stop follow}). The difference is that for
massive $\chi$ the ``falling out of phase'' can occur without the need
for large variances. The reason is the following.  After $\phi_0$
passes through zero and starts growing, it tries to pull $\chi_0$
along, attempting to keep it equal to $\tilde\chi_0$ prescribed in
Eq.~(\ref{eq: chi_0}). However, if $m_\chi$ is large enough so that
$\chi_0$ has time to oscillate while $\phi_0$ is small, then a phase
mismatch can occur and prevent $\chi_0$ from growing and following
$\tilde\chi_0$. This will be illustrated with a simple model below.
When such a mismatch occurs the infrared $\chi$ modes grow very fast
since for these modes the frequency squared is negative, and their
potential corresponds to an inverted harmonic oscillator.  The growth
exponent $\mu_{\rm eff} = \chi_k^{-1}d\chi_k/dt$ may be as large as
$\sim |q|^{1/2}\omega_\phi$, and the range of momenta which grow can
be as broad as $\Delta k_{\rm phys}^2\sim |g|\Phi^2$. A dramatic effect
indeed, compared to the usual positive coupling parametric
resonance. This is what we mean by {\it negative coupling
instability\/}.  The explosive growth ends once $\chi_0$ grows
sufficiently large to catch up with $\tilde\chi_0$, at which point the
infrared modes resume their normal oscillatory behavior.  An example
of this effect is presented in figure~11(a). For $m^2_\chi=5.4\times
10^{-11}M_{\rm P}^2$ the variance grows with an exponent of order
$\mu_{\rm eff}\approx 14\approx 2|q|^{1/2}m_\phi$.

A quantity which is clearly of interest is the probability that for a
given set of parameters a negative coupling instability, and the huge
particle production associated with it, occurs. As explained above,
the instability can occur if there is a phase mismatch between
$\chi_0$ and $\phi_0$ at the moment when $\chi_0$ should stop
oscillating about the origin according to Eq.~(\ref{eq:
chi_0}). Specifically, the instability can occur at the instant when
the curvature of the $\chi$ potential $V(\chi)$ at the origin changes
sign, which happens when $\phi_0$ becomes larger than $m^2_\chi/|g|$.
This observation can be used to estimate the {\it a priori\/}
probability for an instability to occur, based on the assumption that
the phase of $\chi_0$ is random at the instant the curvature changes
sign. We stress that this assumption is reasonable only when
$m^2_{\chi\;{\rm eff}} \gtrsim |q|^{1/2}\omega^2_{\phi\;{\rm eff}}$.
If this condition is not satisfied $\chi_0$ does not have time to
oscillate while $\phi_0$ is small, and hence no phase mismatch can
occur. In this case $\chi_0$ simply follows $\tilde\chi_0$, just as in
the first part of the evolution shown in figure~2.

One might guess that the {\it a priori\/} probability for the mode
amplitudes to grow by a factor $G$ ought to be of order $G^{-1}$.  This
estimate is supported by the following toy model. Let us take the time
at which the curvature of $V(\chi)$ at the origin changes sign to be
$t=0$. Assume that, while $\phi_0\simeq 0$, $X=\chi_0$ satisfies 
\begin{equation}
X = X_0\cos m_\chi t+
\frac{\dot X_0}{m_\chi}
\sin m_\chi t\,,\quad
\hbox{\rm for}\quad t<0\,.
\label{eq: solution to X II}
\end{equation}
%
When $\phi_0$ becomes larger than $m^2_\chi/|g|$ at $t=0$, $\chi_0$ 
finds itself, at least momentarily, in the potential of an upside down
oscillator. Hence we have 
\begin{equation}
X = \frac{1}{2}\left(X_0-\frac{\dot X_0}{\nu}\right)
{\rm e}^{-\nu t}+
 \frac{1}{2}\left(X_0+\frac{\dot X_0}{\nu}\right)
{\rm e}^{\nu t}
\,,\quad 
\;\hbox{\rm for}\quad t>0\,,
\label{eq: solution to X III}
\end{equation}
where $\nu\equiv \mu\omega_\phi\approx |q|^{1/2}\omega_\phi$.  We are
interested in the possibility of a phase mismatch, by which we mean
that $X$ matches mostly onto the decaying solution and hence the
amplitude of the growing mode at $t=0$ is small, {\it i.e.\/}
\begin{equation}
\frac{1}{2}\left(X_0+\frac{\dot X_0}{\nu}\right)
=\epsilon A_0
\,,\quad   
\;\hbox{\rm where}\quad 
A_0^2=X_0^2+\left(\frac{\dot X_0}{m_\chi}\right)^2
\,,\qquad \epsilon\ll 1
\,.
\label{eq: growing mode amplitude}
\end{equation}
To recover its original amplitude $A_0$, and catch up with $\tilde
\chi_0$, $X$ has to grow by a factor $\exp \nu t\sim \epsilon^{-1}\gg
1$. During this time period the infrared modes are unstable, and the
$\chi$ variance may grow by a factor as large as
$\sim\epsilon^{-2}\approx G^2$.  The question we want to answer is,
how likely is such an event? As mentioned above, the estimate we are
about to present is based on the assumption that $\chi_0$ has a
uniform random phase at $t=0$, which should not be a bad approximation
for $m_\chi> |q|^{1/4}\omega_\phi$.  The appropriate probability
distribution is then the uniform distribution on the circle of phases:
\begin{equation}
d{\cal P}=\frac{d\varphi}{2\pi}\,,\qquad
X_0=A_0\cos\varphi\,,\quad
\frac{\dot X_0}{m_\chi}=A_0\sin\varphi\,.
\label{eq: uniform probability distribution}
\end{equation}
After some simple algebra we obtain the desired probability
\begin{equation}
{\cal P}_\epsilon\equiv
{\cal P}
\left(\left|
\frac{1}{2A_0}
\left[X_0+\frac{\dot X_0}{\nu}\right]\right|\le \epsilon
\right)\approx 
\frac{4}{\pi}
\frac{\epsilon}
{\left[1+\left({m_\chi}/{\nu}\right)^2\right]
^{\frac{1}{2}}}
\,,\quad \epsilon\ll 1
\label{eq: probability I}
\end{equation}
The probability that the variance grows by a factor
$\langle(\delta\chi)^2\rangle/ \langle(\delta\chi)^2\rangle_0\sim G^2$
is then ${\cal P}_\epsilon\simeq (2/\pi)G^{-1}\sim G^{-1}$, as
anticipated above.  Here we used $G\approx\epsilon^{-1}$ for the
amplitude growth factor and the fact that production becomes
impossible when $m^2_\chi\gtrsim |g|\Phi^2$ ({\it cf.\/}
Eq.~(\ref{eq: negative g shut off})), so Eq.~(\ref{eq: probability I})
applies only when $\nu\sim |q|^{1/2}\omega_\phi\gtrsim m_\chi$.
%
%
In figure~11(a) we present a run with $m_\chi^2=5.4\times
10^{-11}M_{\rm P}^2=75m_\phi^2$ in which the $\chi$ variance grows by
a factor $3\times 10^4$ in the interval $t\in [2.21,3.16]\times
10^{6}M_{\rm P}^{-1}$.  According to our estimate the probability for
such a large growth to occur is about $0.8\%$.  Figure~11(b) shows
$\chi_0$ {\it vs.\/} time for three slightly different $\chi$ masses
to illustrate the dramatic effect of a phase mismatch at the instant
when the curvature of the $\chi$ potential at the origin changes sign.
The $m_\chi^2=5.4\times 10^{-11}M_{\rm P}^2$ run matches almost
entirely on the decaying mode, so in that case $\chi_0$ spends a long
time near the origin and the variance grows by a huge factor. As can
be seen in figures~11(a) and~(b), a tiny ($1\%$) change in the $\chi$
mass causes a different matching and changes the resulting variances
by several orders of magnitude \cite{fig12}. This explains the spiky
nature of the curves in figures~7 and~10.  The high sensitivity of the
growth to small changes in the parameters can be also seen in figure~12.
Note that because of the chaotic nature of the production mechanism
one should not take the fact that we have connected our data points in
these figures with line segments too seriously - for all but the most
closely spaced points there could be numerous peaks and valleys in
between.  

The probability ${\cal P}_\epsilon$ in Eq.~(\ref{eq: probability I})
is essentially the probability {\it per inflaton half-period} that the
variance grows by a factor $\epsilon^{-2}$, provided that
$q^{1/4}\omega_{\phi\; {\rm eff}} \lesssim m_{\chi\; {\rm
eff}}\lesssim q^{1/2}\omega_{\phi\; {\rm eff}}$. Note that, since the
$\chi$ variance contributes to $ m^2_{\chi\; {\rm eff}}$, the last
inequality also determines the maximum growth that can be
achieved. This will be discussed in detail below. Here we want to
emphasize that the probability is {\it per inflaton half-period} because a
phase mismatch can occur twice per period when the curvature of
$V(\chi)$ at the origin changes sign from positive to negative. This
means that to obtain the total probability for a negative coupling
instability to occur for a given set of parameters one must multiply $
{\cal P}_\epsilon$ by twice the number of inflaton oscillations that take
place while $q^{1/4}\omega_{\phi\; {\rm eff}} \lesssim m_{\chi\;
{\rm eff}}\lesssim q^{1/2}\omega_{\phi\; {\rm eff}}$ is
satisfied \cite{prob}. Since we are mostly interested in very heavy
particles that could not be produced with $g>0$, we will assume that
the first part of the inequality is satisfied from the start. It is
then easy to calculate how large $|g|$ needs to be in order to have $n$
chances for instability before $q^{1/2}<m_{\chi\;
{\rm eff}}/\omega_{\phi\; {\rm eff}}$ and all production shuts off.  
The minimum $|g|$ for $n$ chances is
\begin{equation}
|g_{\rm n}|  = 
\frac{m^2_\chi}{\Phi_0^2}\left(1+\beta n\frac{\Phi_0}{M_{\rm
P}}\right)^{2}\; ,
\label{eq:g_n}
\end{equation}
where $\beta\approx 3 \pi/4$ for a massive inflaton and
$\beta\approx \pi/2\sqrt 2c$ for a massless inflaton. Since
$\Phi_0$ is also slightly larger in the massive case we see that one
requires somewhat smaller values of $|g|$ in the massless case. 

We will now briefly discuss the ratio of ``peak'' to ``valley''
variances in the scattering regime for massive particles. This was
discussed for massless particles at the end of
section~\ref{sec:masslessII}, where we found distinctly different
behaviors for positive and negative $g$. If the $\chi$ field is
massive the situation is more complex. For $g>0$ the ratio
$\langle(\delta \chi)^2\rangle_{\rm peak}/
\langle(\delta\chi)^2\rangle_{\rm valley}$ is still given by
$\omega_{\rm max}/\omega_{\rm min}\sim q^{1/4}$. Of course this is
only valid if the scattering regime is reached. For $g<0$ two kinds of
behaviors are possible. If the resonance is shut off by the $\chi$
mass, {\it i.e.\/} the scattering regime is not reached, then the
curvature at the origin of the $\chi$ potential remains positive after
resonance shut off and the oscillating $\phi_0(t)$ simply modulates
the frequency of the $\chi$ oscillations. In this case the frequency
of the $\chi$ field changes continuously, so the ``random phase''
approximation discussed at the end of section~\ref{sec:masslessII}
applies, and the ratio is again $\langle(\delta \chi)^2\rangle_{\rm
peak}/ \langle(\delta\chi)^2\rangle_{\rm valley} \sim \omega_{\rm
max}/\omega_{\rm min}$. The difference is that now $\omega_{\rm
max}^2= m_\chi^2+3\lambda_\chi\langle(\delta\chi)^2\rangle_{\rm
valley}\approx m_\chi^2$, and $\omega_{\rm min}^2=
m_\chi^2-|g|\Phi^2+3\lambda_\chi\langle(\delta\chi)^2\rangle_{\rm
peak} \approx m_\chi^2-|g|\Phi^2$.
If, on the other hand, the scattering regime is reached when 
$m_\chi^2\ll |g|\Phi^2$, then the infrared $\chi$ modes can be  
dragged along by $\phi_0$ just as in the massless negative $g$ case,
and hence
\begin{equation}
\frac{\langle(\delta \chi)^2\rangle_{\rm peak}} 
{\langle(\delta\chi)^2\rangle_{\rm valley}}
\sim \left(\frac{\omega_{\rm max}}{\omega_{\rm min}}\right)^2
\sim
\frac{m_\chi^2
+3\lambda_\chi\langle(\delta\chi)^2\rangle_{\rm valley}}
{\omega_\phi^2}\; ,
\label{eq:variance growth massive chi}
\end{equation} 
which reduces to $\sim (m_\chi/\omega_\phi)^2$ when
$m_\chi^2>3\lambda_\chi\langle(\delta\chi)^2\rangle_{\rm valley}$, and
to $\sim |g|^{1/2}\Phi/\omega_\phi$ when
$m_\chi^2<3\lambda_\chi\langle(\delta\chi)^2\rangle_{\rm valley}$.
Here we used $3\lambda_\chi\langle(\delta\chi)^2\rangle_{\rm peak}
\sim |g|\Phi^2$. Note that eventually the expansion of the universe
always reduces $\Phi$ sufficiently so that the condition
$m^2_\chi<|g|\Phi^2$ becomes violated and this case turns into the
one discussed above Eq.~(\ref{eq:variance growth massive chi}). 
 
We are now ready to discuss the specifics of the massless and 
massive inflaton cases.

\subsubsection{Massless Inflaton, $g<0$}
\label{sec:massiveII.4}
In figure~7(a) we show the peak variances reached as a function of
$m^2_\chi$ for various values of $|q_0|$. Figure~7(b) is a blow-up of
the $|q_0|=350$ curve. For these runs
$\omega_\phi^0=c\sqrt\lambda_\phi\Phi_0\approx1.5\times10^{-6}M_{\rm
P}$, so the horizontal range in figure~7(a) corresponds to
$m_\chi/\omega_\phi^0\approx 0.63 - 66$. The range in figure~7(b)
corresponds to $m_\chi/\omega_\phi^0\approx 1.84 - 8$. As discussed in
section~\ref{sec:massiveII.1}, for positive $g$ the scattering regime
would never be reached for these mass ratios with $q_0=350$. Even more
dramatic is the fact that for $m^2_\chi/M^2_{\rm P}\gtrsim 2\times
10^{-11}$, there would be no production at all for the chosen
parameters if $g$ were positive\cite{pos g production}. This
illustrates in concrete terms the power of negative coupling to
produce massive particles.

The spiky nature of the curves was already explained in the previous
section. In figures~8(a) and~(b) we show the variances and occupation
numbers \cite{occup} of a run with a $\chi$ field mass for which there
happen to be no phase mismatches and hence no negative coupling
instability. In figures~9(a) and~(b) we show the same quantities for
a case where the instability does occur. Note that $m_\chi$ is
actually larger for the run in figure~9.

The minimum value of $|q_0|$ for which there is any chance of production
can be immediately obtained from Eq.~(\ref{eq:g_n}) with $n=1$:
\begin{equation}
|q_0^{\rm min}| = \left(\frac{m_\chi}{2\omega_\phi^0}\right)^2
\left(1+\frac{\pi}{2\sqrt 2 c}\frac{\Phi_0}{M_{\rm P}}\right)^2\,.
\end{equation}
With $\Phi_0=1.8M_{\rm P}$ this evaluates to $|q_0^{\rm min}|=2.8$ for
$m_\chi/\omega_\phi^0=1$ and $|q_0^{\rm min}| = 280$ for
$m_\chi/\omega_\phi^0=10$. These values are orders of magnitude less
than the corresponding positive $g$ values derived in
section~\ref{sec:massiveII.1}.

The maximum variance which can be reached is determined by
Eq.~(\ref{eq: negative g shut off}) (with $m^2_\chi \rightarrow
m^2_{\chi\;{\rm eff}}$) to be
\begin{equation}
\langle(\delta\chi)^2\rangle_{\rm max} =
\frac{|g|\Phi^2-m^2_\chi}{3\lambda_\chi} \,.
\label{eq: varmax}
\end{equation}
The appropriate value of $\Phi$ depends on exactly when the growth
occurs. If a phase mismatch and the associated negative coupling
instability occurs very early on (right after the first zero crossing,
say) $\Phi$ can be as large as $0.5M_{\rm P}$.

\subsubsection{Massive inflaton, $g<0$}
\label{sec:massiveII.5}
When discussing the massive inflaton case with negative coupling it is
important to keep in mind that the stability bound $r>1$ always
requires the presence of a $\lambda_\phi\phi^4$ (or higher order) term
in the effective potential. For the runs in this section we have
chosen the inflaton mass such that $m^2_\phi \lesssim
3\lambda_\phi\Phi_0^2$ at the end of slow roll. This clearly
corresponds to mild fine tunning as there is no a priori reason that
these quantities should be of the same order. If one chooses
$\lambda_\phi \ll m^2_\phi/M^2_{\rm P}$ it becomes more difficult to
satisfy $r>1$ and the available parameter space is somewhat
limited. For these reasons we believe that with negative $g$ the more
natural situation is the one with $m^2_\phi \ll
3\lambda_\phi\Phi_0^2$, in which case the anlaysis of the previous
section applies. Nevertheless it is interesting to study the situation
where the inflaton dynamics is dominated by the mass term during the
oscillatory stage, and we do so below.

Figure~10 shows the maximum peak and valley variances reached during
preheating {\it vs.\/} $m^2_\chi$ for $|q_0|\approx 350$ (the exact
parameters are given in the caption). The horizontal range corresponds
to $2.4 < m_\chi/m_\phi < 12$. According to the analysis of
section~\ref{sec:massiveII.2} the scattering regime is never reached
for the same parameter values with $g>0$. Moreover, for
$m^2_\chi/M^2_{\rm P}\gtrsim 10^{-11}$ there is no production at all
for the chosen value of $|g|$ if $g>0$\cite{pos g production}.

As in the massless inflaton case the minimum value of $|q_0|$ for
which there is any chance of heavy particle production is obtained
from Eq.~(\ref{eq:g_n}) with $n=1$:
\begin{equation}
|q_0^{\rm min}| = \left(\frac{m_\chi}{2m_\phi}\right)^2
\left(1+\frac{3\pi}{4}\frac{\Phi_0}{M_{\rm P}}\right)^2\,.
\end{equation}
For $m_\chi/m_\phi=1$ this evaluates to $|q_0^{\rm min}| = 6.9$, and for
$m_\chi/m_\phi=10$ one obtains $|q_0^{\rm min}| = 690$ (here we have taken
$\Phi_0=1.8M_{\rm P}$). Comparison with the results of
section~\ref{sec:massiveII.2} shows that once again the required
values are orders of magnitude less for negative coupling than for
positive coupling. On the other hand comparison with
section~\ref{sec:massiveII.4} shows that for negative coupling the
massless inflaton is actually somewhat more effective in producing
massive particles than the massive inflaton. This is opposite to the
positive coupling situation, although it should be noted that for the
present case the difference arises only from a numerical factor; the
parametric dependence is the same. Finally we point out that the
maximum possible variance is given by Eq.~(\ref{eq: varmax}) also for
the massive inflaton since Eq.~(\ref{eq: negative g shut off}) holds
in both cases.

\section{Baryogenesis}
\label{sec:baryogenesis}

It is well known that grand unified theories generically
predict the existence of baryon number (B) violating
interactions. Much effort has gone into converting this fact into a
viable mechanism for baryogenesis \cite{kolb}. In the standard
scenario, B violating gauge or Higgs bosons fall out of thermal
equilibrium as the temperature of the universe drops below their mass
and subsequently decay, creating net B in the process. This B must
then be protected from being wiped out by sphalerons, 
electroweak B violating processes.  Fortunately
sphaleron transitions conserve B--L, where L is lepton number, so that
any GUT that creates a non vanishing B--L remains a viable
candidate. Examples are $SO(10)$ and $E_6$.

According to the old theory of reheating \cite{old theory}, based on
perturbative decay of the inflaton, GUT baryogenesis seems rather
difficult to implement \cite{chaoticBgenesis}. The reason is simply
that the heavy B violating gauge or Higgs bosons cannot be produced
efficiently. The gauge bosons are expected to have masses of order the
GUT scale ($10^{16}$GeV), while the Higgs bosons may have masses as
low as $10^{14}$GeV. Since the mass of the inflaton $m_\phi$ cannot be
larger than about $10^{13}$GeV in order not to be in conflict with the
measured anisotropy in the microwave background \cite{cobe}, direct
perturbative production of GUT bosons is kinematically forbidden. In
addition, the weak coupling of the inflaton to other fields causes the
reheat temperature, as obtained in perturbation theory, to be so low
that thermal production is also ineffective \cite{reheat}.

Clearly GUT baryogenesis after inflation is resurrected by parametric
resonance. Not only can bosons with masses larger than the inflaton be
produced directly. They are automatically far from thermal
equilibrium, thus satisfying one of the prerequisites for
baryogenesis. The first model following the baryon asymmetry from the
end of preheating to the final decay of the inflaton is presented in
\cite{KolbLindeRiotto}. The authors assume that during the resonance
stage a fraction of the initial inflaton energy is rapidly transferred
to GUT bosons with mass $10m_\phi$. These bosons then decay in a CP
violating manner, producing net baryon number in the process.

While the results of the toy model in \cite{KolbLindeRiotto} are very
encouraging, it has a rather unpleasant feature. As shown in
\cite{KhlebnikovTkachevII} as well as sections~\ref{sec:massiveII.1}
and~\ref{sec:massiveII.2}, for $g>0$ it is quite difficult to produce
particles as massive as $10m_\phi$ during preheating. Specifically,
production of such particles requires a massive inflaton and
$q_0\gtrsim 10^8$. Since $m_\phi\approx 10^{13}$GeV this corresponds
to $g\gtrsim 10^{-3}$, which leads to a fine tunning problem: in order
to reproduce the anisotropies in the microwave background the quartic
self coupling of the inflaton is constrained to be
$\lambda_\phi\lesssim 10^{-12}$, which is unnatural because
$\lambda_\phi$ receives contributions of order $g^2$ at the one-loop
level. As mentioned in \cite{KolbLindeRiotto} this problem does not
arise if the value of $\lambda_\phi$ is protected by a supersymmetric
non-renormalization theorem. However, so far no such model has been
constructed, and in any case it is interesting to explore other
solutions to the naturalness problem.

In light of these considerations our investigation of massive particle
production in section~\ref{sec:massive} is particularly important for
baryogenesis. We have shown that with $g<0$ large amounts of heavy
particles are produced even for moderate values of $q_0$. In
particular the naturalness condition $g^2\lesssim\lambda_\phi$ is
easily satisfied even for $m_\chi \sim 10^{14}$GeV or
more. Additionally, the production of the heavy particles is extremely
rapid, so that it is not affected by competing decay channels. In
fact, as explained in section~\ref{sec:massiveII}, for $g<0$ the
decay into massive particles is much faster than the decay into light
particles provided the mass is such that $\chi_0$ and $\tilde \chi_0$
quickly fall out of phase. Finally, in contrast to the positive $g$
case, with $g<0$ heavy particles are produced by both massive and
massless inflaton fields. For these reasons we feel that negative
coupling baryogenesis after preheating is an exciting possibility.

In order to illustrate how this baryogenesis mechanism works we adopt
the toy model presented in \cite{KolbLindeRiotto}, with some minor
modifications. Before doing so let us briefly mention two potential
problems with the GUT baryogenesis scenario in general. First, there
is the possibility that GUT symmetry is restored after inflation by a
thermal or nonthermal \cite{nonthermal,KasuyaKawasaki} phase
transition. This could lead to the production of cosmologically
dangerous topological defects. Fortunately it seems unlikely that
either type of phase transition would be strong enough to restore the
symmetry, at least for large portions of parameter space
\cite{KasuyaKawasaki,KolbLindeRiotto}. The second potential problem
with the scenario is that gravitinos are overproduced and spoil
nucleosynthesis. This also does not look like a major obstacle because
the late decay of the inflaton has the ability to sufficiently dilute
the gravitino abundance produced early on
\cite{KolbLindeRiotto,gravitino}.

The toy baryogenesis model we will analyze is as follows.  When the
production of massive particles shuts off ($\Phi=M_\chi/\sqrt{|g|}$)
\cite{capitalM}, there is a fraction of energy $\delta_0\equiv
\rho_\chi^0/\rho_\phi^0$ in the GUT boson field $\chi$. As
indicated by our results in section~\ref{sec:massive}, we will take
$\delta_0\ll 1$.  Since the $\chi$ bosons are only moderately
relativistic at creation ($k\sim M_\chi$) they quickly become ``cold''
as the Universe expands and can be approximated as having zero kinetic
energy.  We assume that they decay into light particles with a
constant decay rate $\Gamma_\chi$. The decay of each $\chi -\bar\chi$
pair produces $\epsilon$ baryons, where
$\epsilon$ should be thought of
as an effective CP violating parameter \cite{kolb}. In a realistic
model $\epsilon$ is expected to be small. We also assume that the
decay products thermalize instantly, which is a good approximation if
they are strongly coupled. The decay of what remains of the inflaton
field is modeled by a constant rate $\Gamma_\phi$, and we will assume a
hierarchy of time scales
\begin{equation}
H_0^{-1}\ll\Gamma_\chi^{-1}\ll\Gamma_\phi^{-1}\,,
\end{equation}
where $H_0$ is the
initial Hubble parameter. Finally, we make one important
simplification over the model presented in \cite{KolbLindeRiotto}: it
is assumed that at all times the temperature of the decay products is
much smaller than the mass $M_\chi$. This allows us to compute the
baryon to entropy ratio $n_B/s$ analytically.
The precise criterion for the validity of the inequality
$M_\chi\gg T_{\rm max}$ will be given below.

With these assumptions, the $\chi$ and baryon number densities {\it per
comoving volume\/}, $N_\chi=n_\chi (a/a_0)^3$ and $N_B=n_B (a/a_0)^3$,
evolve according to
\begin{equation}
\dot N_\chi =-\Gamma_\chi N_\chi\,,\qquad 
\dot N_B=-\epsilon\dot N_\chi\,,
\label{eq: baryogenesis eom}
\end{equation}
with initial conditions $N_B(t=0)=0$ and $N_\chi(t=0)\equiv N_\chi^0 =
n_\chi^0$.  Note that we have neglected the production of $\chi$
particles via inverse decays, which is justified because of our
assumption that the temperature $T\ll M_\chi$ at all times. This
condition will be made quantitative below.  The solutions for $N_B$
and $N_\chi$ are
\begin{equation}
N_\chi =N_\chi^0{\rm e}^{-\Gamma_\chi t}\,,\qquad 
N_B=\epsilon N_\chi^0\left(1-{\rm e}^{-\Gamma_\chi t}\right)
\end{equation}
The first step in computing the baryon to entropy ratio is to solve
for the energy density of the baryon fluid, $\rho_{\rm rad}$, which is
related to the entropy density $s$ by $s=
(2\pi^2/45)(30/\pi^2)^{3/4}g_*^{1/4}\rho_{\rm rad}^{3/4}$, since the
baryon fluid is relativistic and thermal. Here $g_*$ is the number of
relativistic degrees of freedom, and we are assuming that the entropy
is dominated by the relativistic particles.  We will consider
separately the {\it massless inflaton\/}, whose energy density scales
as $\rho_\phi\propto a^{-4}$, and the {\it massive\/} inflaton, with
$\rho_\phi\propto a^{-3}$.

\bigskip

For the {\it massless\/} inflaton we will assume that the universe 
is radiation dominated at all times, which requires 
$\rho_\phi^0\gg \rho_\chi^0(H_0/\Gamma_\chi)^{1/2}$.
Then 
\begin{eqnarray}
\frac{d}{dt}\left(\rho_{\rm rad}
\left[\frac{a}{a_0}\right]^4\right) &=&
-\frac{a}{a_0}M_\chi\dot N_\chi
-\frac{d}{dt}\left(\rho_\phi
\left[\frac{a}{a_0}\right]^4\right)
\label{eq: rho rad}\\
\frac{a}{a_0} &=& \left(2H_0 t+1\right)^{\frac{1}{2}}
\,,\qquad H_0^2=\frac{1}{3M_{\rm P}^2}\rho_\phi^0\, .
\label{eq: rho and massless scale factor}
\end{eqnarray}
Note that in order to compute the late
time contribution of the last term on the 
right hand side of (\ref{eq: rho rad})
to $\rho_{\rm rad}$, 
we need not know the details of how the inflaton
decays, {\it i.e.\/} the exact time dependence of
$\rho_\phi (a/a_0)^4$. The relevant information 
is the time scale $\Gamma_\phi^{-1}$ at which the decay takes place. 
We will first study the solution to 
Eq.~(\ref{eq: rho rad})
at early times ($t\ll\Gamma_\chi^{-1}\ll\Gamma_\phi^{-1}$)
to obtain an expression for the maximum temperature. This will allow
us to derive the inequality that needs to be satisfied in order that
$T_{\rm max} \ll M_\chi$. Subsequently we will obtain the solution at 
intermediate ($\Gamma_\chi^{-1}<t<\Gamma_\phi^{-1}$) and 
late times ($t>\Gamma_\phi^{-1}$), which will be  
used to compute $n_B/s$.

Early times ($t\ll\Gamma_\chi^{-1}$) are relevant since, as we will see
in a moment, the temperature reaches its maximum value at $t\sim
H_0^{-1}\ll\Gamma_\chi^{-1}$. Using the fact that in our simple model
$\rho_\phi=(a_0/a)^4 \rho_\phi^0\exp(-\Gamma_\phi t)$ and early on
$\dot N_\chi\approx-\Gamma_\chi N_\chi^0$, we find
\begin{equation}
\rho_{\rm rad}\simeq
\rho_\chi^0\frac{\Gamma_\chi}{H_0}\frac{1}{3}
\frac{(a/a_0)^3-1}{(a/a_0)^4}
+\rho_\phi^0\frac{\Gamma_\phi}{H_0}\frac{1}{2}
\frac{(a/a_0)^2-1}{(a/a_0)^4}
\end{equation}
Note that for $a/a_0\gg 1$ the two terms on the right hand side
scale as $a^{-1}$ and $a^{-2}$, respectively. Hence 
$\rho_{\rm max}$ is reached when $a/a_0$ is
of order {\it one\/}, {\it i.e.\/} $t\sim H_0^{-1}$, and we obtain 
\begin{eqnarray}
\rho_{\rm max}\equiv 
\frac{\pi^2}{30}g_*T^4_{\rm max}\simeq
\cases{
4^{-4/3}\rho_\chi^0\frac{\Gamma_\chi}{H_0}\,, 
\;\;{\rm at}\;\; a=4^{1\over 3}
\,, & when $\rho_\chi\Gamma_\chi>\rho_\phi\Gamma_\phi$\cr
\frac{1}{8}\rho_\phi^0\frac{\Gamma_\phi}{H_0}\,, 
\;\;{\rm at}\;\; a=2^{1\over 2}
\,, & when $\rho_\chi\Gamma_\chi<\rho_\phi\Gamma_\phi$\cr
}
\end{eqnarray}
so that $T_{\rm max}\ll M_\chi$ if
\begin{equation}
\rho_\chi^0\Gamma_\chi\, ,\; \rho_\phi^0\Gamma_\phi\ll g_*M_\chi^4H_0
\,.
\label{eq: consistency check I}
\end{equation}
In this paper we assume that Eq.~(\ref{eq: consistency check I}) is
satisfied.  In addition to Eq.~(\ref{eq: consistency check I}) there
is an upper bound on the baryon to entropy ratio $n_B/s$ which can be
used as a consistency check. Since we assume instant thermalization of
the baryon fluid and $T_{\rm max} < M_\chi$, the number density of
relativistic particles $g_*n_{\rm rad}$ is always {\it larger} than
the number density of decayed $\chi$-$\bar \chi$ pairs,
$n_B/\epsilon$. Since the number density of relativistic particles is
related to their entropy {\it via\/}
$n_{\rm rad}=[\zeta(2)/4\zeta(4)]s\simeq 0.28 s$
(where $\zeta(z)$ is the Riemann zeta function) we hence have 
\begin{equation}
\left(\frac{n_B}{s}\right) \le
\epsilon \frac{\zeta(3)}{4\zeta(4)}\; .
\label{eq: absolute baryon-to-entropy bound}
\end{equation}
We will now derive an expression for the baryon to entropy ratio in
our model.  At intermediate ($\Gamma_\phi^{-1}>t>\Gamma_\chi^{-1}$)
and late times ($t>\Gamma_\phi^{-1}$) the scale factor is to a good
approximation $a\simeq (2H_0t)^{1/2}$. Hence from Eq.~(\ref{eq: rho rad})
\begin{equation}
\rho_{\rm rad}\left(\frac{a}{a_0}\right)^4=
\left(\frac{\pi}{2}\right)^{\frac{1}{2}}
\left(\frac{H_0}{\Gamma_\chi}\right)^{\frac{1}{2}}
\rho_\chi^0+
\left\{
1-\exp\left[
-\frac{\Gamma_\phi}{2H_0}\left(\frac{a}{a_0}\right)^2
\right]
\right\}\rho_\phi^0\,,
\label{eq: intermediate and late times rho I}
\end{equation}
where we used
$\int_0^\infty dt\, t^{1/2}\exp(-t)=\pi^{1/2}/2$.
The second term on the right hand side of 
Eq.~(\ref{eq: intermediate and late times rho I}) 
first grows $\propto a^2$ and then, 
when the inflaton decays
at $t\sim \Gamma_\phi^{-1}$, it attains a constant
value $\rho_\phi^0$. Since we are assuming that $\rho_\phi^0\gg
\rho_\chi^0(H_0/\Gamma_\chi)^{1/2}$ we find the ``final'' baryon to
entropy ratio
\begin{equation}
\frac{n_B}{s}\simeq
\epsilon
\left(\frac{\rho_\chi^0}{g_*M_\chi^4}
\right)^{\frac{1}{4}}
\left(\frac{\rho_\chi^0}{\rho_\phi^0}
\right)^{\frac{3}{4}}\quad\quad {\rm (massless\; inflaton)}
\,,
\label{eq: baryon-to-entropy ratio II}
\end{equation}
where we used $N_B=\epsilon N_\chi^0$ for $t\gg \Gamma_\chi^{-1}$.
For simplicity in this, and in the results below, we ignore
multiplicative numerical constants that change the final result by
less than about $20\%$. The remarkable feature of Eq.~(\ref{eq:
baryon-to-entropy ratio II}) is that the final baryon to entropy ratio
does not depend on the decay constants $\Gamma_\chi$ and
$\Gamma_\phi$.  Why this is so can be understood as follows.  As long
as the temperature never rises close to $M_\chi$, so that the massive
$\chi$ particles are never re-populated by inverse decays, they simply
decay out of equilibrium and create $N_B\simeq \epsilon N_\chi^0$
baryons, which is independent of the decay rate $\Gamma_\chi$.  The
entropy produced per comoving volume, if dominated by the inflaton
decay, is $s(a/a_0)^3\simeq [\rho_\phi (a/a_0)^4]^{3/4} \simeq
(\rho_\phi^0)^{3/4}$, again independent of the decay rate
$\Gamma_\phi$. Hence the final baryon to entropy ratio is also
independent of the decay rates. In particular, even if the decay is
not as simple as the one-particle out-of-equilibrium decay, but a more
complex process like parametric resonance, we expect Eq.~(\ref{eq:
baryon-to-entropy ratio II}) to be valid as long as the inflaton decay
time is much greater than the massive particles decay time,
$\Gamma_\phi\ll \Gamma_\chi$, and the entropy production is dominated
by the late time inflaton decay. This concludes our discussion of the
massless inflaton, and we now turn to the massive case.

\bigskip

When the inflaton is {\it massive\/}, its number density per comoving
volume decays as $N_\phi\equiv n_\phi (a/a_0)^3
=N_\phi^0\exp(-\Gamma_\phi t)$. The evolution equation for the
radiation density is then
\begin{eqnarray}
\frac{d}{dt}\left(\rho_{\rm rad}
\left[\frac{a}{a_0}\right]^4\right) &=&
-\frac{a}{a_0}\left(M_\chi\dot N_\chi+M_\phi\dot N_\phi\right)
\\
\frac{a}{a_0} &=& \left(\frac{3}{2}H_0 t+1\right)^{\frac{2}{3}}
\,,\qquad H_0^2=\frac{1}{3M_{\rm P}^2}\rho_\phi^0\,.
\end{eqnarray}
Note that because of our assumption that $\rho_\phi^0\gg\rho_\chi^0$
the universe will remain matter dominated until the inflaton decays at
$t\sim\Gamma_\phi^{-1}$. At early times ($t\ll\Gamma_\chi^{-1}\ll
\Gamma_\phi^{-1}$) $\dot N_\chi\approx -\Gamma_\chi N_\chi^0$, and hence
\begin{equation}
\rho_{\rm rad}\simeq
\left[\rho_\chi^0\frac{\Gamma_\chi}{H_0}+
\rho_\phi^0\frac{\Gamma_\phi}{H_0}
\right]
\frac{2}{5}
\frac{(a/a_0)^{5/2}-1}{(a/a_0)^4}\,.
\end{equation}
This peaks at $a/a_0=(8/3)^{2/5}\simeq 1.48$ ($t\simeq 0.52
H_0^{-1}$), leading to a maximum
temperature
\begin{eqnarray}
\frac{T_{\rm max}}{M_\chi}\simeq
\left(
\frac{\rho_\chi^0}{g_*M_\chi^4}\frac{\Gamma_\chi}{H_0}+
\frac{\rho_\phi^0}{g_*M_\chi^4}\frac{\Gamma_\phi}{H_0}
\right)^{\frac{1}{4}}\,,
\end{eqnarray}
%
Thus  $T_{\rm max}\ll M_\chi$ requires
\begin{equation}
\rho_\chi^0\Gamma_\chi+ \rho_\phi^0\Gamma_\phi
\ll g_*M_\chi^4H_0\,,
\label{eq: consistency check II}
\end{equation}
which is the same constraint as in the massless inflaton case,
Eq.~(\ref{eq: consistency check I}).

At intermediate times
($\Gamma_\phi^{-1}>t>\Gamma_\chi^{-1}$)
the scale factor is
approximately $a/a_0\simeq (3H_0t/2)^{2/3}$ and hence
\begin{equation}
\rho_{\rm rad}\left(\frac{a}{a_0}\right)^4=
\left(\frac{2}{3}\right)^{\frac{1}{3}}
\Gamma\left(\frac{2}{3}\right)
\left(\frac{H_0}{\Gamma_\chi}\right)^{\frac{2}{3}}
\rho_\chi^0+
\frac{2}{5}
\frac{\Gamma_\phi}{H_0}\rho_\phi^0
\left(\frac{a}{a_0}\right)^{\frac{5}{2}}\,,
\end{equation}
where we used $\int_0^\infty dt \,t^{2/3}\exp(-t)=
(2/3)\Gamma(2/3)\simeq 0.90$. This leads to a baryon to entropy ratio
\begin{equation}
\frac{n_B}{s}\simeq 
\epsilon \left(\frac{\rho_\chi^0}{g_*M_\chi^4}
\right)^{\frac{1}{4}} \left(\frac{\Gamma_\chi}{H_0}
\right)^{\frac{1}{2}} \left[
1+\frac{2}{5\Gamma(2/3)}\left(\frac{3}{2}\right)^{\frac{1}{3}}
\frac{\rho_\phi^0}{\rho_\chi^0}
\left(\frac{\Gamma_\chi}{H_0}\right)^{\frac{2}{3}}
\left(\frac{\Gamma_\phi}{H_0}\right)
\left(\frac{a}{a_0}\right)^{\frac{5}{2}} \right]^{-\frac{3}{4}}\, .
\end{equation}
%
While the second term in square
brackets is  subdominant, $n_B/s$ is to a good approximation constant
and determined by the entropy release from the massive particle
decay. Eventually the second term takes over and $n_B/s\propto
a^{-15/8}$ until the inflaton decays at $t\sim \Gamma_\phi^{-1}$.

At late times ($t>\Gamma_\phi^{-1}$) the radiation energy density is
\begin{equation}
\rho_{\rm rad}\left(\frac{a}{a_0}\right)^4=
\left(\frac{2}{3}\right)^{\frac{1}{3}}
\Gamma\left(\frac{2}{3}\right)
\left(\frac{H_0}{\Gamma_\phi}\right)^{\frac{2}{3}}
\rho_\phi^0
\,,
\end{equation}
leading to a ``final'' baryon to entropy ratio
\begin{equation}
\frac{n_B}{s}\simeq
\epsilon\left(\frac{\rho_\chi^0}{g_*M_\chi^4}
\right)^{\frac{1}{4}}
\left(\frac{\Gamma_\phi}{H_0}
\right)^{\frac{1}{2}}
\left(\frac{\rho_\chi^0}{\rho_\phi^0}
\right)^{\frac{3}{4}}\quad\quad {\rm (massive\; inflaton)}.
\label{eq: baryon-to-entropy ratio IV}
\end{equation}
This means that for a massive inflaton the final value of the baryon
to entropy ratio is suppressed by a factor $(\Gamma_\phi/H_0)^{1/2}$
in comparison to the massless case, Eq.~(\ref{eq: baryon-to-entropy
ratio II}). Note also that $n_B/s$ is dependent on the decay rate
$\Gamma_\phi$, unlike in the massless case, but still independent of
$\Gamma_\chi$.  Intuitively, this dependence can be understood as
follows.  For a massive inflaton the energy density scales as
$a^{-3}$, which is slower than for the massless baryonic fluid
($\propto a^{-4}$).  Hence, later inflaton decay leads to more
abundant entropy production relative to the entropy of the baryonic
fluid. 

Before giving an estimate of $n_B/s$ based on the equations derived
above we would like to emphasize that the assumptions $T_{\rm max}\ll
M_\chi$ and $\rho_\phi^0\gg\rho_\chi^0$ are in no way necessary for
successful GUT baryogenesis after preheating. These conditions were
imposed only to enable us to obtain simple analytical expressions for
the baryon to entropy ratio. For a more sophisticated numerical
treatment in which both these assumptions are relaxed see
\cite{KolbLindeRiotto}.

In order to obtain estimates for the baryon to entropy ratios that
could be produced at preheating in a theory with negative coupling we
need to know $\rho_\chi/\rho_\phi$ at the time when the massive
particle production shuts off. To this end we ran our code for a range
of ``natural'' $g$ values with $m\chi=10^{14}$GeV. The results are
shown in figures~12(a) for a massless inflaton and~12(b) for a massive
inflaton. We used the COBE values $\lambda_\phi=3\times10^{-13}$ in the
former case and $\lambda_\phi=3\times10^{-13}$, $m_\phi=10^{13}$GeV in
the latter case.  We point out that for natural values of $g$ there is
no production with these parameters if $g>0$. The coupling
$\lambda_\chi$ was chosen so that $r=10$ in all of the runs.

The ratio of energy densities $\rho_\chi^0/\rho_\phi^0$ can be easily
obtained from figure~12. To see this recall that for the massive
nonrelativistic $\chi$ particles the variance is simply related to the
energy density by
$\rho_\chi=M_\chi^2\langle(\delta\chi)^2\rangle$. 
Now if there is any
growth at all we find that $\langle(\delta\chi)^2\rangle$ does not
{\it decrease} until the time at which the massive particle production
is shut off. Hence the values in figure~12 are proportional to
$\rho_\chi^0$, the energy density at the instant when production
ceases. The corresponding value $\rho_\phi^0$ can be obtained by
estimating $\Phi$ at this time from 
Eq.~(\ref{eq: negative g shut off}). 
Note that this last step is a good approximation only for very
heavy particles such as the ones we are considering here. For lighter
$\chi$ particles the $3\lambda_\chi\langle(\delta\chi)^2\rangle$
contribution to $m^2_{\chi\; {\rm eff}}$ dominates over $m^2_\chi$
when the production shuts off. 

Here we will concentrate on the massless inflaton case and obtain an
estimate for $n_B/s$ based on Eq.~(\ref{eq: baryon-to-entropy ratio
II}) \cite{B massless}. Using
$\rho_\chi^0=M_\chi^2\langle(\delta\chi)^2\rangle_0$ and 
$\rho_\phi^0=\lambda_\phi(\Phi^0)^4/4\approx \lambda_\phi
M^4_\chi/4|g|^2$
we obtain
\begin{equation}
\frac{1}{\epsilon}\frac{n_B}{s}\simeq 
g_*^{-\frac{1}{4}}
\left(\frac{4g^2}{\lambda_\phi}\right)^{3/4} 
\frac{\langle(\delta\chi)^2\rangle_0}{M_\chi^2}\;.
\label{eq: baryon-to-entropy ratio V}
\end{equation}
From figure~12(a) we see that the largest valley values are about
$\langle(\delta\chi)^2\rangle_0/M_{\rm P}^2 \approx 5\times10^{-10}$.
It is easy to check that for these values the constraint Eq.~(\ref{eq:
consistency check I}) is indeed satisfied provided
$\Gamma_\phi,\,\Gamma_\chi\ll H_0$.  Using this value for
$\langle(\delta\chi)^2\rangle_0$ in Eq.~(\ref{eq: baryon-to-entropy
ratio V}) we find
\begin{equation}
\frac{1}{\epsilon}\frac{n_B}{s}\sim 10^{-3}\;,
\end{equation}
where we have set $g_*=100$.  This leaves plenty of parameter space
for baryogenesis consistent with the observed value $(n_B/s)_{\rm
observed}\sim 4-7\times 10^{-11}$.

\section{Summary and Concluding Remarks}
\label{sec:conclusion}

In this work we investigate the preheating dynamics in chaotic
inflationary models with a negative cross-coupling between the
inflaton and a second scalar field. Such couplings are completely
natural in theories with more than one scalar multiplet. The details
of the inflaton decay are extremely complicated, and an accurate
treatment requires that one solve the full nonlinear
problem. Fortunately the dynamics is dominated by states with large
occupation numbers which admit a semiclassical description, so the
problem can be treated by solving the classical equations of motion
numerically on the lattice. We use a general renormalizable two field
scalar theory that respects a ${\cal Z}2\times {\cal Z} 2$ discrete
symmetry, {\it i.e.\/} the potential contains quadratic and quartic
terms as in Eq.~(\ref{eq:V}).  Perhaps our most important finding is
that the negative coupling can catalyze the inflaton decay into a
heavy scalar field, opening a new window for baryogenesis at the grand
unified scale with a natural choice of parameters.

In order to make our presentation self-contained, we review some of
the important properties of inflaton decay for massless fields with
positive coupling in section~\ref{sec:massless}. In the old theory of
reheating with perturbative inflaton decay \cite{old theory} the
relevant quantity was assumed to be the reheat temperature.  Based on
the reheat temperature one can infer whether the grand unified
symmetry is restored, whether there is a monopole problem, whether GUT
baryogenesis is feasible, etc.  In the new theory of reheating the
inflaton decay products are typically low energy, far from equilibrium
excitations, so that one cannot meaningfully assign a temperature to
this infrared condensate.  Instead, the relevant measure of the
inflaton decay efficiency is the maximum variance the decay products
reach since, with the variance known, all of the problems mentioned
above can be addressed.  Hence we focus on obtaining estimates of the
maximum variances reached during preheating.  The numerical results
for cross-coupling $g>0$ are presented in figures~3 and~5, and
estimates for the maximum variances are given in
Eqs.~(\ref{eq:variance II}) - (\ref{eq:variance Ib}). (Recall that
$\phi$ is the inflaton and $\chi$ is a scalar field coupled to
it). The estimates are based on the observation that for much of the
relevant parameter space the backreaction of the produced particles
terminates the exponential growth when the variances reach a certain
magnitude, which can be understood intuitively in terms of the
stability chart of the Mathieu equation, shown in figures~1(a)
and~(b). Once the critical variance is reached, a slowly varying
state, which we call the scattering regime, sets in.  At this stage of
the evolution the slow particle production cannot keep up with the
expansion of the universe and the variances start decreasing. As
explained in section~\ref{sec:masslessII} and \cite{ProkopecRoos}, the
fact that our nonlinear lattice calculations take scatterings into
account is crucial for the determination of the correct maximum
variances. An important consequence of the existence of the scattering
regime for large values of $g$ and $\lambda_\chi$ is the fact that the
inflaton decays more efficiently for smaller couplings. For example, a
massless inflaton decays fastest and produces the largest variances if
the cross-coupling satisfies $\lambda_\phi\le g \le 10^2 \lambda_\phi$
(provided the self coupling of the decay product $\chi$ is not too
large, $\lambda_\chi\le g^{3/2}/\lambda_\phi^{1/2}$).  The explanation
of this rather counter intuitive result -- that larger couplings lead
to less efficient decay -- is simply that larger couplings lead to
more efficient scatterings and stronger backreaction, and consequently
lower maximum variances. (For completeness we note that for a massive
inflaton the most efficient decay into massless $\chi$ particles
occurs when $q_0 \sim q_0^{\rm min}$ as given in
Eq.~(\ref{eq:qmin,massless}), which translates into $10^{-7}\lesssim
g\lesssim 10^{-6}$ in the case when $\lambda_\chi$ is small. Here we
took for the inflaton mass $m_\phi\approx 2\times
10^{13}\hbox{GeV}$.)

We begin our investigation of the negative cross-coupling case by
considering the instability chart of the Mathieu equation (see
Eqs.~(\ref{eq:mathieu}) and (\ref{eq:A0,q0}), and figure~1), which is
essentially an oscillator equation with a time dependent
frequency. The Mathieu equation is a good starting point since it
describes the simplest possible case, namely the linearized mode
equations of the second field in a static space-time when the inflaton
is massive. Its usefulness goes far beyond that simple case,
however. The reason is that the main features of the stability chart
are extremely robust and carry over to both other models (such as a
massless inflaton) as well as the expanding universe. The Mathieu
equation contains the two parameters $A$ and $q$ which parametrize the
frequency squared of the oscillator. The third important parameter is
the instability index $\mu$.  The meaning of $\mu$ is simply that in
one oscillation of the inflaton, the amplitude of an unstable $\chi$
mode increases by a factor ${\rm e}^{2\pi\mu}$. In figures~1(a)
and~(b) we show the regions of stable and unstable solutions as a
function of $A$ and $q$, as well as some curves of constant $\mu$. The
only features of these charts which we need in order to obtain
analytical estimates for the maximum variances are (i) that the
maximal distance to the first instability band above the line $A=2|q|$
is approximately $\sqrt q$, and (ii) that $\mu$ is a rapidly
decreasing function of $A$ above the line $A=2|q|$. As stated above,
these features are extremely robust, and from them one can derive the
resonance shut-off condition Eq.~(\ref{eq: delta A}) on which all of
our estimates, for both positive and negative coupling, are based.

For a quartic interaction term $g\phi^2\chi^2/2$ with a positive
cross-coupling $g$, the phase space is limited to $A\ge 2|q|$. As a
consequence the instability exponent $\mu\lesssim 0.3$ for any $q$,
which can be seen nicely in figures~1(a) and~(b). Notice also that in
the first unstable band above $A=2|q|$ $\mu$ may peak at a value between
$\mu\approx 0.1$ and $\mu\approx 0.3$, depending on the exact value of
$|q|$.  On the other hand, for a negative cross-coupling, all of the
parameter space above $A= -2|q|$ is accessible. This means that 
$\mu\le (4/\pi)|q|^{1/2}\gg 1$ ({\it cf.\/} Eq. (\ref{eq: mu for large
A q})), and that there is a wide range of unstable modes: $\Delta k_{\rm
physical}\sim 2|q|^{1/2}\omega_\phi$.  The possibility of an
extremely rapid decay into a broad range of momenta motivated
us to study the negative cross-coupling case in detail. 

The following are our main observations and findings. As explained
above, naively the maximum value of the instability coefficient is
$\mu_{\rm max}\simeq (4/\pi)|q|^{1/2}$. However, such an instability
occurs only if the expectation value of the second field is zero at
the end of inflation, {\it i.e.\/} $\chi_0=0$. In a realistic
situation this is not the case since for $g$ negative the true minimum
of the $\chi$ potential energy is displaced from the origin (note that
in order to have a potential which is bounded from below the stability
condition Eq.~(\ref{eq: r}) must be satisfied). Our analytical
analysis and numerical simulations show that quite generically the
dynamics of the fields during inflation will drive $\chi_0$ towards
its true minimum. This implies that the natural initial condition at
the end of inflation is $\chi_0=\tilde\chi_0$, where $\tilde\chi_0$ is
the ($\phi_0$ dependent) location of the minimum of the $\chi$
potential, given by Eq.~(\ref{eq: chi_0}). Using this initial
condition we studied the case of massless fields with negative
coupling in section~\ref{sec:massless}. We found that during the early
stages of the post inflationary evolution the unstable modes grow just
as in the positive coupling case, with instability coefficient
$\mu\sim 0.2$. This can be observed in figures~2 and~4. The reason for
this behavior is as follows. As long as the $\chi$ zeromode follows
its true minimum $\chi_0=\tilde\chi_0$, no $\chi$ modes encounter
genuine {\it negative coupling instability\/} (negative frequency
squared), and hence they grow only via the standard positive $g$
resonance mechanism, with $q_{\rm eff}\simeq 2|q|$. This picture
breaks down once the backreaction of the created $\chi$ particles
becomes large enough to disrupt the evolution of $\chi_0$ so that it
stops following $\tilde\chi_0$.  When this happens the mode amplitudes
become unstable and grow very fast, quickly reaching values large
enough to terminate the instability, and the slowly varying scattering
regime sets in. During this stage we find that the dynamics of the
negative cross-coupling model differs profoundly from that with a
positive cross-coupling.  Perhaps the most striking difference is that
the $\chi$ variance peaks at a significantly larger value -- by a
factor $4|q|^{1/2}$ larger than in the positive $g$ case -- leading to
a much more complete inflaton decay into $\chi$ particles. This
difference can be understood from the instability chart in
figure~1. To shut off the resonance in the positive $g$ case, one
needs a backreaction of order $\delta A\sim q^{1/2}$, while in the
negative $g$ case a much larger shift $\delta A\sim 4|q|$ is
required. Estimates for the maximum variances with a negative
cross-coupling are given in Eqs.~(\ref{eq:variance IV})
and~(\ref{eq:variance V}). Another notable difference between $g>0$
and $g<0$ is the way the variances fluctuate between ``peak'' and
``valley'' values as the inflaton oscillates in the scattering regime.
We present a simple explanation of this phenomenon using the matching
of two oscillator solutions with different frequencies (see
Eqs.~(\ref{eq: toy model for X}) -- (\ref{eq: peak-valley II})).  The
peak to valley ratio in the case of massive fields is discussed at the
end of section~\ref{sec:massiveII.3}.
 
In section~\ref{sec:massive} we study in detail the decay of the
inflaton into massive particles ($m_\chi \neq
0)$. Sections~\ref{sec:massiveII.1} and~\ref{sec:massiveII.2} discuss
the positive coupling case for a massless and massive inflaton,
respectively. The main conclusion is that the estimates for the
maximum variances obtained in section~\ref{sec:masslessII}
(Eqs.~(\ref{eq:variance II}), (\ref{eq:variance III}),
and~(\ref{eq:variance I})) hold also for a massive $\chi$ field {\em
provided} that $q_0$ is larger than a critical value $q_0^{\rm min}$
given by Eqs.~(\ref{eq: minimum q massless}) and~(\ref{eq: minimum q
massive}). (see also \cite{var eq massive}.)

For $g<0$ the situation is very different.  The criterion for the {\it
complete\/} particle production shut-down is $m_\chi^2>|g|\Phi^2$
({\it cf.\/} Eq.~(\ref{eq: negative g shut off})), where
$\Phi=\Phi(t)$ is the inflaton amplitude, which decreases due to the
expansion of the Universe. As explained in
section~\ref{sec:massiveII.3}, particles can be produced only at the
moment when $m_\chi=|g|\phi_0^2(t)$, and while $\phi_0(t)$ is
growing. The minimum condition to encounter {\it any\/} massive
particle production is hence $|g|\ge m_\chi^2/\Phi_1^2$, where
$\Phi_1\approx 10^{18}\hbox{GeV}$ is the inflaton amplitude at the
first extremum in the oscillatory regime. For
$m_\chi=10^{14}\hbox{GeV}$, this gives $|g|\gtrsim 10^{-8}$, which agrees
well with our numerical results in figures~12(a) and~(b). To encounter
$n$ chances to grow, the required value of $|g|$ increases as given in
Eq.~(\ref{eq:g_n}).  Note that this value is many orders of magnitude
{\it smaller\/} than the value $g\gtrsim 10^{-3}$ required for the
production of equally massive particles in the positive $g$ case. The
particle production mechanism is very different from the usual
positive $g$ resonance, and we call it {\it negative coupling
instability\/}. The details are complicated, but the main mechanism
can be readily understood and is explained in
sections~\ref{sec:massiveII.3} -~\ref{sec:massiveII.5}. Characteristic
features of this new process are that massive particles are produced
in extremely rapid bursts over a broad range of momenta (much faster
and broader than in the positive $g$ case). This is illustrated in
figures~9 and 11(a). Another important feature are the sharp spikes in
the production of massive $\chi$ particles as a function of the
parameters, seen in figures~7, 10, and 12. This behavior is explained
with the aid of a simple analytical model in
section~\ref{sec:massiveII.3} (see Eqs.~(\ref{eq: solution to X II})
-- (\ref{eq:g_n})). The system may be considered ``chaotic'' in the
sense that it is quite impossible to predict the exact position and
amplitude of the spikes as a function of the parameters.

The main result of section~\ref{sec:massive} can be re-stated as
follows.  For a realistic choice of couplings, which in chaotic
inflationary models is constrained by the COBE satellite measurements
to be $\lambda_\phi\approx 3\times 10^{-13}$ or $m_\phi\lesssim
2\times 10^{13}\hbox{GeV}$, one can produce massive particles with
$m_\chi\sim 10^{14}\hbox{GeV}$ (as required by GUT baryogenesis
models), provided $|g|\gtrsim 10^{-8}$ and $\lambda_\chi > 3\times
10^{-4}$ (for stability) (see figures~12(a) and~(b)). This means that
the parameter space available for production of massive particles with
``natural'' coupling constants is appreciable: $10^{-6}\gtrsim
|g|\gtrsim 10^{-8}$, $1 > \lambda_\chi > 3\times 10^{-4}$, leaving
plenty of opportunity for baryogenesis model building.  This is to
be contrasted with the positive $g$ case for which $g\gtrsim 10^{-3}$ is
required. Such a large value leads to an unpleasant fine tuning
problem since the small value of $\lambda_\phi$ needs to be protected
against radiative corrections.

To obtain an estimate of the baryon asymmetry that could be produced
during preheating we study a simple toy model in
section~\ref{sec:baryogenesis}.  In short, the model can be described
as follows.  Initially a certain amount of energy is transferred from
the inflaton to a heavy GUT scalar field {\it via\/} the
nonperturbative mechanisms described in this paper. The result is a
cold (far from equilibrium) fluid of massive particles (with a mass
$M_\chi$), which we assume decays in a B and CP violating manner into
light degrees of freedom that instantly thermalize. In order to treat
the problem analytically we further assume that at all times the
temperature $T$ of the light relativistic particles is $\ll M_\chi$,
which requires that Eq.~(\ref{eq: consistency check I}) is
satisfied. The main results of this investigation are the final baryon
to entropy ratios given in Eqs.~(\ref{eq: baryon-to-entropy ratio II})
and~(\ref{eq: baryon-to-entropy ratio IV}) for a massless and massive
inflaton, respectively. These equations are obtained assuming a
``natural'' hierarchy of time scales,
$H_0\gg\Gamma_\chi\gg\Gamma_\phi$.  As can be seen from Eq.~(\ref{eq:
baryon-to-entropy ratio II}), under these conditions the baryon to
entropy ratio for a massless inflaton has the pleasant property that
it is independent of the decay rates $\Gamma_\chi$ and
$\Gamma_\phi$. We believe that this feature, which can be understood
intuitively and is explained in section~\ref{sec:baryogenesis}, lends
more credibility to our result since the final estimate is somewhat
independent of the details of the model. Eqs.~(\ref{eq:
baryon-to-entropy ratio II}) and~(\ref{eq: baryon-to-entropy ratio
IV}) say that the final baryon to entropy ratio is proportional to the
initial energy density in $\chi$ particles.  Since the energy density
of the (nonrelativistic) $\chi$ particles is proportional to their
variance ($\rho_\chi\approx M^2_{\chi}\langle(\delta\chi)^2\rangle$),
the spikes in figure~12 are directly imaged to spikes in 
baryon production. For a massless inflaton the final baryon to entropy
ratio is expressed in terms of the maximum variance in Eq.~(\ref{eq:
baryon-to-entropy ratio V}) \cite{B massless}. Our numerical results
indicate that it is not hard to obtain values in the range $n_{\rm
B}/s \sim \epsilon\times 10^{-3}$, where $\epsilon$ is the effective
$CP$ violating parameter in our model.  Clearly more realistic models
for GUT baryogenesis at preheating will have to be constructed to
verify the viability of this scenario, but the preliminary results
look promising.

Another interesting question that deserves further study is whether it
is possible to construct a model such that the {\em negative coupling
instability} discussed in this paper does not require any ``phase
mismatch'' to become operational. In other words, the aim is to obtain
natural initial conditions at the end of inflation that immediately
allow the inflaton to decay rapidly into a heavy scalar field. The
preliminary investigation of one realization of such a scenario is
underway. 
 
Close to completion of this work we received a preprint
[hep-ph/9704452] \cite{KofmanLindeStarobinskiiNew} in which the
preheating dynamics in an expanding universe is also studied in
detail. In this very interesting paper the authors focus exclusively
on the {\em massive inflaton}, {\em positive} $g$, {\em small}
$\lambda_\chi$ case, for which their conclusions partially overlap
with ours. A detailed comparison will be deferred to future
publications.

\acknowledgements This research was conducted using the resources of
the Cornell Theory Center, which receives major funding from the
National Science Foundation (NSF) and New York State, with additional
support from the Advanced Research Projects Agency (ARPA), the
National Center for Research Resources at the National Institutes of
Health (NIH), IBM Corporation, and other members of the center's
Corporate Partnership Program. We would like to thank Andrei Linde for
useful comments. BRG acknowledges partial funding from the NSF and
Alfred P. Sloan Foundation.  TP acknowledges funding from the NSF, and
hospitality of Columbia University, where part of this work was done.

\pagestyle{empty}

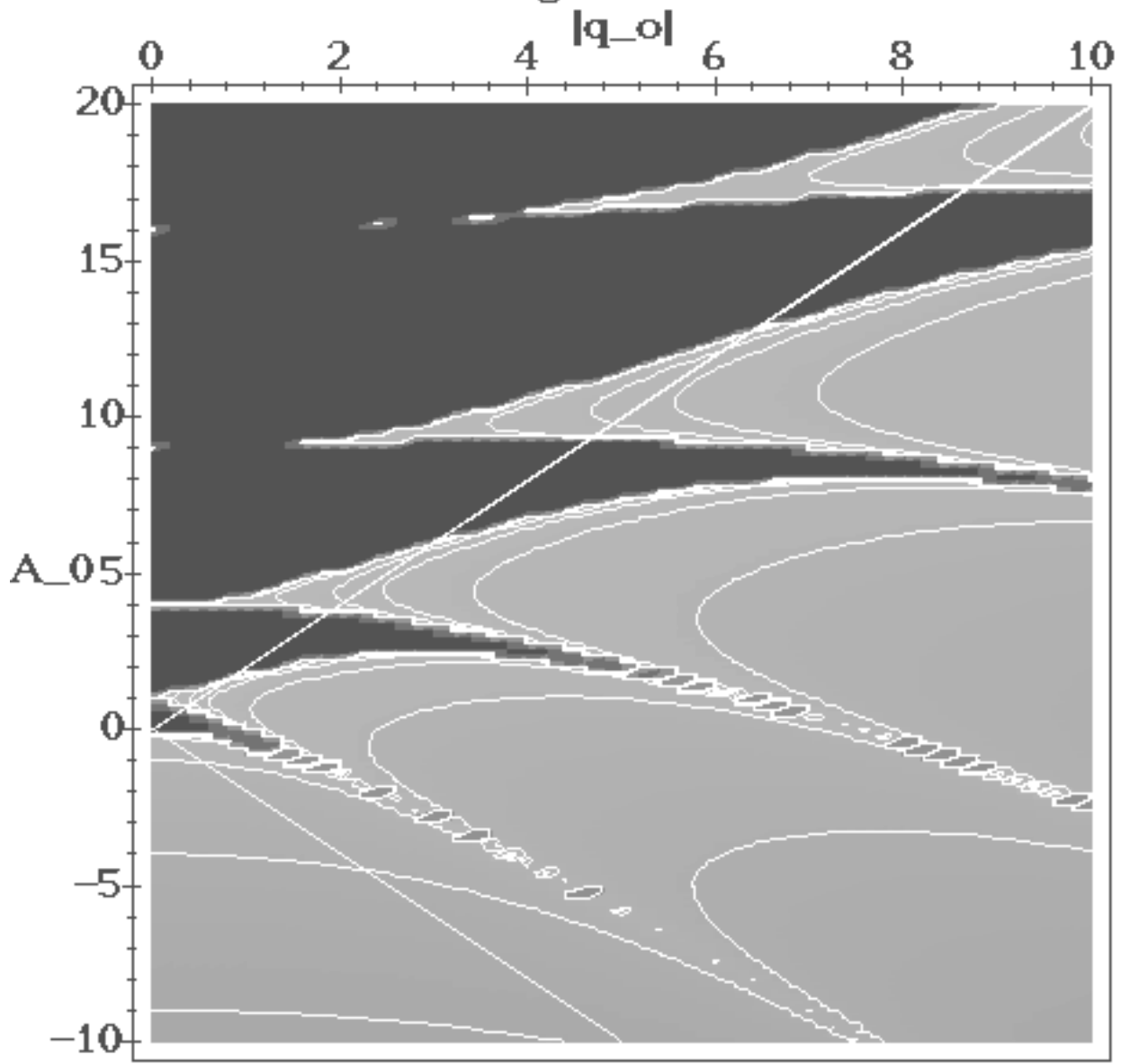
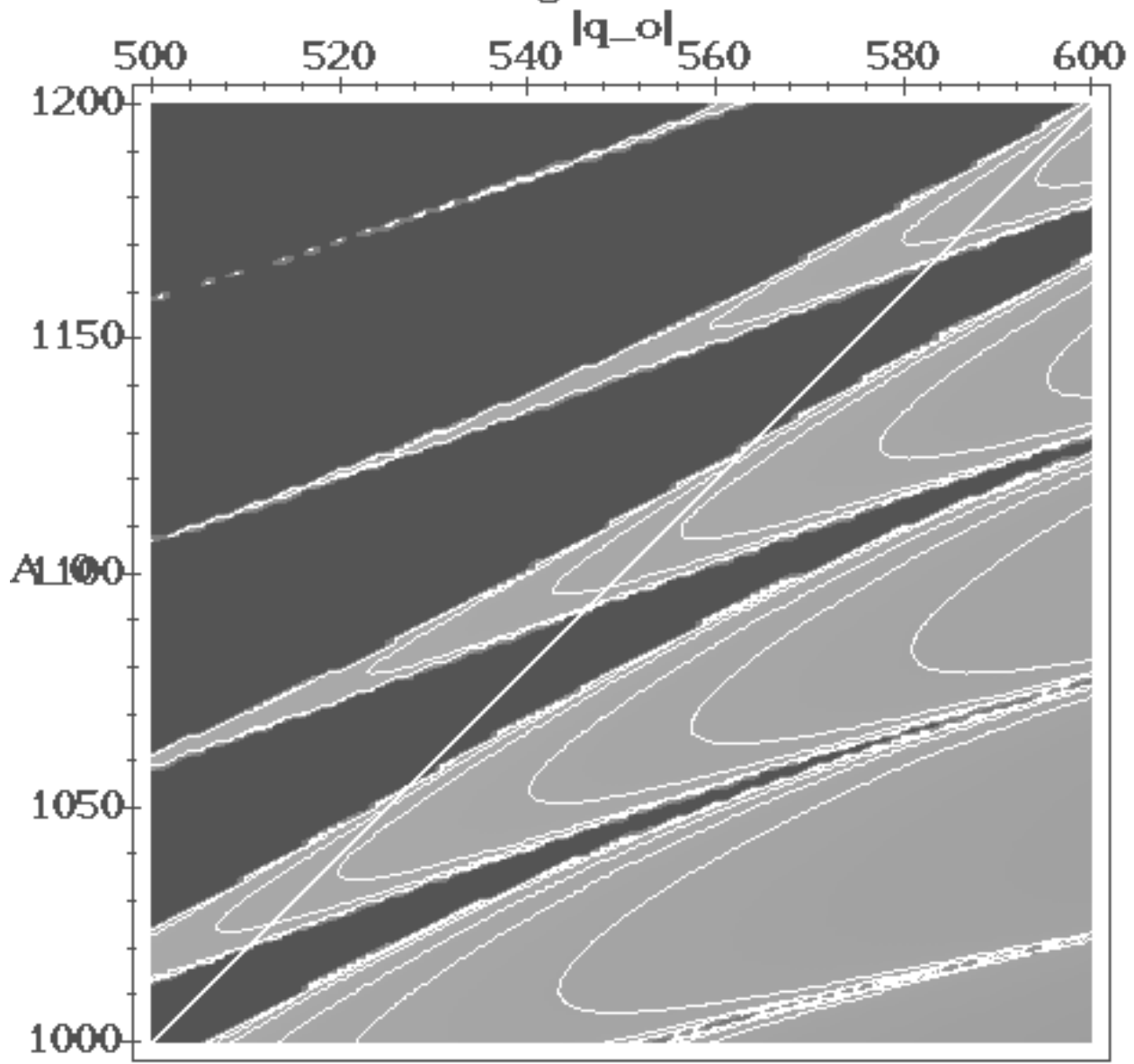
\begin{figure}
\caption{(a) : The stability chart of the Mathieu equation,
Eq.~(\protect\ref{eq:mathieu}).  The dark regions correspond to stable
solutions while the light regions correspond to exponential
instabilities. We also show contours of constant $\mu$, where $\mu$ is
the instablity index. The curve $\mu=0$ divides the parameter space
into stable and unstable regions. The contours shown are $\mu=$ \{0,
0.1, 0.2, 0.3, 0.5, 1.0, 2.0, 3.0\}. The lines $A_0=2|q_0|$ and
$A_0=-2|q_0|$ are also plotted. (The plot was generated numerically,
and in order to keep the file size small we used a fairly coarse
grid. The instability bands really extend all the way to $q_0=0$ at
the points $A_0=n^2$, $n=1,2,3,\dots$. Also, for negative $A_0$ the
narrow regions of stability form continuous bands rather than the
``island'' structure shown.)}
\label{fig:1}
\end{figure} 

\setcounter{figure}{0}

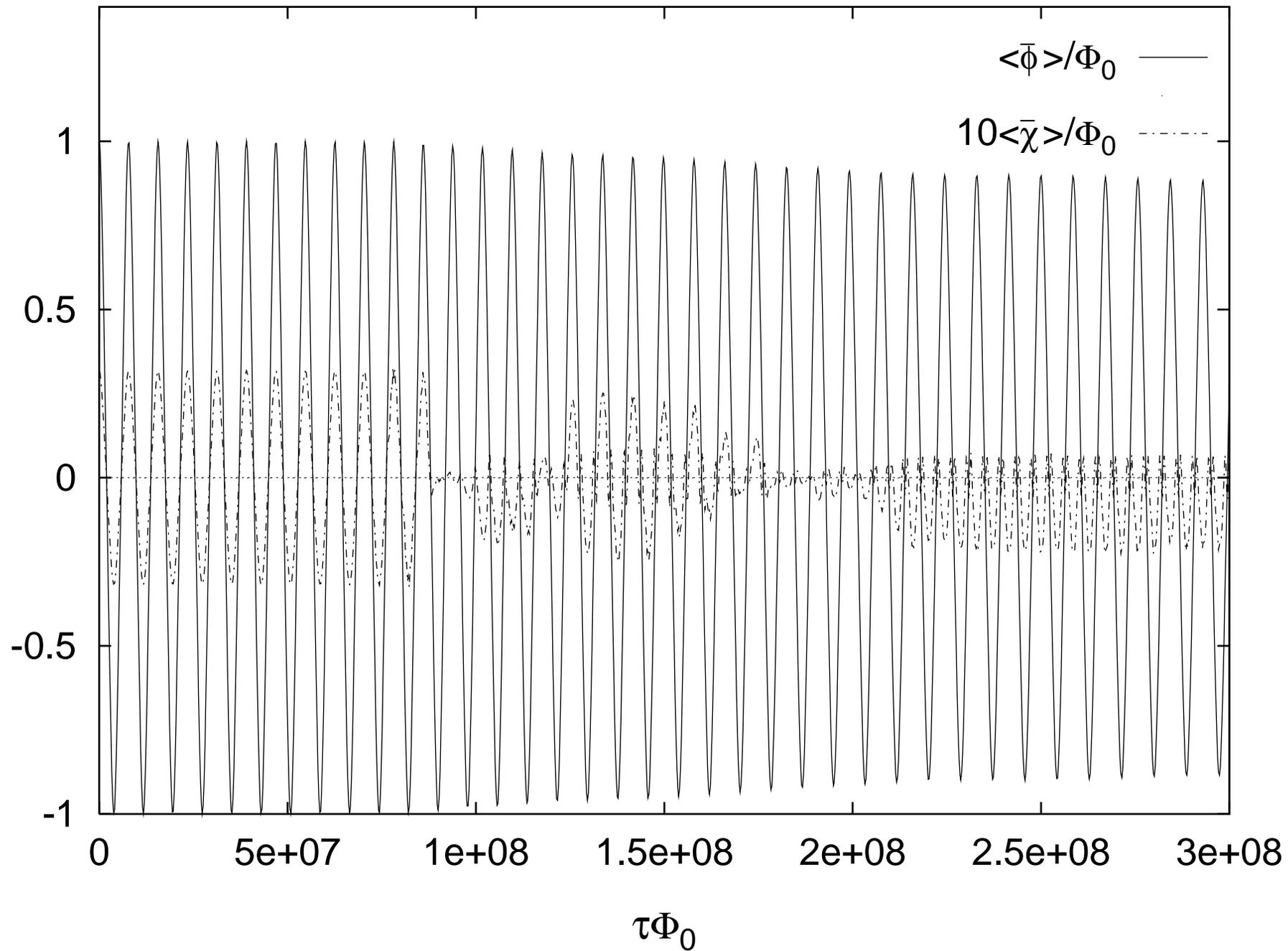
\begin{figure}
\caption{(b) : The stability chart of the Mathieu equation for
large $A_0$ and $q_0$. As in figure 1(a) the dark regions correspond
to stable solutions while the light regions correspond to exponential
instabilities.  The contours shown are $\mu=$ \{0, 0.1, 0.2, 0.3, 0.5,
1.0 \}, and the line $A_0=2|q_0|$ is also plotted. Notice that the
distance between resonance bands above $A_0=2|q_0|$ for fixed $q_0$ is
$\delta A_0 \approx |q_0|^{1/2}$. Notice also that
the instability index
$\mu$ decreases rapidly with increasing $A_0$.}
\end{figure} 
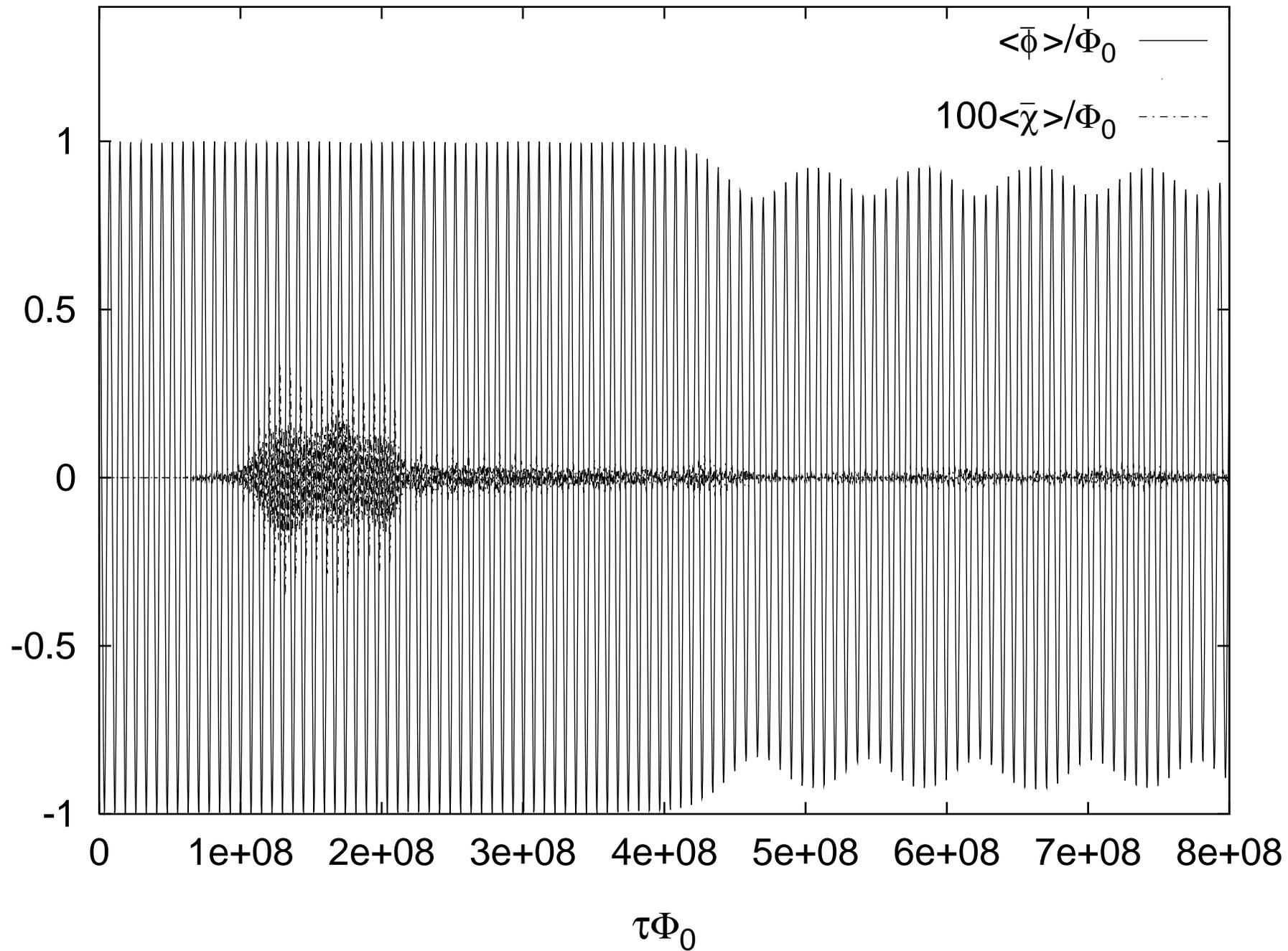
\begin{figure}
\caption{The expectation values of the fields as a function of time for
negative coupling, $|q_0|\approx 35$, $r=10$ ($m_\chi = m_\phi = 0$,
$\lambda_\phi=10^{-12}$, $\lambda_\chi=10^{-7}$, $g=-10^{-10}$).}
\end{figure}
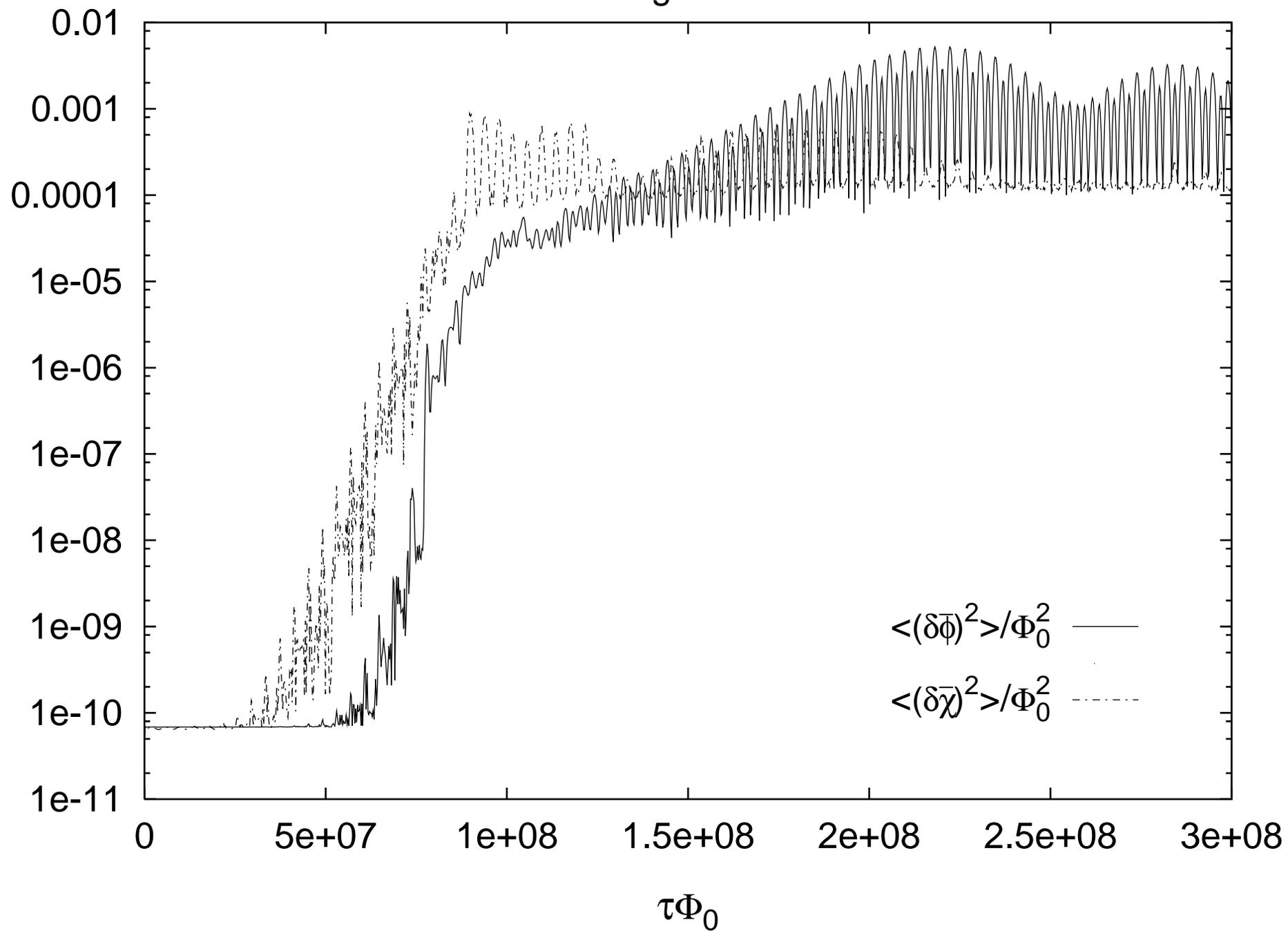
\begin{figure}
\caption{The expectation values of the fields as a function of time for
positive coupling, $q_0\approx 35$, $r=10$ ($m_\chi = m_\phi = 0$,
$\lambda_\phi=10^{-12}$, $\lambda_\chi=10^{-7}$, $g=10^{-10}$).}
\end{figure} 
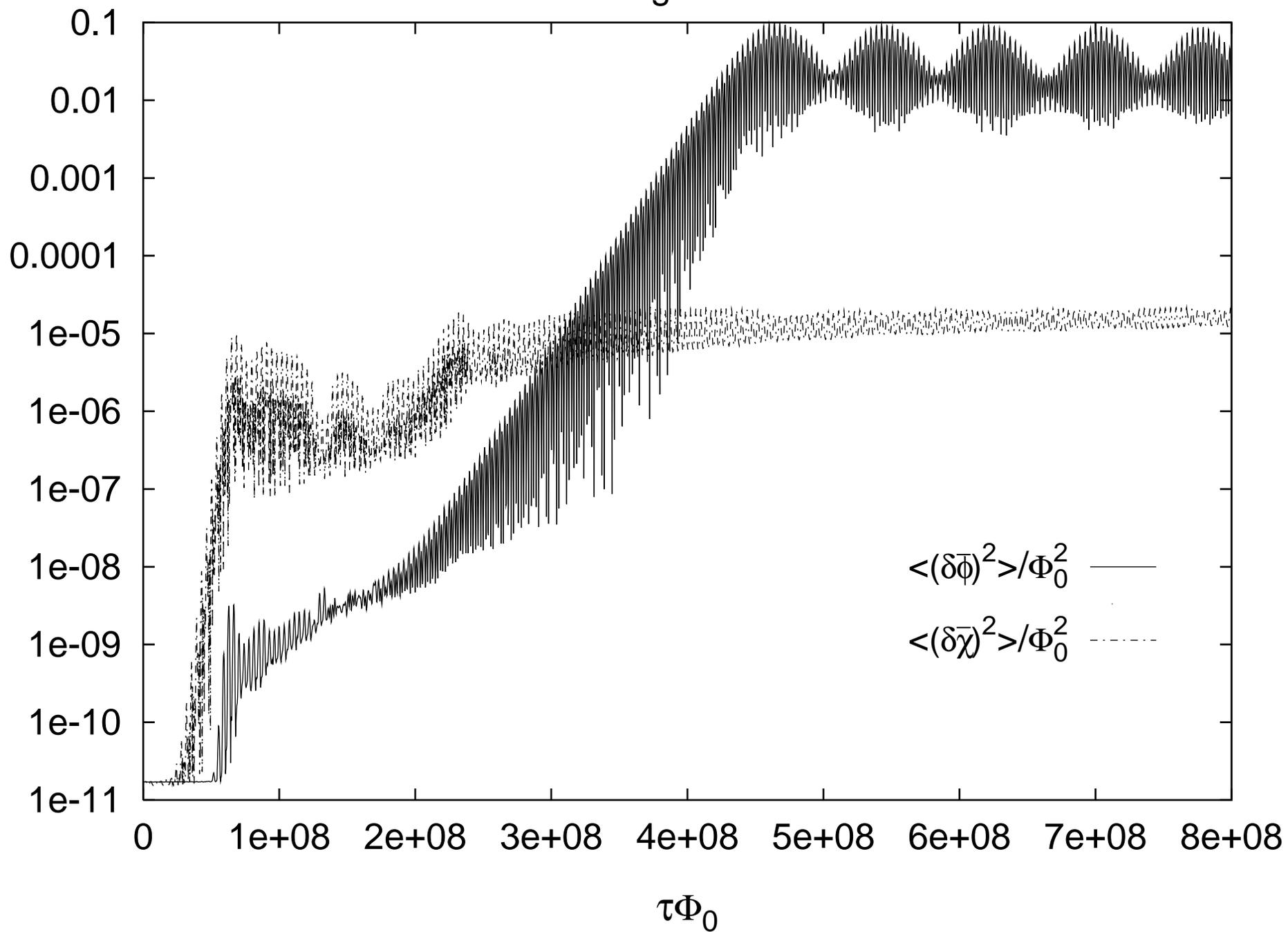
\begin{figure}
\caption{The variances of the fields as a function of time for
negative coupling, $|q_0|\approx 35$, $r=10$ ($m_\chi = m_\phi = 0$,
$\lambda_\phi=10^{-12}$, $\lambda_\chi=10^{-7}$, $g=-10^{-10}$).}
\end{figure}
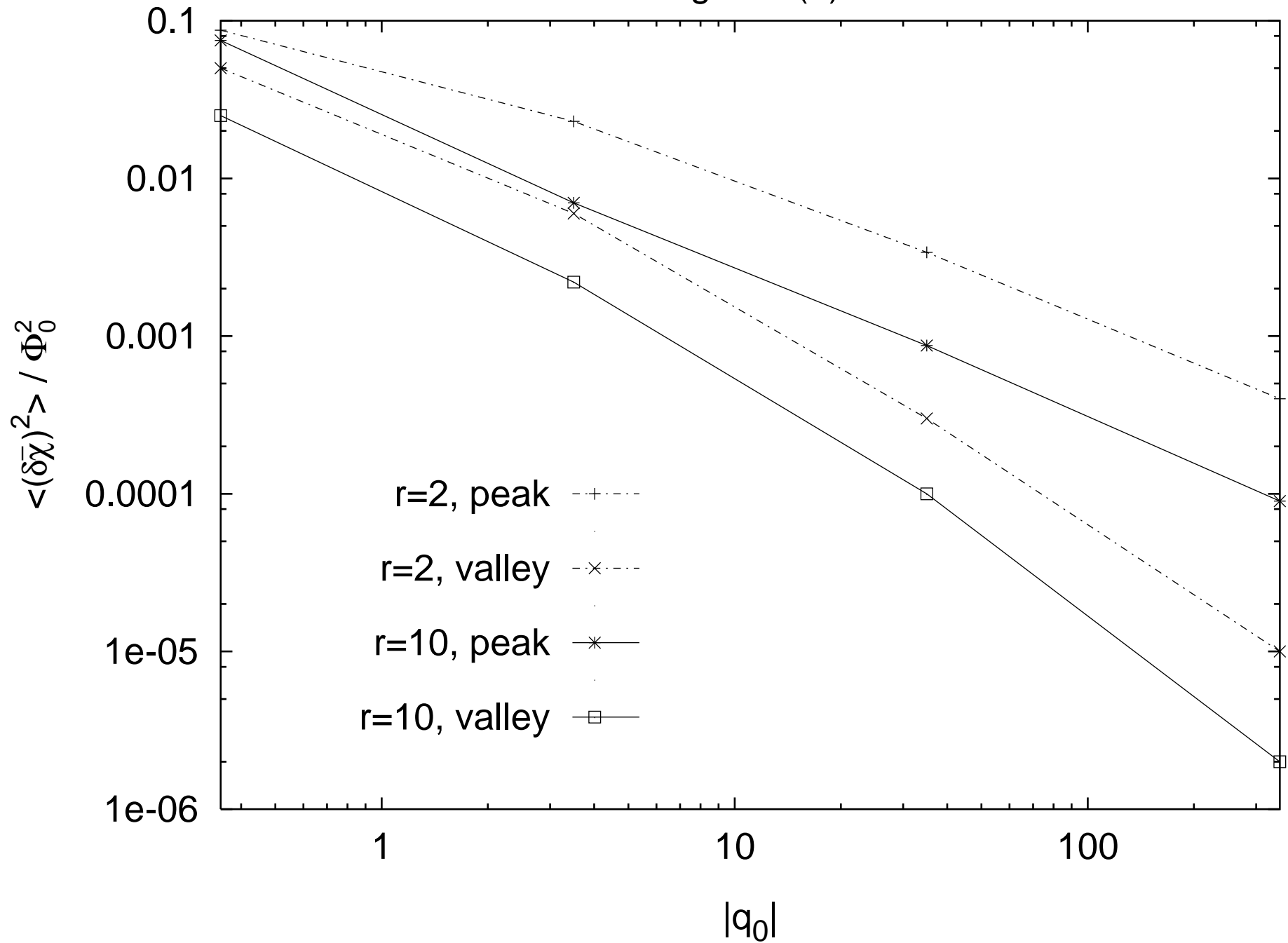
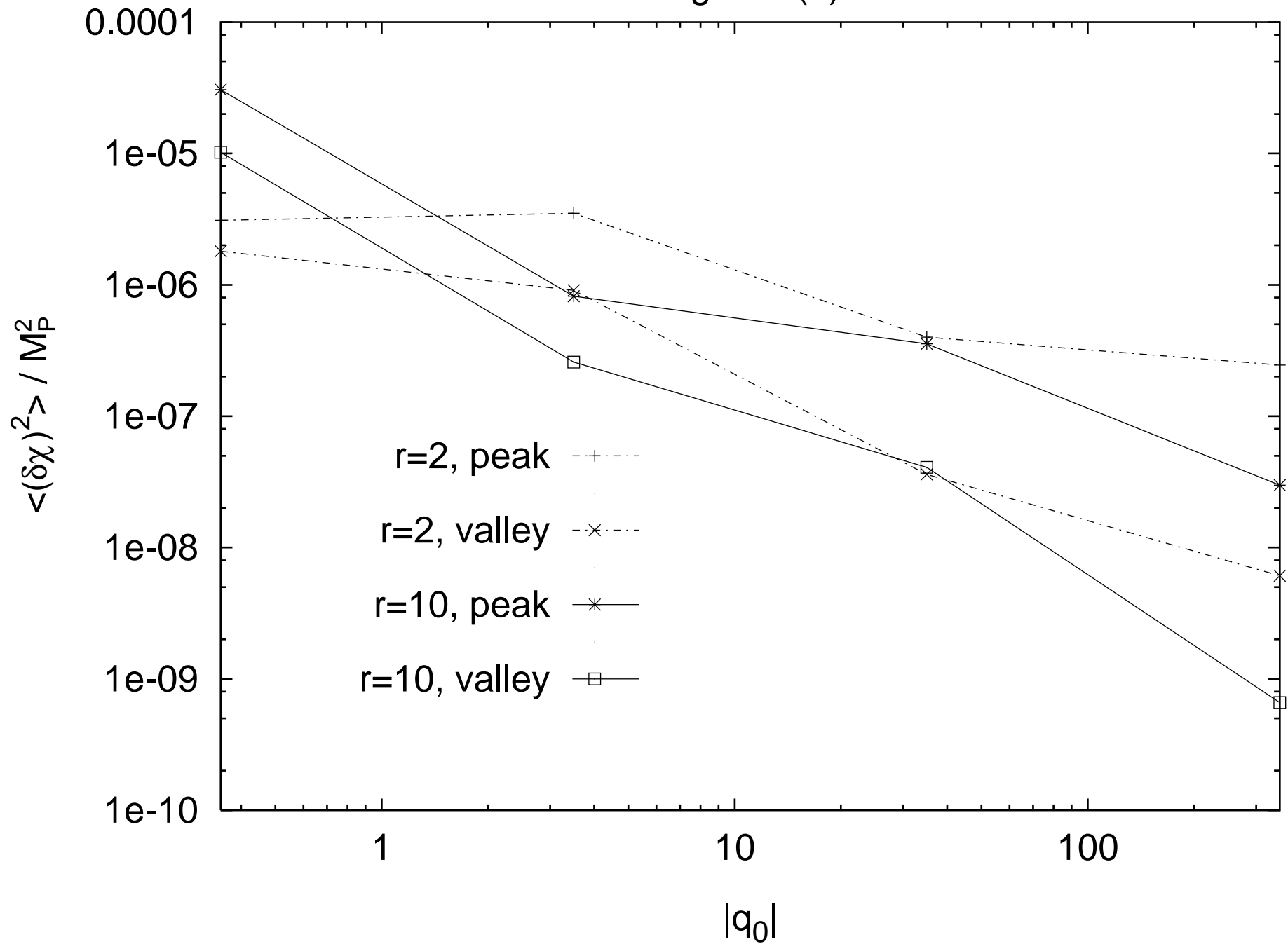
\begin{figure}
\caption{The variances of the fields as a function of time for
positive coupling, $q_0\approx 35$, $r=10$ ($m_\chi = m_\phi = 0$,
$\lambda_\phi=10^{-12}$, $\lambda_\chi=10^{-7}$, $g=10^{-10}$).}
\end{figure} 
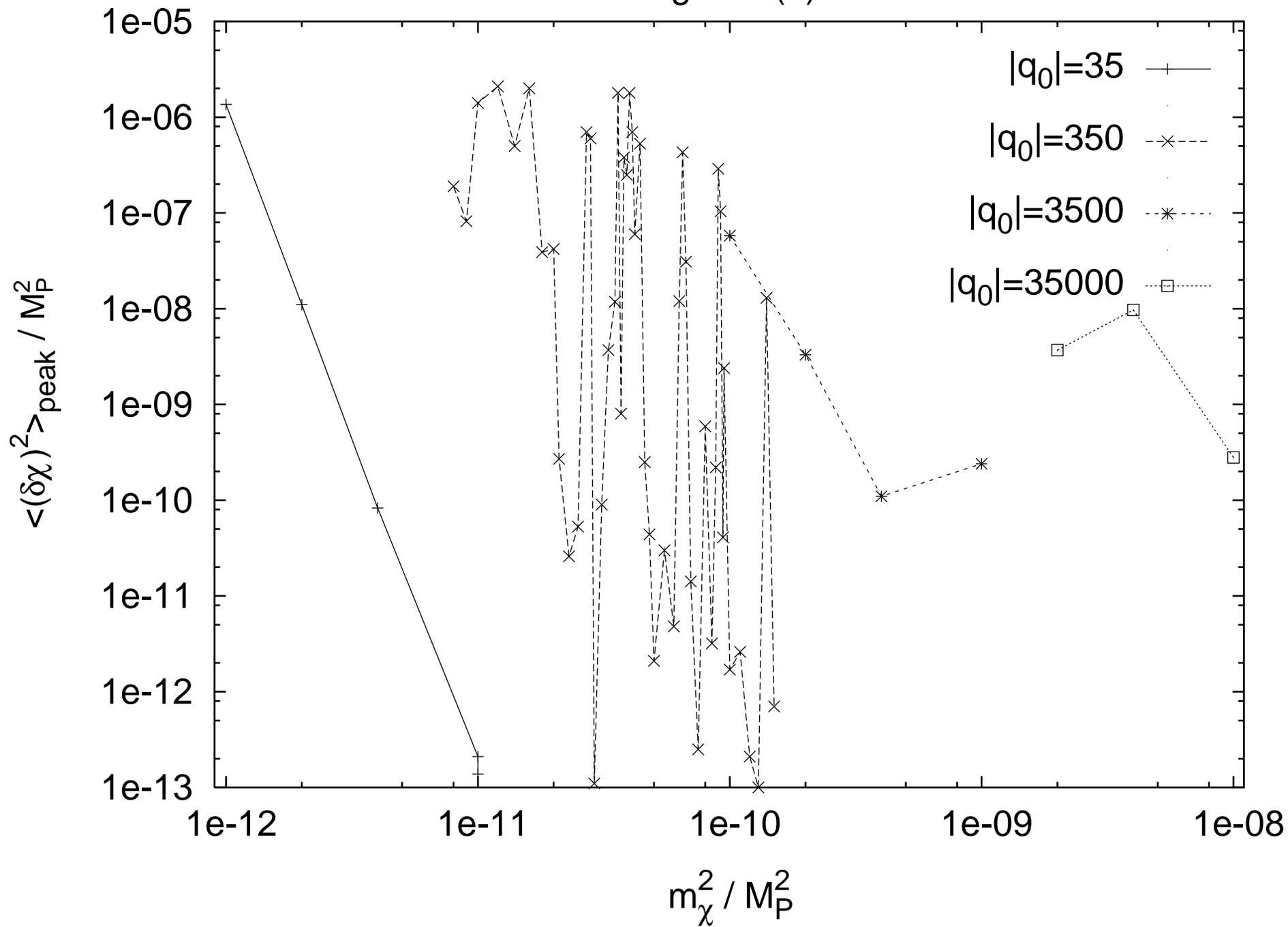
\begin{figure}
\caption{(a): Peak and valley lattice variances at the beginning of
the scattering regime as a function of $|q_0|$ for negative coupling
and two values of $r$.}
\end{figure}
\setcounter{figure}{5}
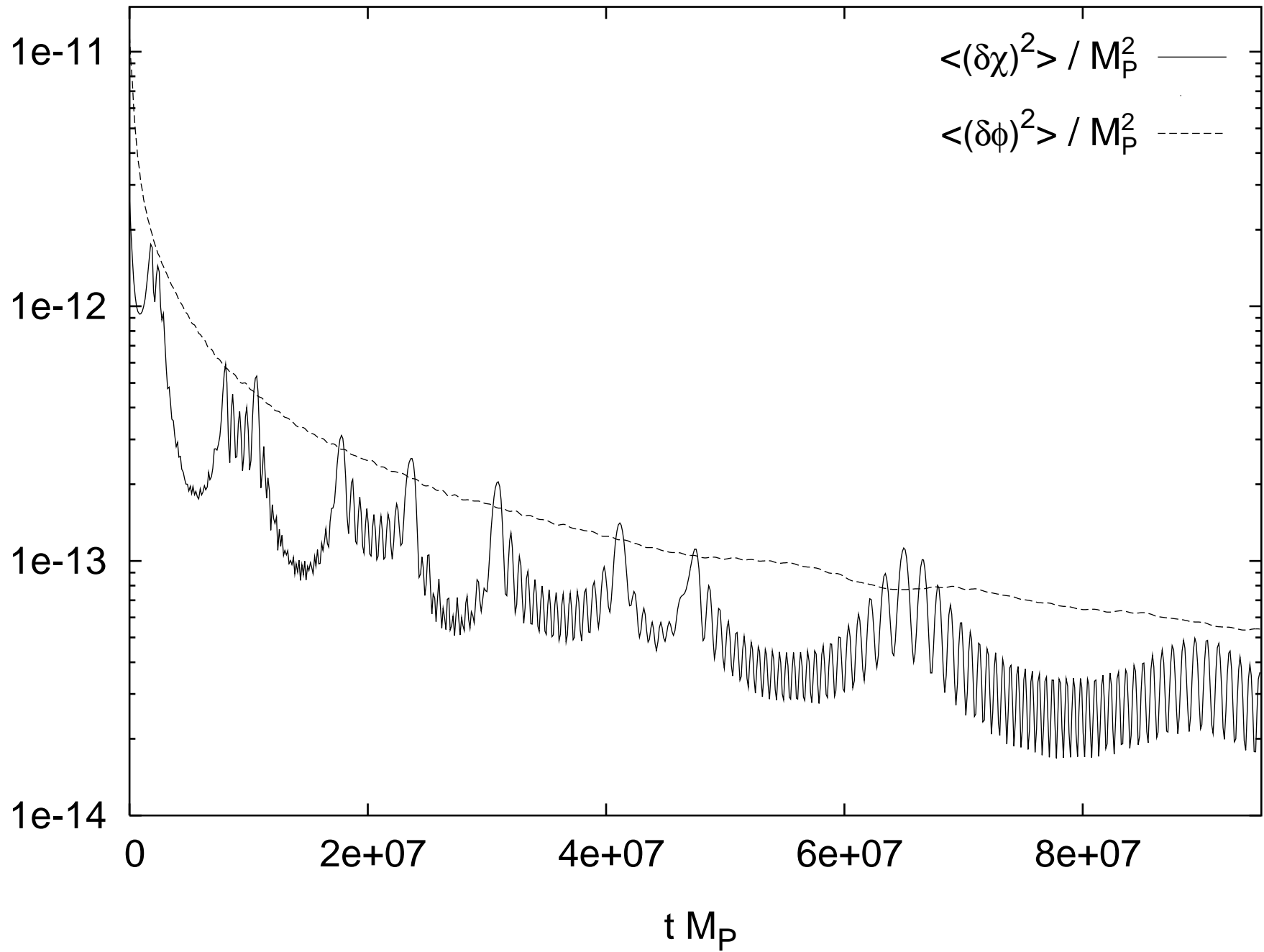
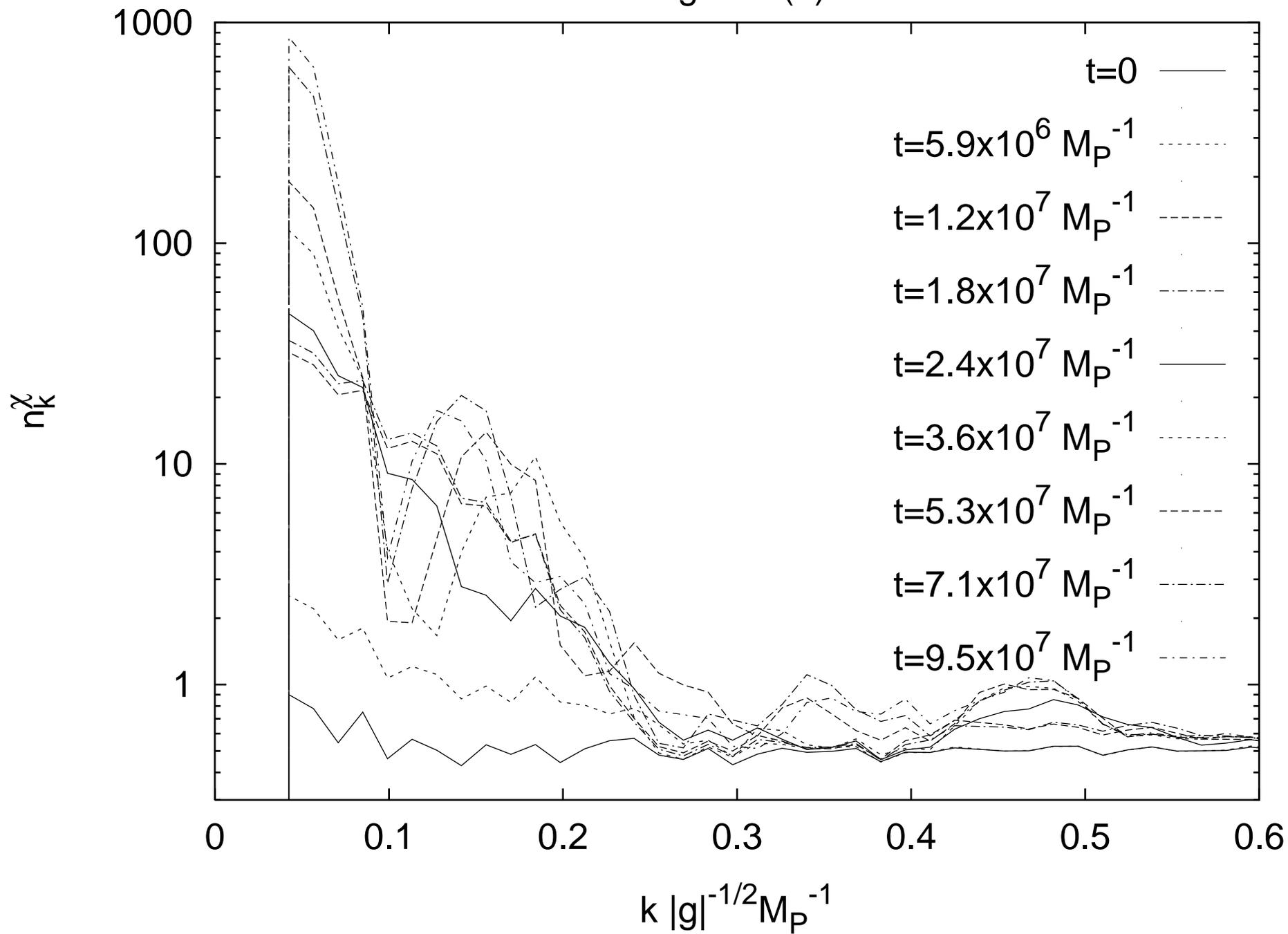
\begin{figure}
\caption{(b): Maximum physical peak and valley variances reached in
the expanding universe as a function of $|q_0|$ for negative coupling
and two values of $r$.}
\end{figure} 
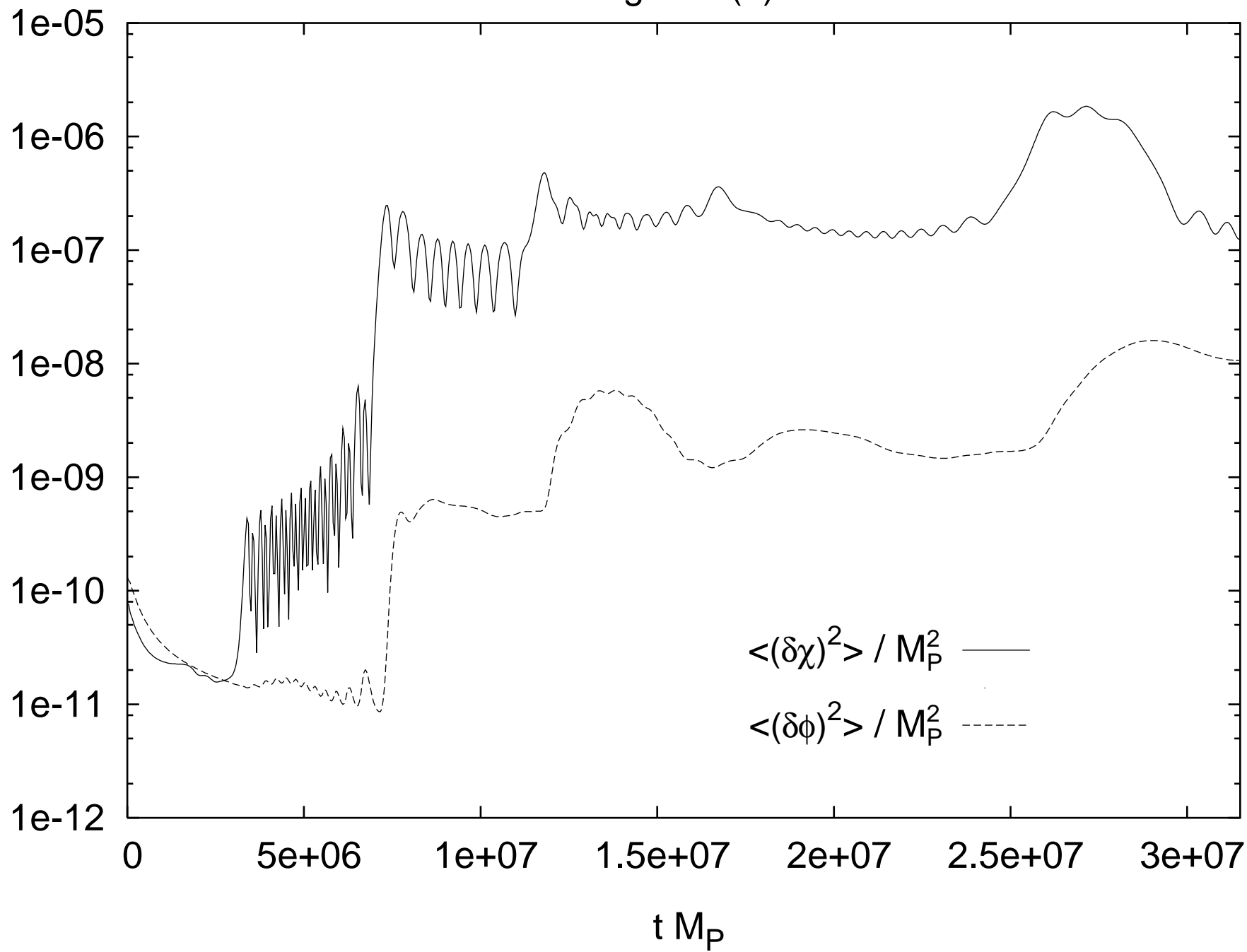
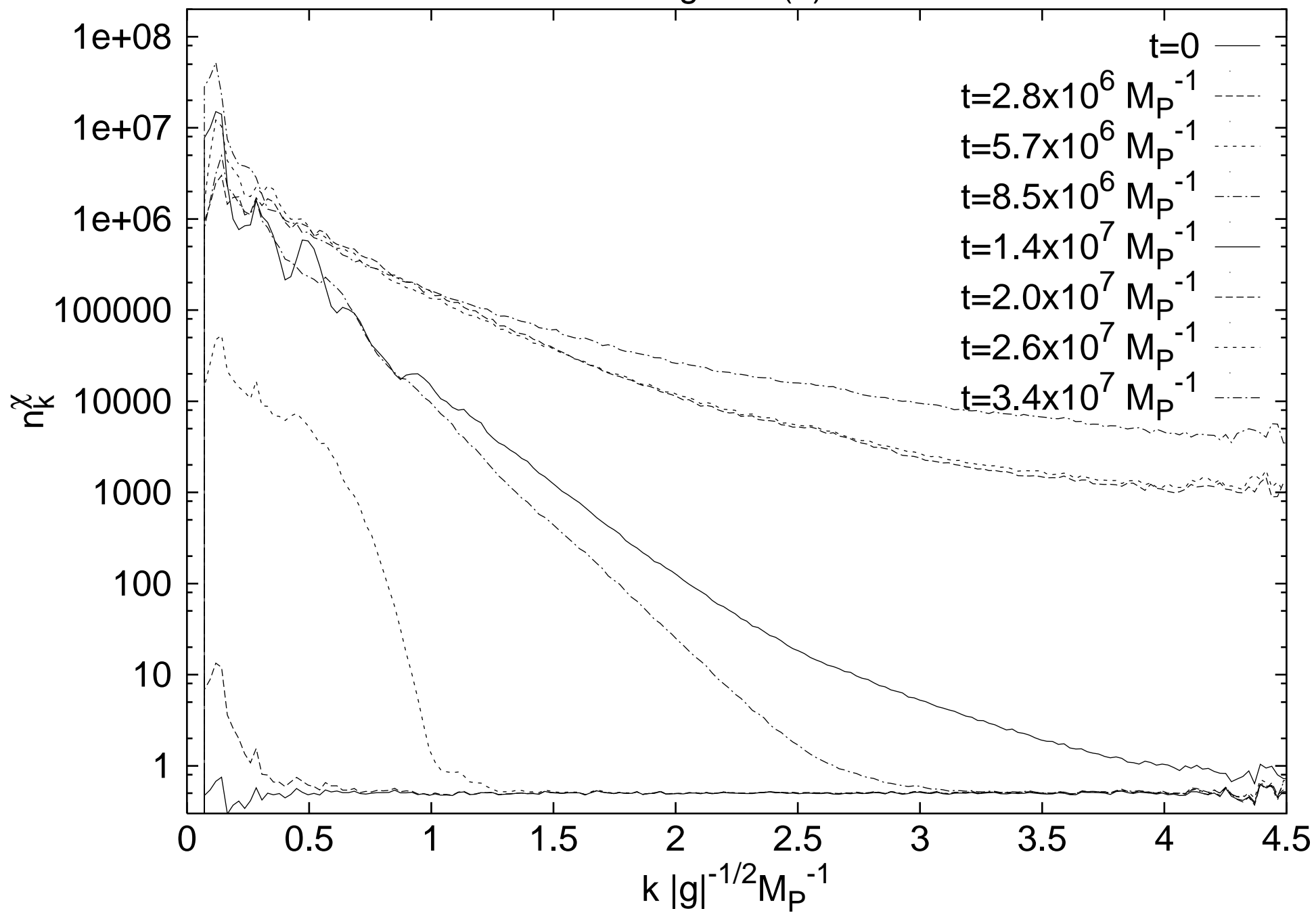
\begin{figure}
\caption{(a): Maximum variances as a function of $m_\chi$ for several
values of $|q_0|$, with $r=10$ and $g<0$. The inflaton is massless
($m_\phi=0$) and $\lambda_\phi=10^{-12}$ throughout.}
\end{figure}
\setcounter{figure}{6}
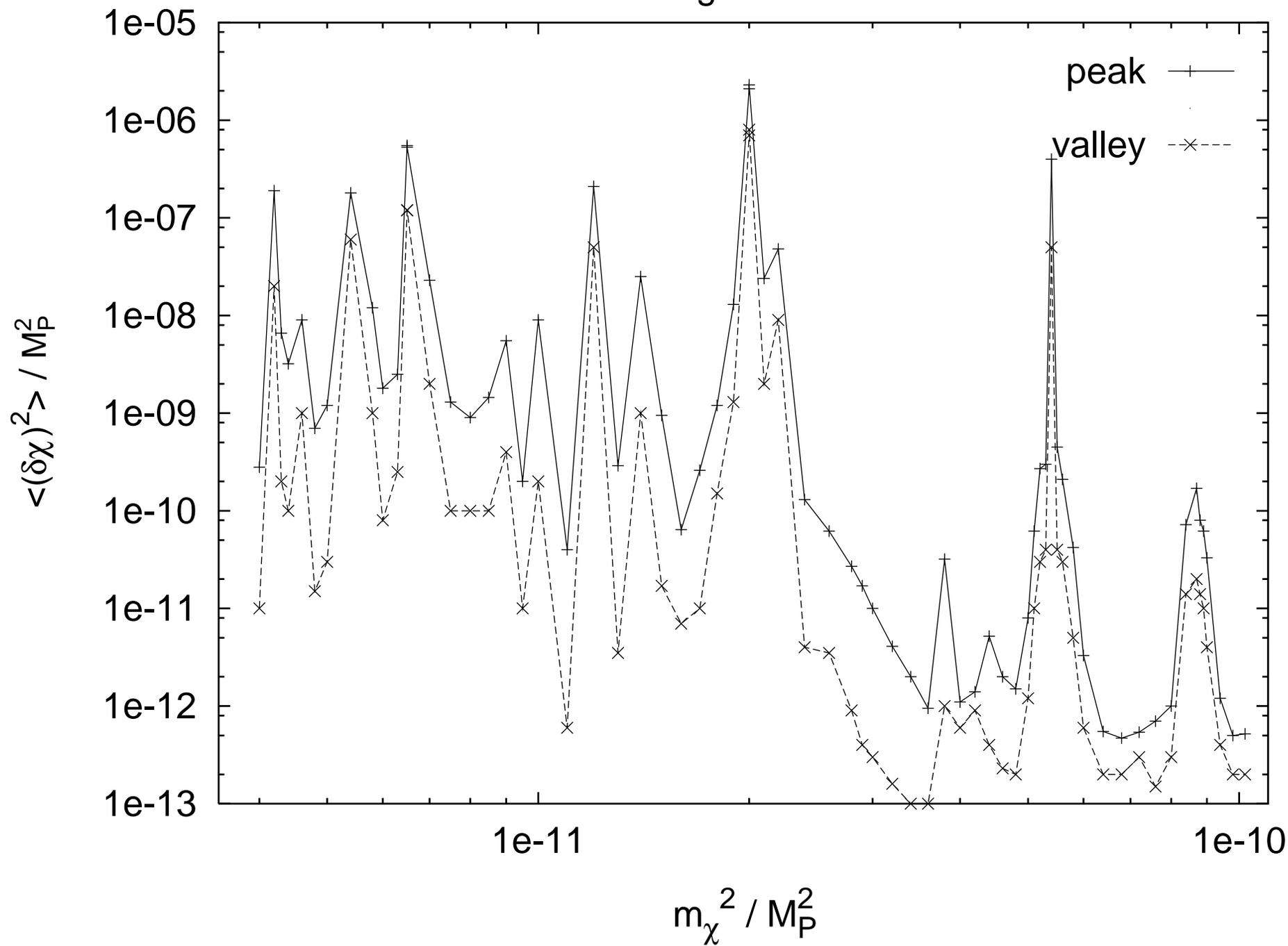
\begin{figure}
\caption{(b): A blow-up of the $q_0=-350$ curve shown in figure~7(a).}
\end{figure} 
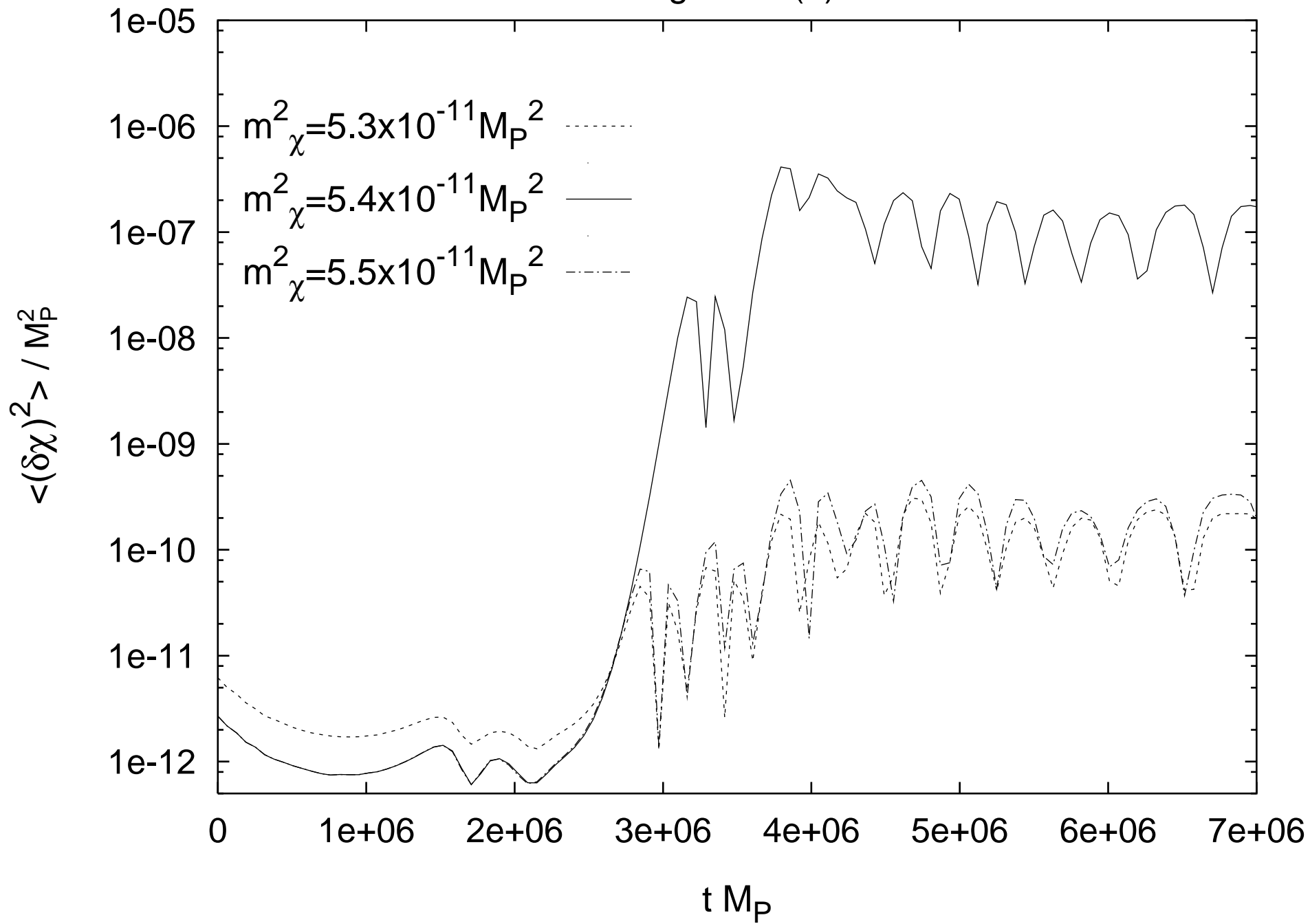
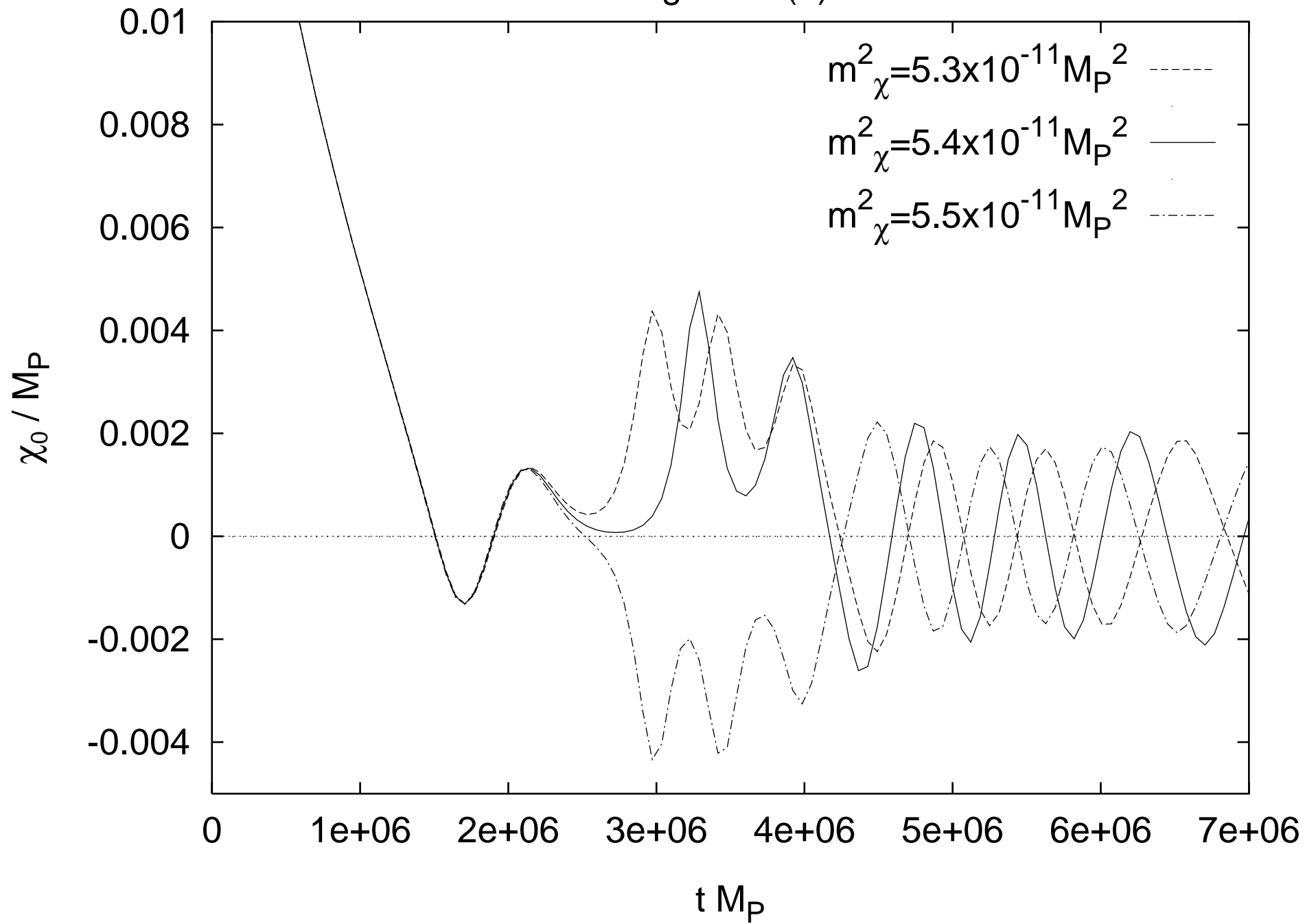
\begin{figure}
\caption{(a): The variances for a set of parameters for which there is
little particle production. $|q_0|=350$, $r=10$ ($m^2_\phi = 0$, $m^2_\chi
= 2.9\times 10^{-11}M^2_{\rm P}$, $\lambda_\phi=10^{-12}$,
$\lambda_\chi=10^{-5}$, $g=-10^{-9}$).}
\end{figure}
\setcounter{figure}{7}
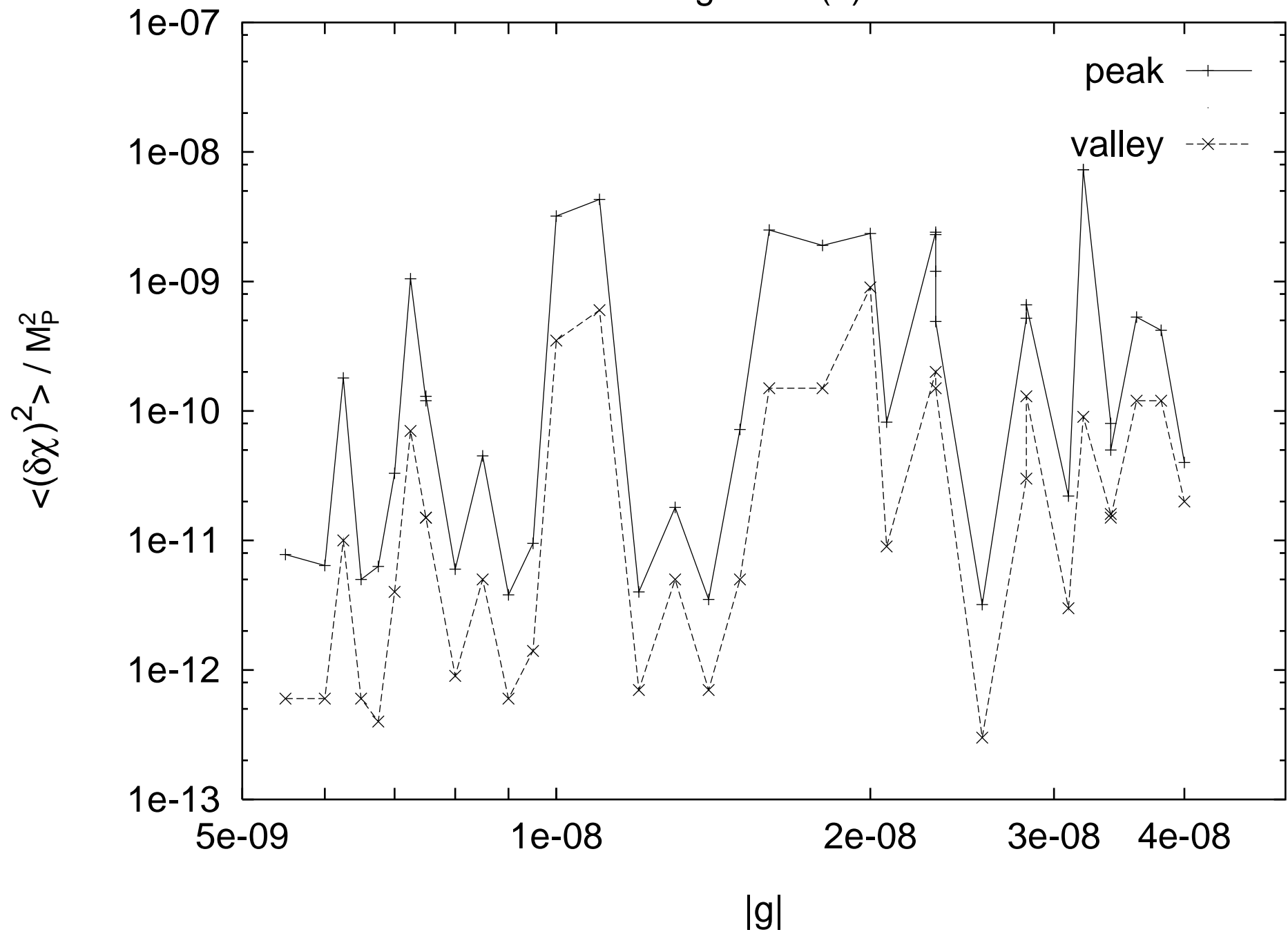
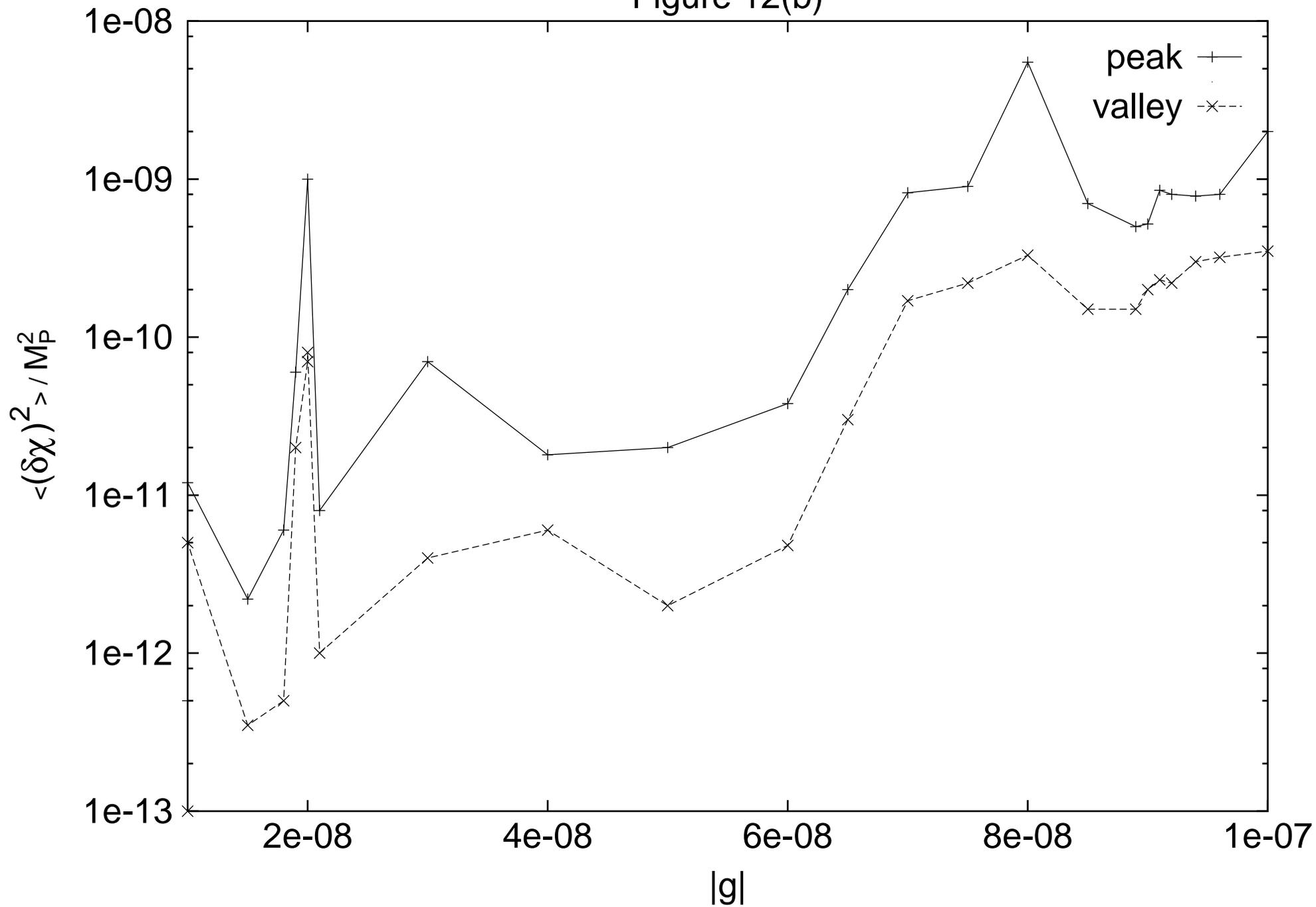
\begin{figure}
\caption{(b): The $\chi$ field occupation numbers at various times for
the run of figure~8(a).}
\end{figure} 
\begin{figure}
\caption{(a): The variances for a set of parameters for which there is
significant particle production. $|q_0|=350$, $r=10$ ($m^2_\phi = 0$, 
$m^2_\chi = 4\times 10^{-11}M^2_{\rm P}$, $\lambda_\phi=10^{-12}$,
$\lambda_\chi=10^{-5}$, $g=-10^{-9}$).}
\end{figure}
\setcounter{figure}{8}
\begin{figure}
\caption{(b): The $\chi$ field occupation numbers at 
various times for
the run of figure~9(a).}
\end{figure} 
\begin{figure}
\caption{Maximum variances as a function of $m_\chi$ for $|q_0|\approx
350$, $r=10$ and $g<0$. The inflaton is massive ($m^2_\phi=7.2\times
10^{-13}M^2_{\rm P}$, $\lambda_\phi=10^{-12}$, $\lambda_\chi=10^{-5}$,
$g=-10^{-9}$).}
\end{figure} 
\begin{figure}
\caption{(a): The $\chi$ field variances for three runs with slightly
different $\chi$ masses ($m^2_\phi=7.2\times 10^{-13}M^2_{\rm P}$,
$\lambda_\phi=10^{-12}$, $\lambda_\chi=10^{-5}$, $g=-10^{-9}$).}
\end{figure}
\setcounter{figure}{10}
\begin{figure}
\caption{(b): The evolution of the $\chi$ field zeromode
$\chi_0$ for the
three runs with slightly different $\chi$ masses shown in figure~11(a)
($m^2_\phi=7.2\times 10^{-13}M^2_{\rm P}$, $\lambda_\phi=10^{-12}$,
$\lambda_\chi=10^{-5}$, $g=-10^{-9}$).}
\end{figure} 
\begin{figure}
\caption{(a): Maximum peak and valley variances as a function of $|g|$
for $m_\chi=10^{14}$GeV. The inflaton is massless ($m_\phi=0$),
$\lambda_\phi=3\times 10^{-13}$, $g<0$ and $\lambda_\chi$ is adjusted
to keep $r=10$.}
\end{figure}
\setcounter{figure}{11}
\begin{figure}
\caption{(b): Maximum peak and valley variances as a function of $|g|$
for $m_\chi=10^{14}$GeV, $m_\phi=10^{13}$GeV, and $\lambda_\phi=3\times
10^{-13}$. Here $\lambda_\chi$ is adjusted to keep $r=10$ and $g<0$.}
\end{figure}


\begin{thebibliography}{99}
\bibitem{oscillation}{This is the case when
$m_\phi^2\gg\lambda_\phi\Phi_0^2$, where $\Phi_0$ is the oscillation
amplitude. In the opposite limit the oscillations are given by
elliptic functions, but for the purposes of this illustration the
difference is immaterial.}
\bibitem{negg mu}{A simple estimate can be obtained 
as follows. Note that $\Omega_\chi^2\equiv A_0-2q_0\cos (2z)$
oscillates in the range $[-4|q_0|,0]$ along the line $A_0=-2|q_0|$.
Eq.~(\ref{eq:mathieu}) then reduces to an inverted harmonic
oscillator with a time dependent frequency.
Naively one can write $\Omega_\chi^2=-2|q_0|$, its average value 
over one oscillation, to obtain $\mu\sim (2|q_0|)^{1/2}$ which
is close to $\mu=(4/\pi)|q_0|^{1/2}$,
the value obtained from the exact expression 
Eq.~(\ref{eq: mu for large A q}).}
\bibitem{ic} The initial conditions for the inhomogeneous modes are
chosen as follows. Each mode with momentum $k$ is treated like a
quantum harmonic oscillator in its ground state. The initial  
amplitudes of the mode and its velocity are then picked randomly from
gaussian distributions with variances proportional to $1/(4\omega_k)$
and $\omega_k/4$, respectively. Here $\omega_k^2=k^2+m^2$, $m^2 =
V^{\prime\prime}(\phi_0,\chi_0)$ (the derivatives are with respect to
the appropriate field), and $\phi_0$ and $\chi_0$ are the initial
values of the field zero modes. See \cite{ProkopecRoos} for details.
\bibitem{why-not chi-not}{
This can be seen as follows. 
Since any inhomogeneities are quickly
redshifted, one can neglect the gradient terms 
in the EOM for the fields during inflation.
In addition the second order time derivatives can be 
neglected during the slow roll. 
In order to investigate what is the deviation of $\chi_0$ from
$\tilde\chi_0$, we write $\chi_0=\tilde\chi_0+\delta\chi$, and
hence $3H\dot{\tilde\chi_0}+(m_\chi^2+g\Phi_0^2+
3\lambda_\chi\tilde\chi_0^2)\delta\chi=0$. 
This leads to $\delta\chi/\tilde\chi_0\simeq 0.1|q|^{-1}$ 
for $r\gg 1$ and $m_\chi\ll (|g|/\lambda_\phi)^{1/2}\omega_\phi$.
Note that this last condition reduces to
$m_\chi\ll |g|^{1/2}\Phi$ for a massless inflaton.
In section~\ref{sec:massiveII} we 
show that particles with mass $m_\chi>|g|^{1/2}\Phi$
cannot be produced, and so are not interesting for the purpose at hand.
}
\bibitem{rho} We define the end of slow roll by the time at which
$\ddot\phi_0 = H\dot\phi_0$, where $\dot\phi_0$ is determined by its
slow roll value $-(3H)^{-1}dV(\phi_0,\tilde\chi_0)/d\phi$.  The ratio
$\phi_0/M_{\rm P}$ is actually dependent on $|g|/\lambda_\chi$, but
for our runs $|g|/\lambda_\chi<1/2$ and the effect on the value of
$\phi_0/M_{\rm P}$ is negligible.
\bibitem{large q_0} We speak of a scattering regime when the resonance
is shut off while most of the energy is still in the inflaton
zeromode. The result is a state with slowly evolving variances and
smooth power spectra characteristic of efficient scattering
\cite{KhlebnikovTkachevI,ProkopecRoos,KhlebnikovTkachevIII}. For
massless fields, positive $g$ and $q_0^{1/2}r < 1$ the scattering
regime is reached when $q_0\gtrsim 25$. For $q_0^{1/2}r \gtrsim 1$ the
condition becomes $q_0r \gtrsim 2$. For negative $g$ the scattering
regime is reached when $|q_0|^{2/3}r>1$. All of these conditions are
derived by comparing the energy in fluctuations at the time the
maximum variances are reached to the energy remaining in the zeromode
of the inflaton. The maximum variances reached are derived in
section~\ref{sec:masslessII}.
\bibitem{mu-decay} We found in \cite{ProkopecRoos} that for the first
few resonances above $A=2|q|$ the maximum $\mu_k$ decays approximately
as $\mu_k\propto \exp {[-\alpha (k^2+m_{\chi\,,{\rm eff}}^2) /\Delta
k_{\rm res}^2]}$, where $\Delta k_{\rm res}^2\simeq
\sqrt{g}\omega_\phi\Phi_0/2 =q^{1/2}\omega_\phi$ is the distance
between the resonances, and $\alpha \simeq 1.1$. The width of the
resonances is also decreasing exponentially. See \cite{yoshiandco} for
analytical estimates.
\bibitem{decaystop} At first sight one may think that the shift
required to kill the first resonance above $A=2|q|$ is given by the
width of this first band and not by its distance to the one above
it. It must be remembered, however, that the parameters that enter $q$
are changing due to the decay (as well as due to the expansion in the
massive case, see section~\ref{sec:massive}), shifting $q$ to smaller
values. This means that the position of the first band constantly
sweeps through a range of order $\sqrt q$ above $A=2q$. The shift
required to shut off this band is hence $\sqrt q$.
\bibitem{t max var} $\Phi_{\rm s}$ is always the amplitude of the
inflaton zeromode oscillations at the time when
$\langle(\delta\chi)^2\rangle$ reaches its maximum in an expanding
universe. As discussed in the text, depending on the parameters $\phi$
can decay into its own fluctuations, so $\langle(\delta\phi)^2\rangle$
may still grow after $\langle(\delta\chi)^2\rangle$ reaches its
maximum.
\bibitem{phi variance} A simple estimate based on a WKB approximation
can be obtained as follows: $\langle(\delta\phi)^2\rangle_{\rm scatt}
=\int [d^3k/((2\pi)^3\omega_k)]n_k\sim
(2\pi^2)^{-1}\omega_\phi^2/\lambda_\phi\simeq
0.05\Phi^2(\tau^\phi_{\rm s})$, where we assumed that the scattering
regime is reached when the occupation numbers $n_k\sim
\lambda_\phi^{-1}$, and that the variance is dominated by momenta of
order $\omega_\phi=c\sqrt\lambda_\phi\Phi$. This estimate is in
reasonable agreement with the maximum variances reached in our
runs. The amplitude $\Phi(\tau)$ at the time $\tau^\phi_{\rm s}$ at
which the maximum variance is reached can be estimated from
$\Phi\propto a^{-1}$, $a=(1+H_0\tau)$, and $n^\phi_{\rm res} \approx
\exp{2\mu_\phi\omega_\phi^0\tau}$. With $n(\tau^\phi_{\rm
s})\sim\lambda_\phi^{-1} $ and $H_0\sim \sqrt\lambda_\phi M_{\rm P}$
one obtains $\Phi(\tau^\phi_{\rm s})\sim 2c\mu_\phi
\Phi_0/\ln(\lambda_\phi^{-1})$. For $\lambda_\phi=10^{-12}$,
$\Phi_0=1.8M_P$, and $\mu_\phi\sim 0.04$ one finds
$\langle(\delta\phi)^2\rangle_{\rm scatt} \sim 10^{-6}M_{\rm P}^2$.
\bibitem{energy in flucts} The fraction of the total energy in $\chi$
fluctuations at the beginning of the scattering regime is of order
$0.2\times q^{-1/2}$ for small $\lambda_\chi$, and of order $0.2\times
g/\lambda_\chi$ for large $\lambda_\chi$. These estimates apply to
both massive and massless particles. For a massless inflaton the last
expression can be rewritten as $\sim 0.1\times (rq)^{-1}$.  In
deriving these expressions, we used the fact that the energy density
in fluctuations and the variance are at the beginning of the
scattering regime dominated by the same momenta, and hence
$\rho_\chi\sim (k_{\rm res}^\chi)^2\langle (\delta \chi)^2\rangle_{\rm
peak}$.
\bibitem{chi 0} In our runs we observe that even after $\chi_0$ stops
following $\tilde\chi_0$ its dynamics is still strongly influenced by
$\phi_0$ and it often oscillates with frequency $n\omega_\phi/2$, where
typically $n=3,4$ or 5.  For $n$ odd, $\chi_0$ develops an expectation
value, indicating symmetry breaking. Here we will not investigate
further the physical implications of this observation.
\bibitem{phi chi equipartition} Equipartition of energy implies that
$\rho_{\delta\phi}\sim\rho_{\delta\chi}$.  The energies can be related
to the variances as follows.  In terms of the occupation numbers, the
variances can be written as: $\langle (\delta\chi)^2\rangle_{\rm
scatt} \sim (2\pi^2)^{-1}n_k^\chi(k\Delta k)_\chi \sim
(2\pi^2)^{-1}n_k^\chi(2\vert q\vert^{1/2}) \omega_\phi^2$ and $\langle
(\delta\phi)^2\rangle_{\rm scatt} \sim (2\pi^2)^{-1}n_k^\phi(k\Delta
k)_\phi \sim (2\pi^2)^{-1}n_k^\phi\omega_\phi^2$, where we have
assumed that the variances are dominated by the momenta of order
$k_\chi\sim\Delta k_\chi\sim k_{\rm res}\simeq
(4|q|)^{1/4}\omega_\phi$ and $k_\phi\sim\Delta k_\phi\sim
\omega_\phi$, respectively. At early stages of the scattering regime
the same momenta dominate the energy densities, so we can write
$\rho_\chi\sim (2\pi^2)^{-1}n_k^\chi(k^3\Delta k)_\chi \sim
(2\pi^2)^{-1}n_k^\chi(4\vert q\vert)\omega_\phi^4\sim 2\vert
q\vert^{1/2}\omega_\phi^2 \langle (\delta\chi)^2\rangle$ and
$\rho_\phi\sim (2\pi^2)^{-1}n_k^\phi(k^3\Delta k)_\phi \sim
(2\pi^2)^{-1}n_k^\phi\omega_\phi^4\sim \omega_\phi^2 \langle
(\delta\phi)^2\rangle$.  This then leads immediately to $ \langle
(\delta\phi)^2\rangle\sim 2\vert q\vert^{1/2} \langle
(\delta\chi)^2\rangle$, implying Eq.~(\ref{eq:variance V}).
\bibitem{phi chi resonance} The maximum unstable momentum of the
$\chi$ field is approximately $k_{\rm res}^\chi\sim \omega_\phi q^{1/2}$
in the negative $g$ case, while $k_{\rm res}^\phi\sim
\omega_\phi$. The ratio hence goes like $q^{1/2}$, which requires
excellent IR resolution in order to capture both accurately. This must
be combined with sufficient ultraviolet (UV) range to accommodate the
created particles and model their scattering towards higher
momenta. Since the $\chi$ particles scatter more efficiently for
larger $r$ the UV cutoff must be increased as $r$ is increased.
\bibitem{n_scatt} Our numerical results indicate that the dependence
of $n_{\rm scatt}$ on the parameters is actually more complicated, but
the expressions given in the text are roughly correct. Since the
dependence of $\tau_{\rm s}$ on $n_{\rm scatt}$ is anyway only logarithmic
and since as discussed in the text there is significant variation in
the relevant value of $\mu$ for small changes in the parameters the
uncertainty in the value of $n_{\rm scatt}$ need not concern us here.
\bibitem{stop follow} When $g\phi_0^2$ is large, $\chi_0$ is trapped
in the deep minimum at $\tilde\chi_0$. However, as $\phi_0$ approaches
zero it looses its grip on $\chi_0$ and $\chi_0$ wants to oscillate
about the origin with frequency $\omega_\chi \approx
(3\lambda_\chi\langle(\delta\chi)^2\rangle)^{1/2}$. In the beginning
$\omega_\chi$ is very small and $\chi_0$ does not have time to
complete a significant portion of an oscillation in the interval when
$\phi_0$ is small (more precisely, when
$g\phi_0^2<3\lambda_\chi\langle(\delta\chi)^2\rangle$). It is easy to
show that when $3\lambda_\chi\langle(\delta\chi)^2\rangle \approx
2g^{1/2}\omega_\phi\Phi$ the $\chi$ zeromode has enough time to
complete half an oscillation. It is then completely out of phase with
$\tilde\chi_0$ when $\phi_0$ becomes large again, and consequently
stops following its minimum. Note that even before this complete phase
mismatch occurs there are small mismatches each time $\phi_0$ grows
after crossing zero, once $\langle(\delta\chi)^2\rangle$ becomes
appreciable. This is the reason why the $\chi$ variance can grow
larger than its positive $g$ value given in
Eq.~(\ref{eq:variance Ib}) before $\chi_0$ stops following
$\tilde\chi_0$.  We point out that the mechanism of $\chi_0$ falling
out of phase with $\tilde\chi_0$ will play an important role when we
discuss the massive $\chi$ case in section~\ref{sec:massive}. As we
will see, in that case the effects can be even more dramatic.
\bibitem{tau massless} As we will see in section~\ref{sec:massive},
this is no longer correct when the $\chi$ field is massive.
\bibitem{random phase} This ``random phase'' approximation is clearly
only sensible if there is enough time between matchings for the
solution to ``oscillate''. This requires that the $n^{\rm th}$ time
interval $\Delta t_n$ between matchings satisfies $\Delta t_n\gtrsim
1/\omega_n = {\rm e}^{n\epsilon}/\omega_{\rm max}$. The minimum total
time required is then $T=\sum_{n=0}^{N-1} \Delta t_n = \omega_{\rm
max}^{-1}[(\omega_{\rm max}/\omega_{\rm min}-1)/({\rm
e}^{\epsilon}-1)]$. In the text we are interested in the limit when
$\epsilon$ is small and the ratio of frequencies is large, so
$T\simeq (\epsilon\omega_{\rm min})^{-1}$. Now from Eq.~(\ref{eq:freq toy})
we found $\omega_{\rm min}^2\sim q^{1/2}\omega_\phi^2$, and the total
time for the frequency to change between $\omega_{\rm max}$ and
$\omega_{\rm min}$ is a quarter period of $\phi_0$,
$T=\pi/(2\omega_\phi)$. This yields $\epsilon \sim q^{-1/4}$,
and so the approximation in the text is reasonable for large $q$.
\bibitem{quantum var} We point out that for the run in figure~8 the
production is so low that the $\chi$ variance barely rises above the
level of background fluctuations as set by the initial
conditions. This contribution is proportional to the ultraviolet
cutoff and therefore has no physical relevance. It scales with the
expansion of the Universe as $a^{-2}$ since it is dominated by the
relativistic $UV$ modes. For typical runs we present,
the background variance is negligible in comparison to the variance
from resonant production. There are, however, a few runs with very
little production, and for these the variance we record on
the figures is dominated by the background fluctuations. Examples are
a few of the lowest points on figures 7, 10, and 12 with typical
values of order $10^{-13} - 10^{-12}M_{\rm P}^2$.
\bibitem{planckmass} Note that reference \cite{KhlebnikovTkachevIII}
uses $M_{\rm Pl}\equiv G^{-1/2}$ while in this paper $M_{\rm P}\equiv
(8\pi G)^{-1/2} \simeq 2.4\times 10^{18}\hbox{GeV}$.
\bibitem{fig12} The reader may have noticed that initially the
variances of the $m_\chi^2=5.4\times10^{-11}M_{\rm P}^2$ and
$5.5\times10^{-11}M_{\rm P}^2$ runs overlap almost perfectly while the
variance of the $5.3\times10^{-11}M_{\rm P}^2$ run is slightly
larger. This is simply due to the fact that the lattice spacing, and
hence the ultraviolet cutoff of the theory, is slightly different for
the latter run.
\bibitem{prob} Since we are interested in large growth factors
for which the probability is small we can replace $1-(1-{\cal
P}_\epsilon)^n$ with $n{\cal P}_\epsilon$. We will also not
distinguish between growing ``at least once'' {\it vs.\/} growing
``exactly once'', {\it etc.\/}
\bibitem{pos g production} This is easily established by comparing the
maximum positive $g$ resonant momentum $k_{\rm res}^{\rm phys}(g>0)\approx
q^{1/4}m_\phi$ to $m_\chi$ at the time of the first zero crossing of
the inflaton zeromode $\phi_0(t)$ (which is the first time that
particles could be produced in the positive $g$ case).
\bibitem{chaoticBgenesis} We are assuming here the same chaotic
inflationary framework as in the rest of the paper.
\bibitem{reheat}
In the old theory the reheat temperature is of order $T_{\rm r}\sim
\sqrt{\Gamma M_{\rm p}}$, where $\Gamma$ is the total decay rate of
the inflaton. Imposing naturalness conditions on the couplings of the
inflaton to other fields in order to stabilize the small value
$\lambda_\phi\approx 3\times10^{-13}$ against radiative corrections
then implies $T_{\rm r}\lesssim 10^{10}$GeV.
\bibitem{capitalM}
We will use $M_\chi$ from now on for the mass of the GUT boson. This is
to remind the reader that the field in question is not quite the real
scalar field $\chi$ with mass $m_\chi$ used in our numerical
simulations.
\bibitem{B massless} The analysis is equally simple for the massive
case, but as discussed previously we prefer the massless inflaton
because a $\phi^4$ term is anyway required if $g<0$ and setting
$m^2_\phi \sim \lambda_\phi\Phi_0^2$, as we have done by choosing the
COBE values for both parameters, actually corresponds to mild
tunning. Additionally, unlike Eq.~(\ref{eq: baryon-to-entropy ratio
IV}), Eq.~(\ref{eq: baryon-to-entropy ratio II}) has the nice feature
that $n_B/s$ is independent of the unknown decay rates $\Gamma_\phi$
and $\Gamma_\chi$.
\end{thebibliography}
\end{document}